\documentclass[fleqn,usenatbib]{mnras}

\usepackage{newtxtext,newtxmath}
\usepackage[T1]{fontenc}
\usepackage{ae,aecompl}
\usepackage{lscape}
\usepackage{rotating}
\usepackage{multirow,multicol}
\usepackage{longtable}

\usepackage{graphicx}	% Including figure files
\usepackage{amsmath}	% Advanced maths commands
\usepackage{amssymb}	% Extra maths symbols

\providecommand{\e}[1]{\ensuremath{\times 10^{#1}}}
\newcommand{\Msun}{\ensuremath{M_{\odot}}}

\title[Total IR light from clusters]{Measuring the total infrared light from galaxy clusters at z=0.5--1.6: connecting stellar populations to dusty star formation}

\author[Alberts et al.]{Stacey Alberts$^{1}$\thanks{E-mail: \href{mailto:salberts@arizona.edu}{salberts@arizona.edu}},
Kyoung-Soo Lee$^{2}$,
Alexandra Pope$^{3}$, 
Mark Brodwin$^{4}$, 
\newauthor Yi-Kuan Chiang$^{5}$,
Jed McKinney$^{3}$,
Rui Xue$^{6}$,
Yun Huang$^{2}$, 
Michael Brown$^{7}$,
\newauthor Arjun Dey$^{8}$, 
Peter R. M. Eisenhardt$^{9}$,
Buell T. Jannuzi$^{1}$, 
Roxana Popescu$^{3}$,
\newauthor Vandana Ramakrishnan$^{2}$,
Spencer A. Stanford$^{10}$,
Benjamin J. Weiner$^{11}$
\\
$^{1}$Steward Observatory, 
University of Arizona, 
933 N. Cherry
Tucson, AZ 85721 USA\\
$^{2}$Department of Physics and Astronomy, 
Purdue University, 
525 Northwestern Avenue, 
West Lafayette, IN 47907, USA\\
$^{3}$Department of Astronomy, University of Massachusetts, 710 North Pleasant Street Amherst, MA 01003, USA\\
$^{4}$Department of Physics and Astronomy, University of Missouri, 5110 Rockhill Road, Kansas City, MO 64110, USA\\
$^{5}$Center for Cosmology and AstroParticle Physics (CCAPP), The Ohio State University, Columbus, OH 43210, USA\\
$^{6}$Department of Physics and Astronomy,  The University of Iowa, 203 Van Allen Hall, Iowa City, IA 52242, USA\\
$^{7}$School of Physics \& Astronomy, Monash University, Clayton, Victoria 3800, Australia\\
$^{8}$NSF's National Optical-Infrared Astronomy Research Laboratory, 950 N. Cherry Ave., Tucson, AZ 85719, USA\\
$^{9}$Jet Propulsion Laboratory, California Institute of Technology, Pasadena, CA 91109, USA\\
$^{10}$Department of Physics, University of California, Davis, One Shields Avenue, Davis, CA 95616, USA\\
$^{11}$MMT/Steward Observatory, 933 N. Cherry St., University of Arizona, Tucson, AZ 85721, USA
}

\date{Accepted XXX. Received YYY; in original form ZZZ}

\pubyear{2020}

\begin{document}
\label{firstpage}
\pagerange{\pageref{firstpage}--\pageref{lastpage}}
\maketitle

\begin{abstract}
Massive galaxy clusters undergo strong evolution from $z\sim1.6$ to $z\sim0.5$, with overdense environments at high-$z$ characterized by abundant dust-obscured star formation and stellar mass growth which rapidly give way to widespread quenching.  Data spanning the near- to far- infrared (IR) spectrum can directly trace this transformation; however, such studies have largely been limited to the massive galaxy end of cluster populations.  In this work, we present ``total light" stacking techniques spanning $3.4-500\,\mu$m aimed at revealing the total cluster infrared emission, including low mass members and any potential intracluster dust.  We detail our procedures for {\it WISE}, {\it Spitzer}, and {\it Herschel} imaging, including corrections to recover the total stacked emission in the case of high fractions of detected galaxies.  We apply our stacking techniques to 232 well-studied massive (log $M_{200}/\Msun\sim13.8$) clusters across multiple redshift bins, recovering extended cluster emission at all wavelengths, typically at $>5\sigma$.  We measure the averaged near- to far-IR radial profiles and spectral energy distributions (SEDs), quantifying the total stellar and  dust content.  The near-IR radial profiles are well described by an NFW model with a high ($c\sim7$) concentration parameter.  Dust emission is similarly concentrated, albeit suppressed at small radii ($r<0.2\,$Mpc).  The measured SEDs lack warm dust, consistent with the colder SEDs expected for low mass galaxies.  We derive total stellar masses consistent with the theoretical $M_{\rm halo}-M_{\star}$ relation and specific-star formation rates that evolve strongly with redshift, echoing that of massive (log$\,M_{\star}/\Msun\gtrsim10$) cluster galaxies.  Separating out the massive galaxy population reveals that the majority of cluster far-IR emission ($\sim70-80\%$) is provided by the low mass constituents, which differs from field galaxies.  This effect may be a combination of mass-dependent quenching and excess dust in low mass cluster galaxies.

\end{abstract}

\begin{keywords}
galaxies:clusters -- infrared:galaxies -- galaxies:evolution -- galaxies:star formation -- techniques:stacking
\end{keywords}

%%%%%%%%%%%%%%%%%%%%%%%%%%%%%%%%%%%%%%%%%%%%%%%%%%

%%%%%%%%%%%%%%%%% BODY OF PAPER %%%%%%%%%%%%%%%%%%

\section{Introduction}

Local environment plays a fundamental role in shaping galaxy properties, an effect which can be studied in its extremes in massive galaxy clusters.  In the high density cluster environments in the nearby Universe, galaxy properties differe from those of their counterparts in the field: clusters host more massive galaxies \citep{kau04, col09, vdb13}, with a strong preference for lower star formation rates \citep[SFRs;][]{bal98, lew02, pen10} and early-type morphologies \citep[i.e.][]{dre80}. These differences, well established by a wealth of detailed studies in the local and low redshift Universe, have prompted the often challenging task of pushing into the high redshift cluster and proto-cluster regimes.  This is necessary in order to construct an evolutionary picture of environmental effects on galaxy populations, which ties into the broader question of large scale structure formation.

Optical and near-infrared (IR) observations of clusters beyond the local Universe have found that cluster populations have colors, luminosity functions, and stellar ages consistent with a model of very early ($z\gg2-3$), vigorous stellar mass growth followed by largely passive evolution \citep[i.e.][]{sta98, bla06, vanD07, eis08, mei09, man10}.  Mid- and far-IR studies have shown this picture to be incomplete, however, with many intermediate redshift ($z\sim1-2$) massive clusters hosting heavily obscured star forming galaxies (SFGs), indicating a deviation from the local SFR-density relation at this epoch \citep[i.e.][]{coo06, tra10, hil10, hay11, fas11, fas14, tad11, zei13, bro13, alb14, bay14, san14, san15, ma15, alb16}. Deep {\it Spitzer Space Telescope} and {\it Herschel Space Observatory}\footnote{{\it Herschel} is an ESA space observatory with science instruments provided by European-led Principal Investigator consortia and with important participation from NASA.} observations of rich $z\sim1-2$ clusters find (dust-obscured) galaxies with SFRs and specific-SFRs (SSFRs) comparable to the field down into the cluster cores at $z\gtrsim1.4$ \citep{bro13, alb14, alb16} followed by rapid evolution which establishes the largely quenched cluster populations at $z \lesssim 1$ \citep{muz08}. Combined with studies of coeval quenched cluster populations \citep[i.e.][]{nan17, cha19}, it has emerged that $z\sim1-2$ is a transitional era for rich, massive clusters from abundant (obscured) star formation and AGN activity \citep{mar13, alb16} to widespread quenching.

The importance of obscured activity at intermediate redshifts extends organically into the proto-cluster regime, where there is increasing evidence that overdensities of dusty star forming galaxies \citep[DSFGs;][]{cas14} are often the signposts of  proto-cluster candidates \citep{cas16}.  These early, potentially coeval DSFGs \citep{kat16, cle16, cas16, ume15, ume17, ote18, gre18, lew18, arr18, lac19, har19, che19} are natural candidates for the precursors to the massive end of later cluster populations, particularly massive ellipticals \citep[i.e.][]{hop08}.

Taken together, these high redshift cluster and proto-cluster studies indicate that infrared-emitting dust is clearly an important tool for studying the evolution of (proto-)clusters. However, IR studies $-$ and cluster galaxy studies in general $-$ have been mostly confined to looking at individual members at the bright, massive end of the cluster populations.  A full accounting of the dust emission in (proto-)clusters requires a different approach to constrain the faint contributors: low mass cluster members and potential emission from intracluster dust \citep[ICD;][]{dwe90} embedded in the hot intracluster medium (ICM).   Statistical methods that average multiple clusters through stacking \citep{dol06} have been used on local and low redshift samples to characterize total cluster properties and place upper limits on the infrared emission from ICD \citep{kel90, mon05, gia08}.  Recently, this technique has been expanded to the proto-cluster regime: \citet{Planck15} obtained targeted {\it Herschel}/SPIRE follow-up and stacked 220 cluster candidates at $2 \lesssim z \lesssim 4$ from the {\it Planck} catalog of compact sources \citep[PCCS;][]{Planck2014}, finding a strong detection of extended far-IR emission on the spatial scales associated with proto-clusters \citep{chi13}.  \citet{kub19} stacked 179 proto-cluster candidates at $z\sim3.8$ selected from the Hyper Suprime-Cam Subaru Strategic Program \citep[HSC-SSP;][]{air18} at multiple wavelengths, including imaging from the {\it Wide-Field Infrared Survey Explorer} {\it (WISE)}, {\it Herschel}, and {\it Planck}.  This study revealed intense star-forming environments with warm stacked spectral energy distributions (SEDs) suggestive of AGN activity.

%The \citet{Planck15} and \citet{kub19} 
The \citet{Planck15} and \citet{kub19} results demonstrate the power of statistical stacking analyses for studying total cluster infrared emission up to the high redshift, proto-cluster regime.  In this work, we develop multi-wavelength stacking procedures to probe the ``total light" in clusters and proto-clusters for wide-field datasets such as the all-sky {\it WISE} and wide-area {\it Herschel} SPIRE surveys.  Here ``total light" refers to the (background subtracted) summation of all light in a sample of clusters without the identification of individual constituents.  Stacking near-IR to far-IR datasets allows us to probe both the existing stellar content and ongoing mass growth over a range of redshifts and halo masses, constraining the dust content, SEDs, and mass assembly from low redshift clusters to clusters at cosmic noon to proto-clusters at $z\gtrsim2$.  In this work, we test our techniques on a well-studied massive cluster sample at $z\sim0.5-1.6$ in the Bo\"{o}tes field \citep{eis08}, taking advantage of ancillary data, and providing the first analysis of the total stellar content and dust emission in massive clusters into the era of active star formation at $z\sim1-2$.

\S~\ref{sec:data} describes our cluster sample and the datasets used in this work.  In \S~\ref{sec:method}, we describe our stacking techniques, including image preprocessing and photometry.  We additionally discuss the applicability of our technique to other (proto-)cluster samples, presenting a correction factor for datasets that lack cluster member information.  \S~\ref{sec:results} tests our stacking technique on clusters in the Bo\"{o}tes field, analyzing the average radial profiles and total photometry in multiple redshift bins. We further build ``total light" SEDs and compare the total output to the output from massive cluster members only.  \S~\ref{sec:disc} discusses our results and \S~\ref{sec:conc} presents our conclusions.  Throughout this work, we assume a WMAP9 cosmology ($\Omega_{M}, \Omega_{\Lambda}, h$) = (0.28, 0.72, 0.70) \citep{hin13} and a \citet{cha03} IMF.

\section{Data} \label{sec:data}

\subsection{Cluster sample} \label{sec:sample}

The IRAC Shallow Cluster Survey \citep[ISCS;][]{eis08} contains over 300 galaxy cluster candidates from $0.1<z<2$, with $>100$ at $z>1$, in the B\"{o}otes field ($\alpha,\delta$ = 14:32:05.7,+34:16:47.5).  Cluster identification was performed using a wavelet search algorithm to find galaxy overdensities in the rest-frame near-infrared in three dimensional space (RA, Dec, photometric redshift) using flux-limited 4.5$\,\mu$m imaging from the IRAC Shallow Survey \citep[][]{eis04} and full photometric redshift probability distribution functions derived from combined deep optical $B_WRI$ from the NOAO Deep Wide-Field survey \citep[NDWFS;][]{jan99} and IRAC 3.6 and 4.5$\,\mu$m imaging  \citep{bro06}.  Due to the flux-limited nature of the IRAC Shallow Survey \citep[8.8$\,\mu$Jy, $5\sigma$, at 4.5$\,\mu$m;][]{eis04}, this cluster catalog is approximately stellar mass selected.  Cluster centers are adopted from the centroid of the wavelet detection algorithm and thus trace the distribution of massive galaxies.

\begin{figure}
	\includegraphics[width=3in]{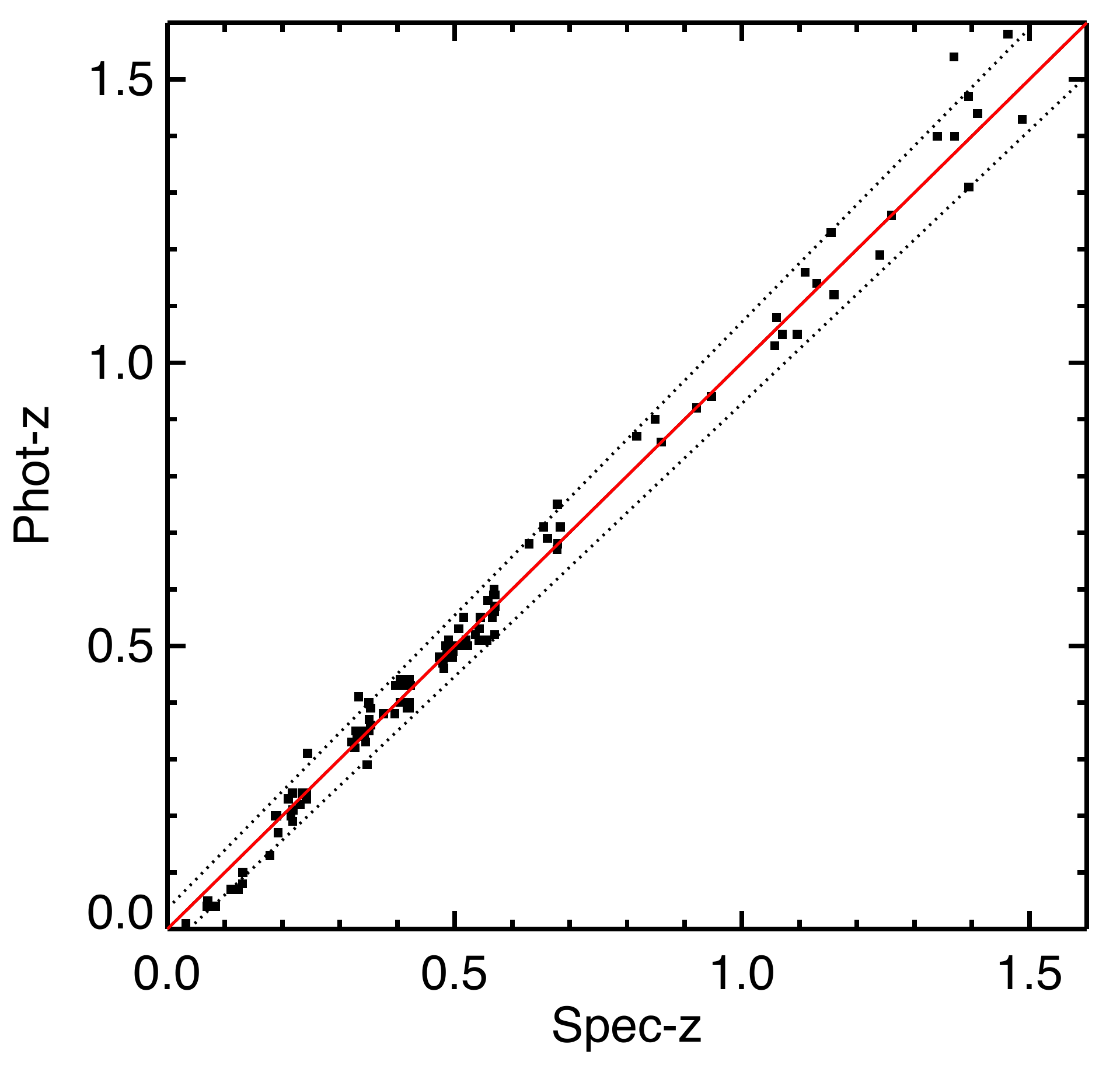}
    \caption{Comparison of the photometric redshifts for clusters in the ISCS sample with spectroscopic confirmation from the MMT \citep{koc12}, Keck \citep{sta05,els06,bro06,eis08}, and {\it HST} \citep{bro11,zei13}.  The dotted lines show the scatter in cluster photometric redshift accuracy, $\sigma = 0.036\,(1+z)$. 
    }
    \label{fig:bands}
\end{figure}

Spectroscopic follow-up of ISCS cluster candidates, shown in Figure~\ref{fig:bands}, by the AGN and Galaxy Evolution Survey \citep[AGES;][]{koc12} has confirmed dozens of clusters at $z<1$.  At $z>1$, over 20 of the ISCS clusters have been spectroscopically confirmed through targeted follow-up with the Low Resolution Imaging Spectrometer (LRIS) on Keck \citep{oke95} or $Hubble$ $Space$ $Telescope$ Wide Field Camera 3 \citep[WFC3;][]{kim08} spectroscopy \citep[see ][]{sta05, els06, bro06, bro11, eis08, zei12, bro13,zei13}.  The photometric redshift accuracy for the confirmed ISCS clusters is $\sigma = 0.036\,(1+z)$. Among the cluster candidates not spectroscopically confirmed, we expect the ISCS to have a $\sim10\%$ false detection rate due to chance line-of-sight projections \citep{eis08}.

ISCS cluster halo masses (quantified throughout as $M_{200}$, the mass interior to $R_{200}$, the radius at which the mean mass density exceeds 200 times the critical density) have been determined through a combination of targeted follow-up of confirmed clusters and statistical arguments.  X-ray observations \citep{bro11, bro16} and weak lensing \citep{jee11} have found that high-richness $z>1$ ISCS clusters have halo masses in the range log $M_{200}/\Msun = 14-14.7$, with the full ISCS sample having typical halo masses of  log $M_{200}/\Msun = 13.8-13.9$, determined using clustering measurements \citep{bro07} and halo mass ranking simulations \citep{lin13, alb14}.  From these works,  it was also determined that there is no significant redshift evolution in the median halo mass of ISCS clusters. This means that due to the selection technique, this cluster sample is not a progenitor sample, but rather provides snapshots of clusters at an approximately fixed halo mass at different epochs.  ISCS clusters at $z\sim0.5$ ($z\sim1.5$) have expected halo masses of $\sim\!1\e{14}\,\Msun$ ($\sim\!3\e{14}\,\Msun$) at $z=0$, with a factor of two scatter \citep{chi13}.

In this work, we utilize 232 ISCS clusters from $0.5<z<1.6$ (Table~\ref{tab:sample}) to perform a ``total light" cluster stacking analysis.  These redshift bounds are chosen to avoid the steep angular size-redshift relation at $z<0.5$ and small number statistics in the cluster sample at $z>1.6$ due to increasing photometric redshift uncertainties.  Given the typical ISCS halo mass, the median virial radius of our cluster sample is $R_{200}\approx1\,$Mpc \citep{bro07}, which we adopt throughout this study.  

Cluster membership was determined for these clusters via spectroscopic redshifts, where available, and full photometric redshift probability distribution functions as described in \citet{eis08, bro13, alb14}.   The photometric redshift catalog used in this work was presented in \citet{alb16}, following the procedure in \citet{chu14}, which incorporates near-infrared observations from the Infrared Bo\"{o}tes Survey\footnote{\url{ https://www.noao.edu/survey-archives/irbootes.php}}. Stellar mass estimates for individual galaxies were calculated using optical-mid-IR SEDs for all sources in the the {\it Spitzer} Deep Wide-Field Survey \citep[SDWFS;][]{ash09}, as described in \citet{bro13}. 

\begin{figure*}
	\includegraphics[width=7in]{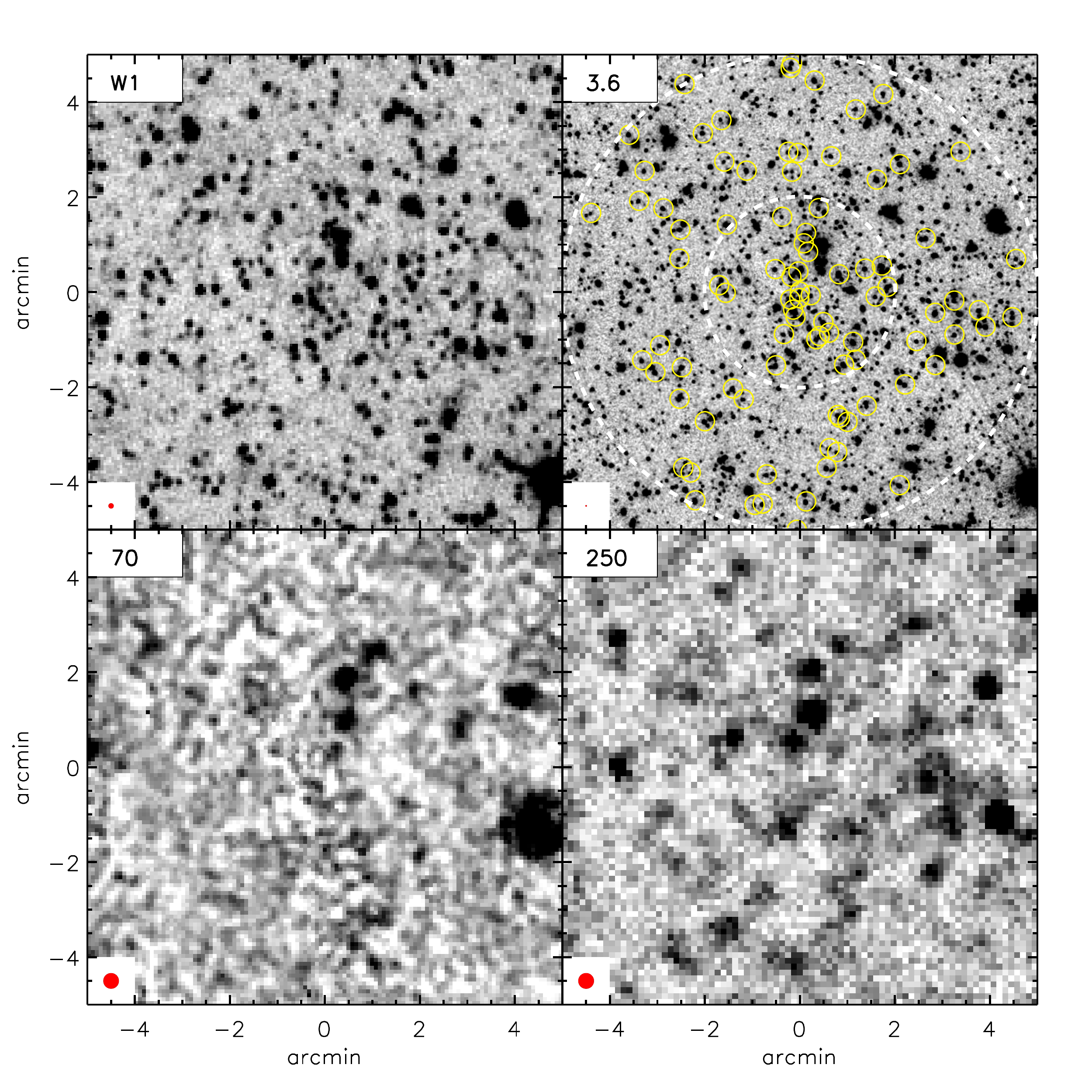}
    \caption{A 10\arcmin$\times$10\arcmin\ image centered around ISCS J1425.6+3403, a galaxy cluster at $z=1.09$,  is shown in the {\it WISE} $W1$ (top left), IRAC 3.6$\,\mu$m (top right), MIPS 70$\,\mu$m (bottom left), and SPIRE 250$\,\mu$m band (bottom right), respectively. Image point spread function FWHM is indicated by a red circle at bottom left corner of each panel. All photometric redshift member candidates are marked as yellow circles in the 3.6$\,\mu$m image, together with 1~Mpc and 2.5~Mpc radius circles around the cluster center. Varying imaging depths and angular resolution are apparent, highlighting the challenges in studying the properties of individual cluster galaxies except for the most luminous ones. 
    }
    \label{fig:onecluster}
\end{figure*}

In Figure~\ref{fig:onecluster} we show representative infrared data centered on a galaxy cluster at $z=1.09$. Photometric redshift cluster member positions are marked therein. 

\begin{table}
	\centering
	\caption{Sample Statistics}
	\label{tab:sample}
	\hspace{-0.4in}
	\resizebox{\columnwidth}{!}{ 
	\begin{tabular}{lclccc} % four columns, alignment for each
	    \hline
		\hline
Sample & Redshift &  \#  & $z_{\rm mean}$ & $z_{\rm median}$ & Avg. $\#$ of photo-$z$ \\
 &  &   &  &  & members per cluster$^\dagger$ \\
		\hline
ISCS & $0.5<z \leq 0.7$ & 61  & 0.57 & 0.55 & 30\\
(Bo\"{o}tes)& $0.7<z \leq 1.0$ & 78 & 0.86 & 0.87 & 33\\
& $1.0<z \leq 1.3$ & 53 & 1.13 & 1.15 & 24\\
& $1.3<z\leq1.6$ & 40 & 1.44 & 1.45 & 11\\
		\hline
%\multicolumn{6}{l}{\dagger \mathrm{For\:cluster\:members\: within\: 1\: Mpc\: with\: log\: $M_{\rm star}/M_\odot \geq 10.1$. }
\multicolumn{6}{l}{$^\dagger$ {\rm For~cluster~members~within~1~Mpc~with~log} $M_{\rm star}/M_\odot \geq 10.1$}
	\end{tabular}
	}
\end{table}

\subsection{{\it Herschel} imaging}
\label{sec:spireimg}

{\it Herschel} SPIRE \citep{gri10} coverage of the B\"{o}otes field was obtained as part of the {\it Herschel} Multi-tiered Extragalactic Survey \citep[HerMES:][]{oli12}.  The {\it Herschel} SPIRE 250, 350, and 500$\,\mu$m imaging was done in a two-tiered design centered on $(\alpha,\delta)$ = (14:32:06,+34:16:48) with a deep survey covering the inner two square degrees, and a shallower survey covering eight square degrees.  The data reduction, presented in \citet{alb14}, was done using the {\it Herschel} Interactive Processing Environment \citep[HIPE v7;][]{ott10} with a particular emphasis on the removal of striping through high order polynomial baseline removal and the removal of glitches. The pipeline-reduced SPIRE maps have a zero mean and are calibrated to units of Jy beam$^{-1}$.  FWHM, pixel scale, and 5$\sigma$ depth for the three SPIRE bands are listed in Table~\ref{datatable}. The $1\sigma$ confusion limits for 250, 350, and 500$\,\mu$m are 5.8, 6.2, and 6.8 mJy \citep{ngu10}. Data reduction, catalog creation, and completeness simulations of the B\"{o}otes SPIRE maps used in this work are described in \citet{alb14}.

%\SA{Note: more about the beam size, confusion, and the problems this creates in stacking, either here or in the stacking sections}

%  10.34     216.539621450   34.061426920   1.09

\subsection{{\it WISE} imaging}

The {\it WISE} mission covered the full sky with four bands centered on 3.4$\,\mu$m ($W1$), 4.6$\,\mu$m ($W2$), 12$\,\mu$m ($W3$), and 22$\,\mu$m ($W4$). The raw images have a field of view $47\arcmin \times 47\arcmin$ which is imaged simultaneously via dichroics on the four detectors at a pixel scale of $2\farcs75$~pix$^{-1}$. The full description of the mission is provided in \citet{wri10}. 
The `ALLWISE' data combined the data taken as part of the four-band cryogenic survey (January 7 - August 6, 2010), 3-Band Cryo phase\footnote{The 3-Band Cryo observations only cover a limited ecliptic longitudinal range, which excluded our survey field.} ($W1$, $W2$, and $W3$ only with the $W3$ band at reduced sensitivity; August 6 - September 29, 2010), and the first year data from the ``Near Earth Object Wide-field Infrared Survey Explorer (NEOWISE) post-cryo'' program \citep[September 29, 2010 $-$ February 1, 2011 with $W1$ and $W2$ only;][]{mai11}. The  NEOWISE reactivataion (NEOWISE-R) program, which began in October 2013, and continues to this day, have thus far produced six separate data releases. 

%The `ALLWISE' data consist of three separate observations, including the four-band cryogenic survey (January 7 - August 6, 2010), 3-Band Cryo phase ($W1$, $W2$, and $W3$ only with the $W3$ band at reduced sensitivity; August 6 - September 29, 2010), and the ``NEOWISE post-cryo'' program \citep[September 29, 2010 $-$ February 1, 2011 with $W1$ and $W2$ only;][]{mai11}. 

The ALLWISE data release\footnote{\url{ http://wise2.ipac.caltech.edu/docs/release/allwise/expsup}} 
\citep{cutri13} includes ``Atlas Images'', each tile of which covers  $1.56\deg\times 1.56\deg$ in area. These images are resampled to  1\farcs375~pix$^{-1}$, i.e., a half of the native pixel scale, and convolved with the instrument point spread function (PSF). While these smoothing steps help detect isolated point sources from the final coadded images, the procedure (and the resultant blurring) is less desirable while photometering  fainter individual sources near the detection threshold.  

In this work, we make use of a new set of coadds of the {\it WISE} images referred to as the unWISE\footnote{\url{ http://unwise.me}} dataset. While the details of the image processing steps are described in \citet{lan14} and \citet{mei17a}, one of the main differences lies in that it preserves the instrument PSF and the native pixel scale with no additional blurring. We use the unWISE `NEOWISE-R3' images \citep{mei17b} for the $W1$ and $W2$ bands, which combine the first 4 years of WISE observations (including three years of the NEOWISE operations). In particular, we use the `masked' coadds in which outlier pixels are rejected in the image combination.

%\citet{lan14} provided a new set of coadds of the {\it WISE} images\footnote{\url{ http://unwise.me}} -- referred to as the unWISE dataset -- that are not blurred while also preserving the native pixel scale. The details of their image processing steps are described in \citet{lan14} and \citet{mei17a}; the full-depth dataset (combining ALLWISE and NEOWISE) is presented in \cite{mei17b}.  We utilize the unWISE `NEOWISE-R3' images  for the $W1$ and $W2$ bands; in particular, we use the `masked' coadds in which outlier pixels are rejected in the image combination.  

For the $W3$ and $W4$ bands, we use the official AllWISE Atlas images. This is mainly because the unWISE products for these bands are known to contain artifacts around bright sources \citep{lan14}; this is in contrast to the $W1$/$W2$ products which were further improved by adding depth and removing artifacts as more post-cryogenic data were added for processing \citep{mei17b}. We note that ALLWISE and unWISE data  should yield identical results for stacking analyses given everything else equal. Indeed, we have verified that this is the case for the original (cryogenic) mission data \citep{wri10,lan14}.

The PSF widths are $\approx$~6\arcsec\ for the unWISE $W1$ and $W2$ bands, i.e., smaller than 8\arcsec\ for the same band in the ALLWISE data \citep{wri10,mei14}. As mentioned previously, this difference stems from additional convolution performed by the latter.  The ALLWISE and unWISE images share the same pointing center. The ALLWISE tiles are 4095$\times$4095 at the pixel scale $1\farcs375$~pix$^{-1}$ while the unWISE tiles are 2048$\times$2048 at the scale $2\farcs75$~pix$^{-1}$. Adjacent tiles overlap approximately 4\arcmin\ in both datasets. 

A total of 11 tiles (tile numbers: {\tt 2156p348}, {\tt 2170p318}, {\tt 2177p333}, {\tt 2192p348}, {\tt 2201p363}, {\tt 2159p333}, {\tt 2174p348}, {\tt 2182p363}, {\tt 2163p363}, {\tt 2188p318}, and {\tt 2195p333}) cover all of our clusters. The median number of individual exposures is in the range 130 -- 155 in the $W1$ and $W2$ bands, and 21 -- 32 in the $W3$ and $W4$ bands.
Given that exposure time per visit was 7.7~sec ($W1$ and $W2$) and 8.8~sec ($W3$ and $W4$),   the total exposure time per pixel is 16.7 -- 19.9~min and 3.1 -- 4.7~min, respectively. For further information on {\it WISE} depths and pixel scales, see Table~\ref{datatable}.

\subsection{{\it Spitzer}/IRAC and {\it Spitzer}/MIPS imaging}

For {\it Spitzer} IRAC imaging, we make use of the final data release of the SDWFS catalog, which provides deep IRAC imaging over the entire NDWFS field proper from which our clusters are selected. The observations and processing of the data are given in \citet{ash09}. The 5$\sigma$ limiting flux densities within 3\arcsec\ aperture diameters are 2.91, 4.61, 25.35, 28.84 $\mu$Jy
%22.65, 22.32, 21.83, and 22.47~AB 
at 3.6, 4.5, 5.8, and 8.0$\,\mu$m. We make use of the MIPS data taken for the MIPS AGN and Galaxy Extragalactic Survey \citep[MAGES;][]{jan10}, reduced using the MIPS-GTO pipeline \citep{gor05}, with $5\sigma$ point source sensitivities of 0.2 and 31 mJy at 24 and 70$\,\mu$m.
%processed by \citet{vac15}. 
All {\it Spitzer} data are single mosaicked images in units of MJy/sr. The pixel scale is 0.86\arcsec\ for the IRAC bands, 1.245\arcsec\ for the MIPS 24$\,\mu$m band, and 4.925\arcsec\ for the 70$\,\mu$m band (see Table~\ref{datatable} for a summary). 

\begin{table*}
\caption{Key characteristics of data used in this work}
\begin{tabular}{lcccccc}
\hline
\hline
Band Name & Effective Wavelength & Filter Width$^\dagger$ & FWHM & pixel scale & $5\sigma$& Reference\\
 & [$\mu$m] & [$\mu$m] & [\arcsec] & [\arcsec/pix] & [$\mu$Jy] & \\
\hline
unWISE $W1$ & 3.35  & 0.73 & 6.1 &2.75 & 54 & \citet{lan14, mei17a, mei17b} \\
unWISE $W2$ & 4.60 & 1.10 & 6.4 & 2.75 & 71 & -\\           
IRAC 3.6$\,\mu$m & 3.55 & 0.75& 1.66 &0.863 & 2.91 & \citet{ash09} \\           
IRAC 4.5$\,\mu$m & 4.49 & 1.02 & 1.72 &0.863 & 4.61 & -\\           
IRAC 5.8$\,\mu$m & 5.73 & 1.43 & 1.88 &0.863 & 25.35 & -\\           
IRAC 8.0$\,\mu$m & 7.87 & 2.91 & 1.98 &0.863 & 28.84 & -\\  
\hline
Band Name & Effective Wavelength & Filter Width & FWHM & pixel scale & $5\sigma$& Reference\\
 & [$\mu$m] & [$\mu$m] & [\arcsec] & [\arcsec/pix] & [mJy] & \\
\hline
WISE $W3$ & 11.56  & 8.30 & 6.5 & 1.375 & 0.7 & \citet{wri10, cutri13} \\      
WISE $W4$ & 22.08  & 3.50 & 12.0 &1.375 & 5.0 & - \\  
MIPS 24$\,\mu$m & 23.68  & 4.7 & 6.0 &1.245 & 0.2 & \citet{jan10} \\     
MIPS 70$\,\mu$m & 71.42  & 19.0 & 18.0 &4.925 & 31 & -\\ 
SPIRE 250$\,\mu$m & 243  & 78 & 18.1 & 6 & 14-26 & \citet{oli12} \\
SPIRE 350$\,\mu$m & 341 & 106 & 24.9 & 10 & 11-22 & -\\
SPIRE 500$\,\mu$m & 482 & 200 & 36.6 & 14 & 14-26 & -\\
\hline
\end{tabular}
\\
$^\dagger$For the {\it WISE} filters, filter widths are approximately measured as full-width-at-half-maxima of the total response function, which includes quantum efficiency, transmission of optics, beamsplitters, and filters. The IRAC sensitivities refer to those measured in 3\arcsec-diameter circular apertures \citep{ash09}. % while the WISE sensitivities and FWHMs are taken from the ALLWISE data release document  (see footnote \#2) and from \citet{wri10}. 
\label{datatable}
\end{table*}

\section{``Total light" cluster stacking method}
\label{sec:method}

 In this section, we present our ``total light" stacking techniques for {\it WISE}, {\it Spitzer}, and {\it Herschel} imaging.  To reiterate, ``total light" is the summed light from all cluster constituents, including previously unidentified (mostly low mass) cluster members and/or ICD components.  We note that our {\it Herschel} stacking technique is similar to those previously presented in the literature \citep[i.e.][]{bet10, Planck15}; however, {\it WISE} and {\it Spitzer} stacking are more complicated due to the much higher source density (of cluster members and non-members both) and non-uniform sky background. 
 We explore several methods for {\it WISE}/{\it Spitzer} cluster stacking and conclude this section with a discussion on which methods are appropriate given the available data. 

Total light cluster stacking was recently used to analyse proto-cluster candidates in \citet{Planck15} and \citet{kub19}.  In \citet{Planck15}, proto-cluster candidates selected as cold, compact {\it Planck} sources \citep{Planck2014} were used as positional priors to stack {\it Herschel} SPIRE imaging. The technique used, presented in \citet{bet10}, averages SPIRE cutouts centered on the {\it Planck} sources, after cleaning the cutouts of bright sources, to obtain the pixel-wise average flux and create a final stacked image.  

\citet{kub19} performed multi-wavelength stacking on {\it WISE}, {\it IRAS}, {\it AKARI}, {\it Herschel}, and {\it Planck} images of proto-cluster candidates from the HSC-SSP.  For reference, we summarize their stacking technique for the datasets ({\it WISE} and {\it Herschel}) relevant to this work: sky-subtracted images were obtained using sky maps generated by evaluating the local (in scales of $\approx10\arcmin$) sky values after masking bright sources.  Stacking was then performed on the sky-subtracted images as an average stack with $3\sigma$ clipping, with uncertainties measured via bootstrapping.  For {\it WISE} stacking, \citet{kub19} smoothed the sky-subtracted images to the {\it Planck} PSF ($\approx5\arcmin$) prior to stacking; for {\it Herschel}, they carried out stacking on both the original and smoothed maps.

\subsection{Image stacking}\label{sec:image_stacking}

\subsubsection{{\it WISE}/{\it Spitzer} Image Processing}\label{subsec:wise_spitzer_processing}
The varying angular resolution and image depth of our datasets pose a significant challenge in measuring the total infrared flux from clusters. In {\it WISE} and {\it Spitzer} data, the  number of both cluster and non-cluster galaxies detected individually in the image varies significantly across the passbands. The overall source density is much greater at shorter wavelength (e.g., $W1$, $W2$, 3.6$\mu$m, 4.5$\mu$m bands) at which low-redshift galaxies (low-level star-formers and passive galaxies) are intrinsically brighter. Hence, a single bright foreground source can bias the flux measurement along cluster sightlines.  To minimize these effects, we process {\it WISE} and {\it Spitzer} images following a procedure modified from that described in \citet{xue17}. Their procedure was used to measure extended low surface brightness (SB) Ly$\alpha$ emission from distant star-forming galaxies. We provide a full description of the adopted procedure below. 

First, we remove  all individually detected sources from each image. This is done by running the SExtractor software \citep{sextractor} on each science image which outputs a source catalog, sky background map, and segmentation maps. Typically, we use {\tt DETECT\_THRESH} of 3.0 and {\tt DETECT\_MINAREA} of 3, requiring the isophotal signal-to-noise ratio to be 5 or higher.  Second, we use the SExtractor-generated sky background map to perform an additional pass of sky subtraction. The images we use already have their sky background close to zero; the median sky estimated using the iterative sigma-clipping algorithm is 10-20\% of the pixel-wise rms. However, large-scale variations of the sky background are present -- due to instrument defects and bright sources -- which can adversely impact the stacking procedure. To avoid eliminating the cluster signal as background, we set the SExtractor {\tt BACK\_SIZE} to be no less than 10\arcmin\ in all cases.  At the cluster redshift range, 10\arcmin\ corresponds to 4--5~Mpc, much larger than the angular extent of the emission. 

Finally, we use the source segmentation map and create several different versions of the science images. In the first image, which we will refer to as a `{\tt \_masked}'  image, we mask sources detected by SExtractor as indicated by the segmentation map by replacing their pixels with NaN. In the second `{\tt \_unmasked}' image, we unmask the pixels belonging to the photometric-redshift-selected (``photo-$z$") cluster members (\S~\ref{sec:sample}), and repeat the same procedure.  Doing so ensures that all fluxes from the photo-$z$ cluster members are preserved in the image. Other sources remain masked. Cluster members that are identified via spectroscopic redshifts only are not included in order to maintain a uniform stellar mass cut. The third `{\tt \_sub}' image is simply a sky-subtracted image where all sources (cluster and non-cluster) are present. In all three versions, the values of {\it unsegmented} pixels are identical throughout the image.

The procedure of masking and unmasking described above is performed separately on each band. While applying the same set of masks uniformly in all bands may result in a cleaner and more consistent measurement of cluster light, it is neither practical nor feasible to cleanly mask/unmask the cluster members from these images as they increasingly blend with other sources and occupy a significant fraction of the cluster region in the sky at longer wavelength (see Figure~\ref{fig:onecluster}). In \S~\ref{sec:understanding}, we discuss how these masking/unmasking process can impact our measurements. 

%This is necessitated by widely varying beamsizes and depths across the bands. As illustrated in Figure~\ref{fig:onecluster}, it 

%http://wise2.ipac.caltech.edu/docs/release/allsky/expsup/sec4_4h.html

For the {\it WISE} data, the same procedure is repeated after we create a large mosaic by combining the tile images. Doing so ensures the full extent of the data are utilized properly for cluster members that lie close to the edge of a {\it WISE} tile. Further, this streamlines the procedure as the {\it Spitzer} data are already in a single mosaic format. We reproject each {\it WISE} tile using a common tangent point (the center of the tile {\tt 2174p348}), and run the iraf task {\tt mscred.mscstack} to combine them ({\tt offset=world, combine=average}).  Reprojection of the WISE tiles at a new tangent point leads to non-integer shifts of the images. That combined with the overlap between adjacent tiles ($\approx$4\arcmin\ or $\sim$80--90 image pixels) results in lower sky rms values in the mosaicked image in these regions, which is entirely artificial. To circumvent this problem,  we expand the projected bad pixel masks to minimize the image overlap. 

\subsubsection{{\it WISE}/{\it Spitzer} Image Stacking}\label{subsec:wise_spitzer_stacking}

For each cluster, we create a 28\arcmin$\times$28\arcmin\ image for each wavelength centered on its position using the IDL routine {\tt hastrom.pro}. Counts are converted to physical units ($\mu$Jy~pix$^{-1}$) at this stage. For the WISE data, we use the photometric zeropoints given in the image header\footnote{The monochromatic AB magnitudes of Vega are 2.699, 3.339, 5.174, and 6.620 in the $W1$, $W2$, $W3$, and $W4$ bands, respectively, according to the WISE All-Sky Data Release document found at \url{http://wise2.ipac.caltech.edu/docs/release/allsky/expsup/sec4_4h.html.} We note that, for W4, this is within 0.04 mag of the zeropoint found in \citet{bro14}, based on an empirical recalibration which revised the effective wavelength to 22.8$\,\mu$m.}.  For the IRAC and MIPS data, we simply multiply the image by an appropriate scale factor to convert from MJy/sr to $\mu$Jy~pix$^{-1}$.  For a given sample consisting of $N$ clusters, the procedure creates a three-dimensional datacube. Image stacking is then performed by collapsing the datacube along the first dimension by taking pixel-wise mean or median. 

As described in the previous section, three different image stacks are created using the `{\tt \_masked}', `{\tt \_unmasked}', and `{\tt \_sub}' images. For the {\tt \_masked} and {\tt \_unmasked} datacubes, the flagged (NaN) pixels are excluded from the considerations. We repeat this stacking procedure for non-cluster sightlines. For each of the $N$ clusters, we pick a random position 12--24\arcmin\ from the cluster center. In Figure~\ref{fig:wise_imagestack}, we illustrate the stacked {\it WISE} W2 images of the cluster and non-cluster sightlines using four different images, namely, {\tt mean\_unmasked}, {\tt mean\_masked}, {\tt median\_masked}, and {\tt median\_sub}. 

Towards the cluster sightlines, we detect extended emission spanning $\approx$2\arcmin\ across ($0.7-1.0$~Mpc at $z=0.5-1.6$). While the `off-cluster' image stack is clearly devoid of any emission at the center, it exhibits similar noise properties to the cluster stacks, which is reassuring. In Table~\ref{table1_appendix} and  Figure~\ref{fig:wise_imagestack}, it is also evident that the noise properties vary depending on the precise stacking method. The sky noise does not obey a pure Poisson statistic in any of the stacks, and the {\tt mean\_unmasked} stack shows the highest root-mean-square in sky pixels. The correlated non-Poisson noise is likely brought on by the angular distribution of faint sources not flagged by the above procedure. This is consistent with the expectation that the mean pixel combination should be more susceptible to their presence than the median combination.  A full description of our results using different image versions and comparisons is given in Appendix~\ref{appendix_a}. The interpretation of these different stacking schemes is discussed in \S~\ref{sec:understanding}.

\subsubsection{{\it WISE}/{\it Spitzer} Photometry}\label{wise_spitzer_radial_profile}

On the image stack, we determine the centroid of the stacked signal using the IDL routine {\tt gcntrd.pro}. The offset of the centroid from the image center is small -- typically less than 2--3\arcsec. Such an offset has a negligible effect on the flux measurements although it can affect the shape of the radial profile near the peak. 

To measure the total flux from our cluster stack, we perform aperture photometry. First, we estimate sky in annular bins in the range of 150\arcsec--200\arcsec.  This is necessary particularly for the {\tt \_unmasked} stack in which cluster member candidates are unmasked. Because they are identified out to 2.5~Mpc from cluster center \citep{eis08}, doing so artificially increases the source density out to the same distance. Indeed, this effect is clearly visible in several short-wavelength bands where a large fraction of cluster member candidates are individually detected. We stress that the effect is entirely artificial. The median image stack, which does not include the fluxes of individual sources, does not show the same level of rise in the baseline at the same angular range, though some bands do show a much lower level of the same effect, suggesting that galaxies falling in towards the clusters that lie much farther than 2~Mpc may play a minor role.  Setting the baseline at larger radii would increase the flux up to 30\%. 

In all {\it WISE} and {\it Spitzer} bands, we measure the cumulative fluxes in a series of circular apertures with a stepsize of 5.5\arcsec, i.e., twice the native {\it WISE} pixel scale.  The cumulative fluxes are simply a sum of all pixel values (post sky subtraction) within the given aperture. The radial surface brightness is measured as the mean of all pixels in annuli with the same stepsize. 

We examine the intensity distribution of sky pixels, and find that  it is skewed towards the high-end wing. This is expected given that some pixels must contain fluxes of unresolved faint sources even after sky subtraction. The `straight-up' standard deviation of the distribution is larger by up to a factor of $
\approx$2  than a sigma-clipped value, $\sigma_{\rm p}$, i.e., a Gaussian fit to the pixel distribution. However, the result is insensitive to a specific choice of a sky annulus. The level of deviation from a pure Gaussian depends on the underlying flux distribution of all sources and what we define as a `source' (i.e., the specific setting of our source detection with the SExtractor software such as {\tt DETECT\_THRESH} and {\tt MINAREA}). The formal errors for the mean surface brightness and for the cumulative flux are $\Delta S=\sigma_{\rm p}/\sqrt{N_a}$ and $\Delta F=\sqrt{N_c}\sigma_{\rm p}$, where $N_a$ and $N_c$ are the number of pixels within a given annulus and aperture, respectively. However, these estimates are based on a single sky annulus and thus do not account for possible variations in $\sigma_{\rm p}$ within the stacked image.

We test the stability of measured $\sigma_{\rm p}$ values by examining the mean and rms values of the sky on  annuli centered on 500 randomly chosen locations in the off-cluster stack. The relative fluctuation of the  $\sigma_{\rm p}$ values in the off-cluster stack is 2--11\% for the {\it Spitzer} bands, and up to 1\% for the {\it WISE} bands.  In addition, we find that the mean sky varies 1--17\% for the {\it Spitzer} bands and 0.5--4\% for the {\it WISE} bands. The largest fluctuation in both sky mean and rms values occurs in the 8.0$\,\mu$m band. 

Taking these measurements as different realizations of the sky for the cluster signal, we obtain a more realistic estimate of our measurement uncertainties. The total pixel-to-pixel fluctuation is calculated as 
$\sigma_{\rm T}^2 = \sigma_{\rm p}^2 + \sigma_{\rm sky}^2 + \sigma_{\rm rms}^2$ where $\sigma_{\rm sky}$ and $\sigma_{\rm rms}$ are the standard deviation of sky background (i.e., mean sky) and pixel rms, respectively, determined from the 500 off-cluster stack measurements. Including these factors increases the photometric errors by a factor of $\approx$2--5 in the {\it Spitzer}/{\it WISE} bands.

\subsubsection{{\it Herschel}/SPIRE Image Processing}

We do not perform any additional image processing of the SPIRE maps prior to stacking. The pipeline reduced SPIRE maps have a zero mean and thus are already background subtracted.  Source masking is similarly not required.  Unlike {\it Spitzer} and {\it WISE}, the SPIRE imaging is confusion limited  (see \S~\ref{sec:spireimg}) and so the vast majority of sources, both cluster and field, are expected to be individually undetected.   This was confirmed for cluster members in \citet{alb14}, which found that $\lesssim10\%$ have a candidate $5\sigma$ counterpart at 250$\,\mu$m.   This leaves the dominant detected population: field galaxies.  Since field galaxies exist across the zero mean maps, they should not contribute to the stacked signal, which we verify in the next section. %\S~\ref{sec:spirestacking}.

\subsubsection{{\it Herschel}/SPIRE Image Stacking}
\label{sec:spirestacking}

Clusters are stacked in the three SPIRE bands by building a three-dimensional datacube comprised of 28\arcmin x28\arcmin\ (roughly $2\times R_{vir}$ at $z\sim1$) cutouts centered on each cluster.  Each cutout is randomly rotated to avoid systematic offsets.  The SPIRE stacks are then created by taking the variance-weighted mean of each pixel across all cutouts.  As discussed in \citet{alb14}, a variance-weighed mean is appropriate for the two-tiered Bo\"{o}tes survey, which has differing depths and therefore differing noise properties across the map.  

The contribution from the field galaxy population, detected or undetected, is determined by stacking cutouts placed randomly off of any known cluster.  While the stacks toward clusters display clearly detected extended emission, the stacks away from clusters show no stacked signal above the noise.  This test verifies that the stacked signal from the field population is zero, as one would expect in a zero mean map.  Example stacks toward and away from the cluster sight-lines can be seen in Figure~\ref{fig:spire_background} in Appendix~\ref{appendix_spire}. 

We determine the centroid and rough widths of the SPIRE stacks by approximating them as a Gaussian using the IDL code {\tt mpfit2dpeak} \citep{mar09}.  The {\it Herschel} stacks have a approximate FWHM of $\sim600$ kpc at all redshifts; the radial profiles will be more carefully examined in \S~\ref{sec:radial_profiles}. Centroiding is performed on the full set of bootstrap realizations to get the best estimate of the stack centers.  In our initial centroiding of the SPIRE stacks, we discovered a systematic offset of ~1-2 pixels in the x,y center of the extended cluster emission.  As this offset persisted across all cluster sub-sets and simulated data (see \S~\ref{appendix_spire}), we determined it was a systematic of the data, likely the scan pattern, rather than a real effect.  In order to be conservative, we randomly rotate the SPIRE cutouts centered on each cluster while stacking.  This eliminates the offset and allows us to centroid on the stacked cluster centers for aperture photometry.  The resulting radial profiles and aperture fluxes are identical within the uncertainties, regardless of cutout rotation.

\subsubsection{{\it Herschel}/SPIRE Photometry}
\label{sec:spirephot}

As with the {\it WISE} and {\it Spitzer} stacks, the {\it Herschel} stacked emission is clearly extended and so aperture photometry is performed on the stack images (after converting from Jy beam$^{-1}$ to Jy pix$^{-1}$) using the python package {\tt photUtils} \citep{bra19}.

The uncertainties on the flux in a given aperture are determined via bootstrapping with replacement, whereby the cluster catalog is resampled and stacked at random, allowing duplicates, 10,000 times.  Aperture photometry is performed on each bootstrap realization, and the mean and standard deviation of the resulting distribution represent the statistical properties of the clusters being stacked.  As discussed in \citet{alb14}, this method captures not only detector and confusion noise, but also the relative spread in the population being stacked.  The bootstrapped mean is checked for consistency with the stacked mean, which reassures us that the stacked signal is not dominated by a few outliers.

\subsection{Measuring Cluster Light: Understanding the meaning of stacked fluxes} \label{sec:understanding}

As described in \S~\ref{subsec:wise_spitzer_processing} and \S~\ref{subsec:wise_spitzer_stacking}, each {\it Spitzer} and {\it WISE} image is prepared in several different ways in which the treatment of the pixels belonging to detected sources differ. Starting from these images, four different image stacks are created ({\tt mean\_unmasked, mean\_masked, median\_masked, and median\_sub}). While we present a detailed comparison of the measured fluxes and the statistical properties in Appendix~\ref{appendix_a}, here we consider the physical meaning of their differences in the context of cluster light. 

The ISCS dataset used in this work has unique advantages which are typically not shared by a vast majority of other cluster samples. First, the availability of multiple passbands of varying depths and of overlapping wavelength ranges enables us to explore the usefulness of the shallower {\it WISE} data in carrying out similar analyses in the future on other cluster samples. The agreement between {\it WISE} and {\it Spitzer} IRAC bands at similar wavelengths (e.g., $W1$ vs 3.6$\,\mu$m and $W2$ vs 4.5$\,\mu$m) is excellent in both measured fluxes and radial profile shapes, as detailed in Appendix~\ref{appendix_wise_v_irac}. Thus, applying similar stacking techniques to larger cluster samples using only the {\it WISE} bands should yield robust results to constrain the overall stellar content and their internal distributions.

Second, the ISCS cluster sample has superb photometric redshifts  (in addition to extensive spectroscopic redshifts of members), information that is often limited or unavailable for other larger cluster surveys. In this work, we have taken advantage of this photo-$z$ information and `unmasked'  cluster members in the image (`{\tt \_unmasked}')  where all the sources are otherwise masked. This step verifies that the stacked signal is not dominated by unfortuitously positioned non-member galaxies (or stars) that are bright. 

In all the images we have tested, we consistently find that the {\tt mean\_unmasked} stack always yields the largest aperture fluxes; the other three stacks -- namely, {\tt mean\_masked}, {\tt median\_masked}, and {\tt median\_sub} -- show  comparable fluxes to one another, which are consistently lower than  the {\tt mean\_unmasked} stack. The agreement of these three stacks assures us that the statistical background subtraction of non-cluster galaxies is effective, and as a result, we can  recover fluxes of faint cluster members. It also implies that as long as we treat all images consistently (in measuring sky background and in detecting and removing individual sources), we can expect a robust result. This may be in part owing to the fact that the present dataset is relatively uniform in coverage and depth. The possibility of combining heterogeneous datasets (e.g., varying exposure time across a given field) would require further investigation. 

\begin{table}
\vspace{1in}
\begin{center}
\caption{Fraction of total flux above the detection limit, $f_{\rm det}\equiv 1-F_{\rm others}/F_{\tt mean\_unmasked}$ where others refers to {\tt median\_masked}, {\tt mean\_masked}, or {\tt median\_sub}, for {\it WISE} imaging. A circular aperture of 100\arcsec\ in radius is used.  
}
\resizebox{\columnwidth}{!}{
\begin{tabular}{ccccc}
\hline
\hline
Sample  & $W1$  & $W2$ & $W3$ & $W4$ \\
\hline
$z=0.5-0.7$ & $68\pm3$\%    &      $47\pm8$\% &    $18\pm13$\% &  0--6\%  \\
$z=0.7-1.0$ & $62\pm4$\%    &      $32\pm8 $\% &    1--3\% &  1--5\%   \\
$z=1.0-1.3$ & $51\pm6$\%    &      $32\pm9$\% &    2--4\% &  5--11\%  \\
$z=1.3-1.5$  &  17--27\%    &      20--44\% &    - &  -      \\

  \hline
\end{tabular}
}
\label{tab:missing_fraction_wise}
\end{center}
\end{table}
\renewcommand{\arraystretch}{1.0}

\begin{table*}
\vspace{1in}
\begin{center}
\caption{Fraction of total flux above the detection limit, $f_{\rm det}\equiv 1-F_{\rm others}/F_{\tt mean\_unmasked}$ where others refers to {\tt median\_masked}, {\tt mean\_masked}, or {\tt median\_sub}, for {\it Spitzer} imaging.  A circular aperture of 100\arcsec\ in radius is used.  
}
\begin{tabular}{ccccccc}
\hline
\hline
Sample & $3.6\,\mu$m & $4.5\,\mu$m & $5.8\,\mu$m & $8.0\,\mu$m & $M24$ & $M70$  \\
\hline
$z=0.5-0.7$ & $72\pm3$\% & $62\pm4$\% & $31\pm17$\%     & $28\pm16$\% & $26\pm14$\% & 0--20\%    \\
$z=0.7-1.0$  &  $73\pm 3$\% & $64\pm5$\% & $27\pm18$\%     & 3--14\% & $29\pm8$\% & 0--19\%    \\
$z=1.0-1.3$  & $60\pm4$\% & $51\pm12$\% & 19--37\%     & 18--23\% & 0--20\% & $34\pm24$\%    \\
$z=1.3-1.5$  & $52\pm22$\% &  $42\pm18$\% & 0--14\%     & 9--24\% & 0--7\% & -   \\

  \hline
\end{tabular}
\label{tab:missing_fraction_spitzer}
\end{center}
\end{table*}
\renewcommand{\arraystretch}{1.0}

The difference between the {\tt mean\_unmasked} and {\tt mean\_masked} stacks can only be interpreted as the unmasked cluster members positively contributing to the resultant flux in the former. 
The fact that the  {\tt mean\_masked} and {\tt median\_masked} stacks yield similar fluxes strongly suggests that a large fraction of cluster light originates from the numerous faint members that lie  below the detection threshold rather than from a few relatively bright members that eluded photo-$z$ identification. In the latter scenario, the overall pixel distribution would be highly asymmetric resulting in its mean significantly greater than the median. Finally, it is reassuring that the {\tt median\_sub} stack fares  well providing an estimate robust against a range of bright sources present in the images; the fact that it returns a comparable flux to the other two {\tt \_masked} stacks reinforces the notion that numerous undetected cluster members raises the overall counts in the cluster regions. 

In this interpretation, both fluxes convey distinct physical significance. The {\tt mean\_unmasked}-derived fluxes represent  {\it total} fluxes coming from clusters encompassing all member galaxies (and possibly from intracluster dust) regardless of their intrinsic brightness. For the remainder of this work, we will refer to this flux as `total flux' and `total light' with regard to the near- and mid-IR stacks.  Informed masking of the images ensures that all possible members are available for counting while sky counts on off-cluster pixels carry information on statistical background (from non-cluster-members). On the other hand, the remaining stacked fluxes with no photo-$z$ member information represent the cluster fluxes originating from the sources that are too faint to be detected. In the case of the stacks presented in Appendix~\ref{appendix_a}, these faint galaxies are responsible for $\approx 60-70$\% of the total cluster flux (Table~\ref{table1_appendix}). 

The number of cluster members that rise above the detection limit depends on a number of factors: imaging depth, cluster luminosity function, distance (thus, redshift) to the cluster, and wavelength. We compute the fractional contribution to the total cluster flux from detected sources in all {\it Spitzer} and {\it WISE} bands for our four redshift samples, computed as $f_{\rm det}\equiv 1-F_{\rm others}/F_{\tt mean\_unmasked} $ where $F_{\tt mean\_unmasked}$ denotes the flux measured from the  {\tt mean\_unmasked}  stack while, for $F_{\rm others}$, we use the fluxes measured from the {\tt median\_masked}, {\tt mean\_masked}, and {\tt median\_sub}  stacks. 

In Tables~\ref{tab:missing_fraction_wise} and \ref{tab:missing_fraction_spitzer}, we list the range of $f_{\rm det}$ for the {\it WISE} and {\it Spitzer} bands, respectively. We give both the mean (of the three {\tt others} stacks) as well as the typical uncertainties. At longer wavelength and at higher redshift, our estimates tend to become less certain. Whenever the uncertainties become comparable to or larger than the mean correction, we list the range of the $f_{\rm det}$ instead of the mean.  Whenever the fluxes from the median stacks become larger than the fiducial {\tt mean\_unmasked}, we choose not to list the $f_{\rm det}$, as it likely suggests that the signal is too noisy to place a meaningful constraint.  

Comparing the values listed the tables, it is evident that the {\it Spitzer} band has higher $f_{\rm det}$ than that in the {\it WISE} band at similar wavelength. This simply reflects the better sensitivity and angular resolution of the IRAC data compared to the {\it WISE} observations. It is also notable that $f_{\rm det}$ declines precipitously with increasing wavelength such that at 8~$\mu$m,  no more than 10--20\% of the cluster light lies above the detection threshold. 

Our results demonstrate the overall usefulness of the stacking approach in recovering the full extent of the emission originating from clusters. Equally importantly, we stress that it may be possible to use a careful calibration, such as that presented here, to fully utilize the power of all-sky missions such as {\it WISE}. Given that the {\it WISE} coverage is largely uniform across the whole sky, one can use the information given in Table~\ref{tab:missing_fraction_wise} as a correction factor to go from the measured {\it WISE} flux to the total flux for datasets where cluster membership is not known. For example, for the $z=0.5-0.7$ clusters, Table~\ref{tab:missing_fraction_wise} lists $f_{\rm det}=0.68$ for the {\it WISE} $W1$ band. If one makes a measurement of the $W1$ band flux $f$ for a different sample of clusters at similar redshift (using methods similar to our {\tt median\_masked}, or {\tt median\_sub} stacks), the total flux is expected to be $f_{\rm corr}=f/0.68 \approx 1.47f$. 

We intend to investigate this further using a larger sample of clusters and across multiple cosmic epochs in the future.  As the {\tt mean\_unmasked} stacks represent all cluster light (regardless of direct detection of the source), our main results in this work are based on those stacks unless  stated otherwise. 

\section{Results} \label{sec:results}

Using the techniques outlined in \S~\ref{sec:method}, we stack the ISCS clusters in four redshift bins: $0.5<z \leq 0.7$, $0.7<z \leq 1.0$, $1.0<z \leq 1.3$, and $1.3<z\leq1.6$ (Table~\ref{tab:sample}).  
The size of the bins reflects the need to detect a significant stacked signal across a broad wavelength range while ensuring that the angular scale does not change substantially within a given redshift bin.  

\begin{figure*}
	\includegraphics[width=7in]{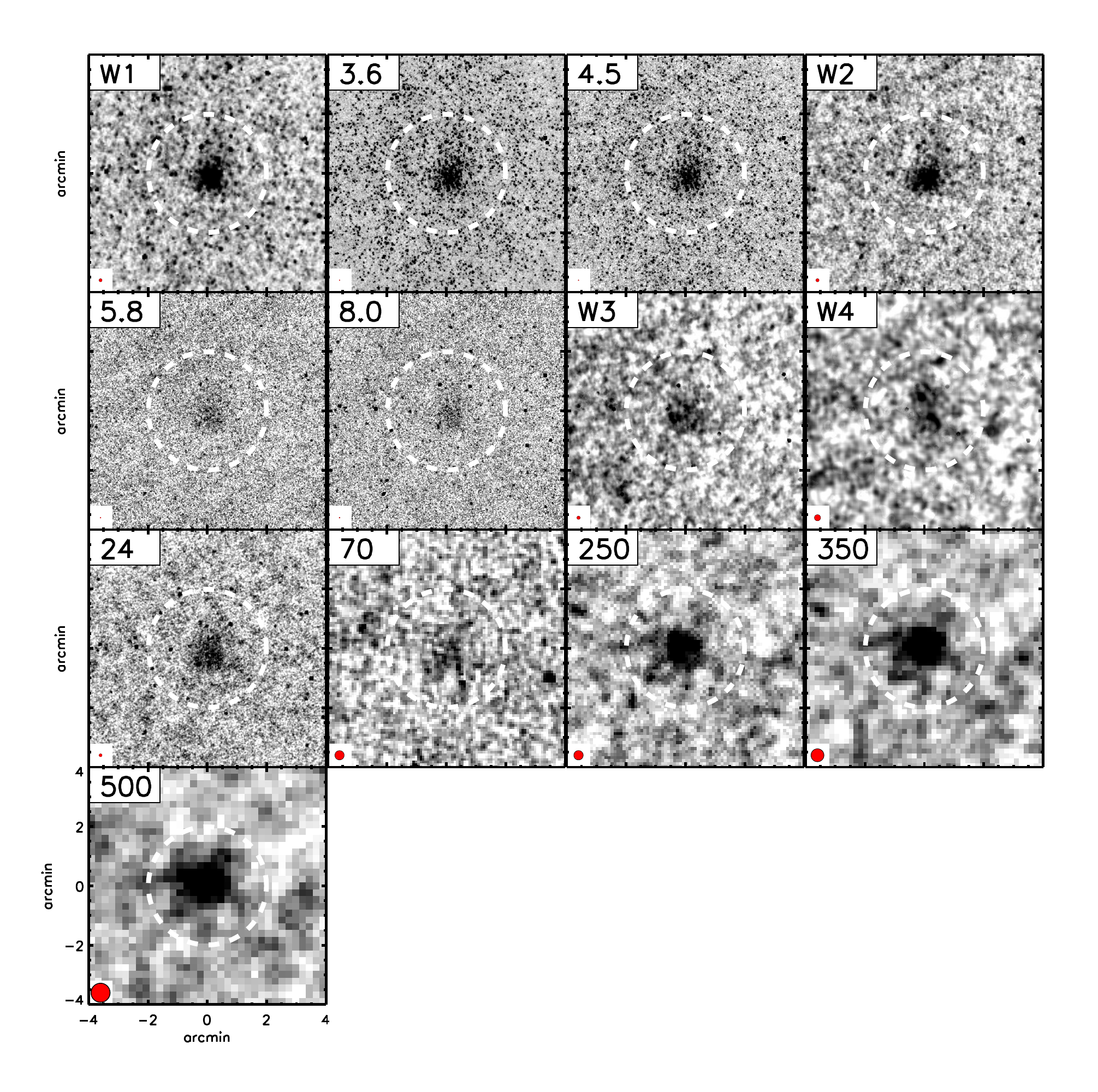}
    \caption{Mean ({\tt mean\_unmasked}) stacks of the ISCS clusters at $z=1.0-1.3$ showing extended cluster emission from $3.4-500\,\mu$m. A higher source density within $\approx 4$\arcmin\ radius is apparent in several bands including the 3.6$\,\mu$m, 4.5$\,\mu$m, and $W1$ bands. 
    The effect is  artificial and is brought on by an enhanced source density around clusters as discussed in \S~\ref{wise_spitzer_radial_profile}. Local sky background is always estimated within this range for accurate sky subtraction. Image point spread function FWHM is indicated by a red circle at bottom left corner of each panel.
    }
    \label{fig:stacks_highz}
\end{figure*}

In Figure~\ref{fig:stacks_highz} we show the $1.0<z \leq 1.3$ cluster stacks at all wavelengths: {\it WISE} $W1$, $W2$, $W3$, $W4$, IRAC 3.6, 4.5, 5.8, and 8.0$\,\mu$m, MIPS 24 and 70$\,\mu$m, and SPIRE 250, 350, and 500$\,\mu$m. For the near- and mid-infrared bands, we use the {\tt mean\_unmasked} images for stacking analyses. These panels are arranged in the order of increasing wavelength starting from top left.  The stacked signal is resolved in all cases, even at 500$\,\mu$m where the beamsize is 36$^{\prime\prime}$ ($\sim$ 300~kpc at $z\sim1.2$).  In the following sections, we analyze the radial profiles of the ISCS clusters as a function of wavelength and redshift and compare to a NFW profile \citep{nav96} as the fiducial cluster profile.  In addition to being commonly used to model Dark Matter haloes (DM), the NFW profile can be used to describe the cluster stellar mass distribution \citep[e.g.][]{lin04, muz07, vdb14, hennig17, lin17}. We then determine the total cluster flux at each wavelength and redshift and build ``total light" cluster SEDs.  In the final section, we measure the stacked far-IR emission from massive galaxies (log $M_{\star}/\Msun\geq10.1$) only, using the techniques from \citet{alb14}, to facilitate a comparison of the behavior of massive cluster galaxies to the total far-IR emission. 

\subsection{Radial flux profiles}\label{sec:radial_profiles}

\begin{figure*}
	\includegraphics[width=7in]{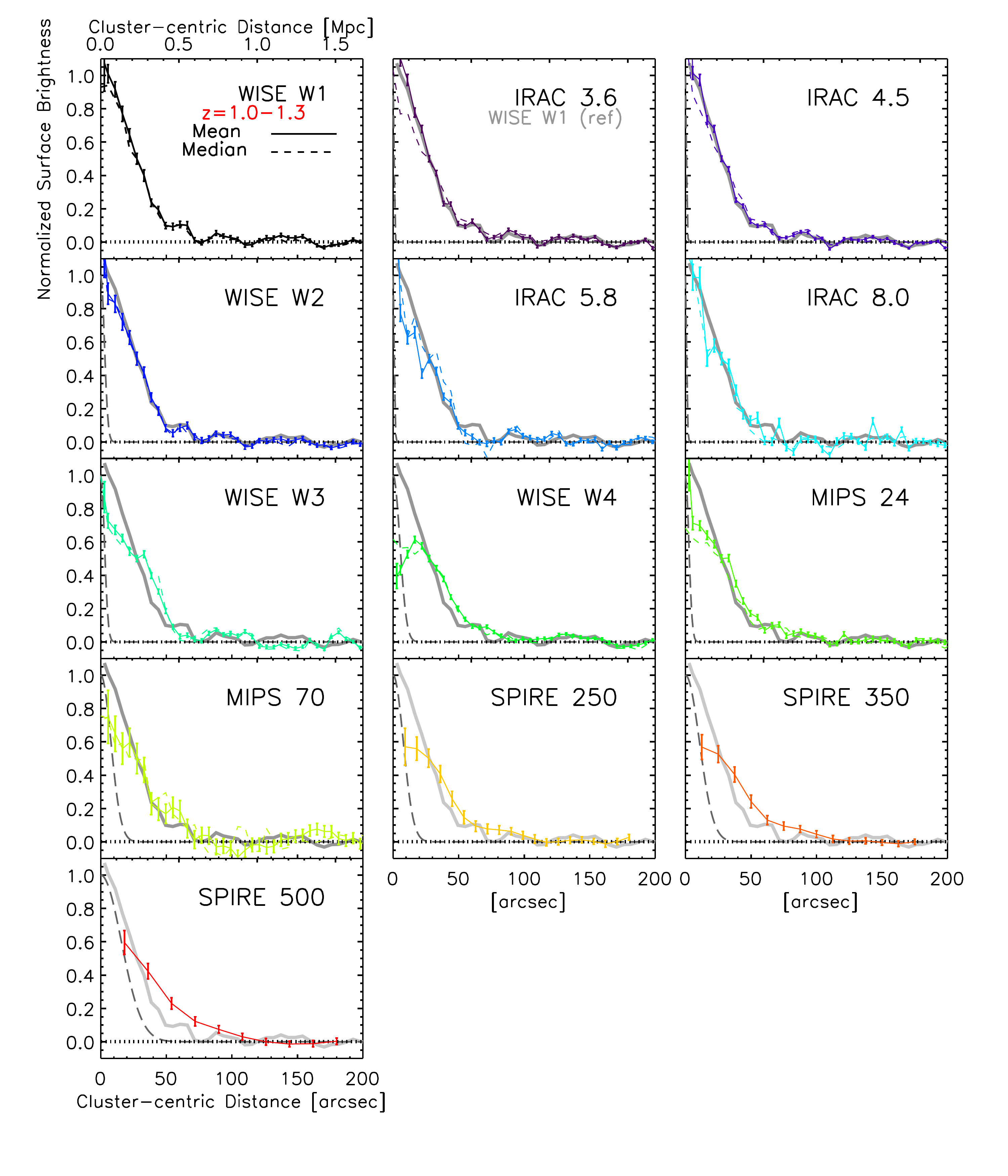}
    \caption{The {\it observed} radial surface brightness profiles of {\it WISE}, {\it Spitzer}, and SPIRE image stacks of the $z=1.0-1.3$ clusters,  arranged in the order of increasing wavelength. All profiles are normalized to have the value 0.5 at 27.5\arcsec. Long dashed lines the PSF size approximated as a Gaussian. In each panel for $W1$ through MIPS 70$\,\mu$m, we show the measurements from both the mean (solid) and median (dashed) stacks. The two profiles are typically consistent with one another except at the inner $\approx$30\arcsec.  The grey solid line in each panel marks the radial profile ({\tt mean\_unmasked}) of the $W1$ band as a reference.  
    }
    \label{fig:wise_radialprofile}
\end{figure*}

In Figure~\ref{fig:wise_radialprofile}, we show the radial surface brightness profile measured for the $z=1.0-1.3$ clusters. For the {\it Spitzer} and {\it WISE} bands, the annuli increase in steps of 5.5\arcsec, i.e., twice the native {\it WISE} pixel scale, 2.75\arcsec. For the {\it Herschel} SPIRE bands, half the FWHM for each band is used as the radial bin size (Table~\ref{datatable}). 
In each panel, the profile is normalized at 27.5\arcsec\ to be at 0.5, a value chosen arbitrarily for visualization. The unWISE $W1$ band profile is additionally overlaid as thick grey line on all panels. 

The overall agreement between the measured profiles in the {\it WISE} and IRAC bands at similar wavelengths is remarkably good considering the range of imaging depth and angular resolution spanned by these data. Similar agreements exist in the other redshift bins. At longer wavelengths, the profiles tend to become shallower than the reference ($W1$ band: grey line); this is particularly the case for the SPIRE bands. Keeping in mind that the SPIRE bands have considerably larger beam FWHMs as well as the pixel scales, a more careful analysis to account for these differences is needed to evaluate the overall similarities of profiles at near- and far-IR wavelengths, which we discuss in  \S~\ref{subsec:spire_profile}. 

The profiles measured from the {\tt mean\_unmasked} and {\tt median\_unmasked} stacks (solid and dashed line, respectively, in each of the {\it WISE} and {\it Spitzer} panels) are in good agreement with each other except for the inner $\approx$30\arcsec. The physical explanation may be that the overall distribution of faint cluster galaxies is similar to that of their brighter cousins while there is an excess of brighter galaxies at the cluster core pulling up the mean (see our discussion in \S~\ref{sec:understanding}). Among those bright galaxies that affect the core profile are the brightest cluster galaxies (BCGs). We do not take the extra step of removing the BCGs from our radial profiles, however, as previous studies have found that their impact on the stellar mass profiles is small \citep{vdb15}.
Regardless of the origin, the agreement is promising for constraining the cluster light profiles based on larger cluster samples for which photometric redshift information is unavailable (discussed in \S~\ref{sec:implications_samples}). 

If the angular extent measured from our image stacks carries physically meaningful information, our stacking approach could potentially provide a powerful tool in tracing past and current star formation activity within clusters encoded at rest-frame near-infrared and far-infrared wavelengths, respectively. Applying this technique to all-sky surveys will prove particularly promising. To further investigate this possibility, we assess the usefulness of such measurements by comparing our radial profile measurements with those obtained by a more conventional method where more detailed information is available, albeit for fewer clusters.

\subsubsection{The effects of beamsize and centroid uncertainties}\label{subsec:radprofile_sim}

In order to compare our radial profiles to a fidicual profile, we need to address observational effects.  For example, Figure~\ref{fig:wise_radialprofile} shows that the profiles are becoming shallower at longer wavelengths. This effect is most obvious in the MIPS 70$\,\mu$m band and {\it Herschel} bands. The decreasing S/N and increasing pixel scales and beamsizes are expected to effectively broaden the profile relative to the intrinsic one. 

Here we address several observational factors that affect our radial profile measurement.  First, unlike cluster studies largely based on spectroscopic members, our determination of cluster centers is only accurate within 15\arcsec. The precision of the ISCS cluster centroids derives largely from the 15\arcsec\ pixel scale of the density maps used for cluster detection. A detailed analysis in the similarly-selected MaDCoWS cluster sample, which used density maps with the same resolution, confirmed rms scatters of $\approx 15$\arcsec\ in both right ascension and declination \citep{gon19}. Comparison of the ISCS centroids with the X-ray centroids for a small number of high-redshift clusters finds consistent offsets (Garcia et al.~in preparation).

Second, a PSF at a given wavelength will systematically move the flux in the inner part outwards, the degree of which depends on the beam size and pixel sampling. Both factors effectively blur the intrinsic profile, resulting in a shallower radial profile than the intrinsic one. 
 
 \begin{figure*}
	\includegraphics[width=7in]{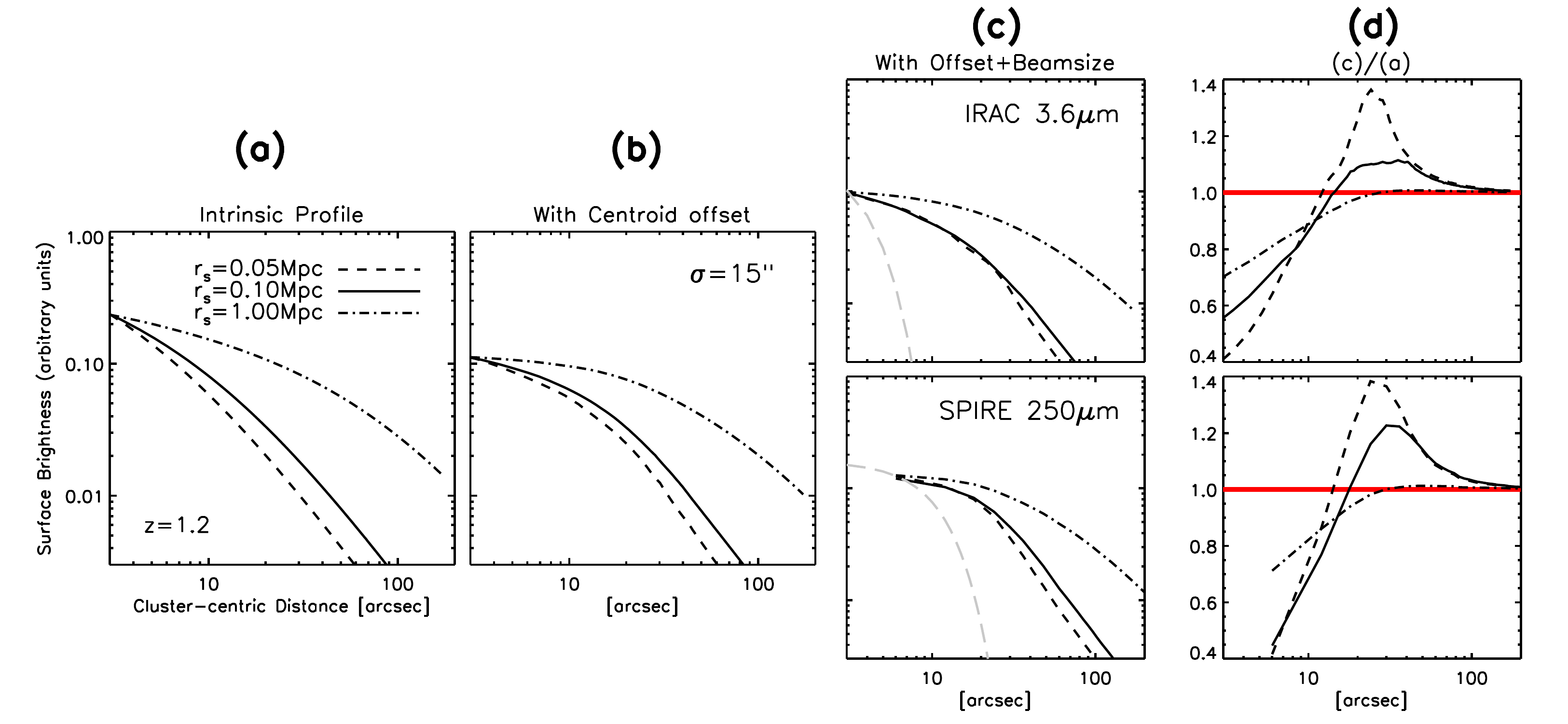}
    \caption{
    The effects of centroiding uncertainties, pixel sizes, and beam (PSF) sizes are illustrated assuming source redshift $z=1.2$. In all but (d) panels, all surface brightness profiles are normalized to match at the smallest radial bin for clarity. \textbf{a)} The intrinsic profile with the scale lengths $r_s=0.05$ (dashed), 0.10 (solid), and 1.00~Mpc (dot-dashed). \textbf{b)} The profiles shown in panel (a) are randomly offset from the true center assuming a Gaussian distribution with $\sigma=15$\arcsec. 100 different realizations are then averaged to create an `observed' profile.  \textbf{c)} The average profile in panel (b) is  convolved with the PSF of the IRAC 3.6$\,\mu$m (top) and with the SPIRE 250$\,\mu$m beamsize (bottom) as indicated by  grey long dashed lines. \textbf{d)} the correction factor to recover the true profile can be estimated by taking the ratio of the intrinsic and mock observed profiles (panel a and c, respectively) at a fixed scale. Thick red lines mark unity.   
    }
    \label{fig:profile_offset}
\end{figure*}

Once these effects are quantified, we can infer the intrinsic profile by making the appropriate correction. We start by creating a simulated cluster whose radial profile follows a projected NFW profile. Since we are only interested in the relative change in surface brightness, our models have a single parameter, i.e., profile scale-length $r_s$, which is related to the concentration parameter ($c\equiv R_{200}/r_s$). At a fixed $r_s$ value and fixed redshift, we create an image positioned at image center, representing the true profile. Then, we create 100 images of the same profile but with a random offset drawn from a normal distribution with a standard deviation of 15$^{\prime\prime}$. A mean stack of these images is created, resampled, and convolved appropriately to match those of a given band. 
We approximate the IRAC PSFs to be a Gaussian with the pre-warm-mission full-width-at-half-maximum values \footnote{\url{ https://irsa.ipac.caltech.edu/data/SPITZER/docs/irac/iracinstrumenthandbook/5/}} given in the IRAC Instrument Handbook. For the {\it WISE} bands, we use the {\it WISE} instrument PSF \citep{wri10,mei14} to represent the unWISE data at the native pixel scale. The {\it Herschel} SPIRE PSFs are modeled as two-dimensional Gaussians with the appropriate FWHM  (Table~\ref{datatable}).

Radial profiles of the simulated images are measured in the same manner as real data (\S~\ref{sec:radial_profiles}).  In Figure~\ref{fig:profile_offset}, we illustrate how the intrinsic surface brightness of three NFW profiles ($r_s$=0.05, 0.10, and 1.0~Mpc at $z=1.2$, panel a) is altered by the uncertainty in cluster centroid determination (panel b) plus pixel sampling and instrument beamsize (PSF; panel c). The relative change -- quantified by the  ratio of {\it observed} to {\it intrinsic} at a given angular scale, i.e., (c)/(a) -- depends on the steepness of the intrinsic profile, which is illustrated in panel (d). In all but panel (d), all profiles are normalized at the smallest radial bin for clarity. It is evident that the observed profile of the more concentrated NFW model ($r_s=0.1$~Mpc) shows a more pronounced depression near the center while showing an excess at 20\arcsec--40\arcsec.  While the centroid errors, pixel scale, and beamsize always play a role, the the factor dominating the correction needed to recover the intrinsic profile depends on the observation parameters.  As expected, the centroid errors are the most important factor in correcting the {\it Spitzer} and {\it WISE} bands, while the beam and pixel size dominate for the SPIRE bands (see Table~\ref{tab:sample}). Finally, we note that the correction factor also has a mild dependence on redshift due to the change in angular scale.

\begin{figure*}
    \centering
    \includegraphics[width=7in]{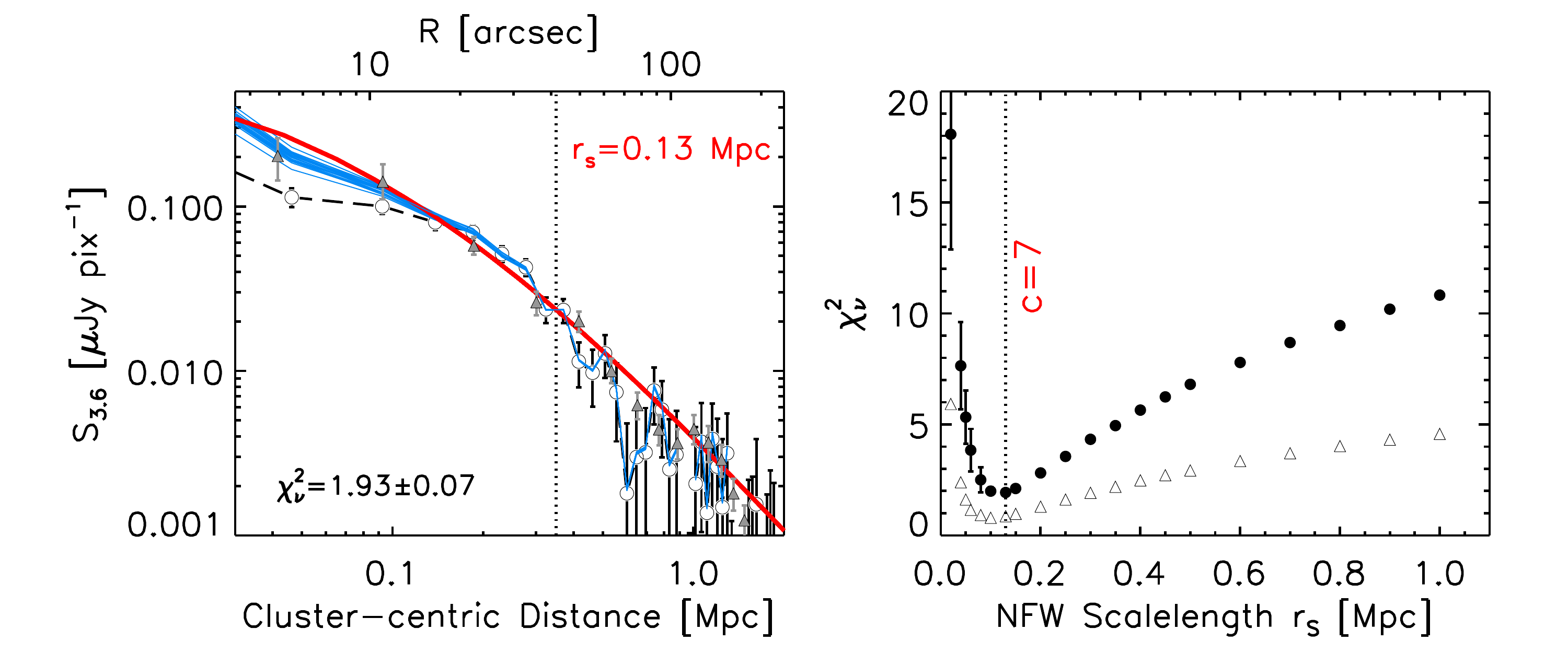}
    \caption{ 
    Our determination of the best NFW model is illustrated. We simulate 20 NFW profiles with scale-lengths ranging $r_s=0.02-1.0$~Mpc, for each of which 20 Monte Carlo realizations are created to account for centroiding errors and beamsize. 
    {\it Left:} the {\it uncorrected} mean surface brightness profile from the {\tt mean\_unmasked} stacks (white circles) in the 3.6$\,\mu$m band is shown for the $1.0<z\leq 1.3$ clusters. Corrections are then applied based on these models to restore its intrinsic profile (blue lines). The corrected profile is then compared to the underlying profile (red line) to obtain the goodness-of-fit $\chi_\nu^2$. Mean and standard deviation of 20 different realizations are indicated on bottom left corner.  We also show \citet{vdb14} measurements for cluster galaxies at $z\sim 1$ (grey triangles), which are in good agreement with our corrected profile realizations (blue lines). All profiles are normalized at $r_{\rm norm}=0.35$~Mpc as indicated by a vertical dotted line.  
    {\it Right:} The computed reduced chi-square values for the 3.6$\mu$m band are shown as filled circles at 20 different $r_S$ values.  Highly concentrated NFW profiles in the range of  $r_S\approx 0.1-1.5$~Mpc  are strongly preferred. The $W1$-band measurements yield similarly low $r_S$ values (open triangles). 
    %Given that the profiles of the $W1$ and 3.6$\mu$m band are nearly identical, so are the best-fit scale-lengths. 
    %\textcolor{red}{Write in text that we get a similar answer using, e.g., $W1$ instead of 3.6 (to be seen). }
    }
    \label{fig:nfw}
\end{figure*}

In Figure~\ref{fig:nfw}, we demonstrate how the observed and intrinsic (corrected) profiles compare in the 3.6$\,\mu$m band using our measurements for the $z=1.0-1.3$ sample. For this, we have assumed 19 NFW profiles with the scale-lengths between  0.02~Mpc and 1.0~Mpc. The virial radius $R_{200}$ at this redshift is $\approx 1$~Mpc \citep{bro07}, and thus, these profiles have concentration parameters $c$ ($\equiv R_{200}/r_s$) in the range 1--50. All curves are normalized such that at $r_{\rm norm}=0.35$~Mpc ($\approx 40$\arcsec\ at $z=1.15$), they match the measured surface brightness given in units of $\mu$Jy~pix$^{-1}$. This normalization radius is chosen such that the cluster-centric distance is not too close to the  range where the correction required for the SB profile is significant ($\lesssim30\arcsec$: see the right panel of Figure~\ref{fig:profile_offset}), but also is not too far such that the signal is still robust. 

We derive the correction factor for the SB profiles as follows. We assume that all 100 simulated clusters lie at the same redshift ($z_{\rm mean}=1.13$) while their centroids offset randomly from the image center in the angular direction. For each NFW model, we create 20 different realizations, thus there are 20 possible ways to correct the observed SB profile. As can be seen in the left panel of Figure~\ref{fig:nfw}, the level of uncertainties in the appropriate correction factor is at best modest and limited to $r\lesssim 10$\arcsec.  We show the observed profile as white circles (and dashed line). For clarity, each of the 20 {\it corrected} profiles are shown as a solid blue line without error bars; the underlying NFW profile is indicated as a red line. The wiggles in the observed SB (white circles) at $\gtrsim 60$\arcsec\ are likely due to a combination of the intrinsic faintness of the cluster signal and the contributions from interlopers. The latter is clearly present in the mean stack of the 3.6$\,\mu$m band shown in Figure~\ref{fig:stacks_highz} as sources lying outside the virial radius.

We determine the best-fit model by evaluating the goodness-of-fit of the corrected surface brightness measures to the model NFW profiles. The $r_s=0.13$~Mpc model provides the best fit as can be seen in the measured $\chi^2_\nu$ values shown in the right panel of Figure~\ref{fig:nfw}. 
The normalization distance has a small impact on the goodness-of-fit. For example, if we normalize at $r_{\rm norm}=0.3$~Mpc, both $r_s=0.13$~Mpc and 0.15~Mpc models yield similar $\chi^2_{\nu}$. However, larger $r_s$ (i.e., lower concentration parameters) models are never favored as they have $\Delta \chi^2_{\nu}=2-10$. We repeat the  fitting procedure for the $W1$ band by comparing the observed $W1$ profile with model simulations with the $W1$ band PSF, and obtain a similar range of $r_S$ as best-fit models as shown in the right panel of Figure~\ref{fig:nfw} (open triangles).  This is not surprising as the two profiles are nearly identical as illustrated in Figure~\ref{fig:wise_radialprofile}. The difference in the PSF size matters little in this case as both profiles are substantially more extended than the respective PSFs. 

Finally, we test the robustness of our results against two possibilities. First, it is possible that the background level (i.e., the zeropoint of the measured profiles) may be overestimated due to the fluxes of infalling galaxies at cluster outskirts. To test this, we have added a constant to each radial bin in increments of of $1\times 10^{-3}$ up to $5\times 10^{-3}$~$\mu$Jy~pix$^{-1}$. As a reference, a typical uncertainty in the radial surface brightness at $\approx 100$\arcsec\ is $\approx (2-3)\times 10^{-3}$~$\mu$Jy~pix$^{-1}$ for the 3.6$\,\mu$m band. Doing so slightly alters the $\chi^2_\nu$ but does not change the best-fit model, which remains in the range $r_s=0.13-0.15$~Mpc. Second, we change the redshift distribution of the clusters. Instead of assuming the mean redshift for all clusters, we populate them at random within the range $z=1.0-1.3$ and scale their overall SB -- according to the cosmological dimming and the change in the angular scale -- before averaging them into a single profile. Once again, doing so does not change the best-fit model. 

In the 3.6$\,\mu$m band, our measurements are in a remarkably good agreement with those measured by \citet{vdb14}, which are shown as grey triangles in Figure~\ref{fig:nfw}. In that work, the stellar mass density of individually detected galaxies was measured ($80\%$ completeness at log $M_{\star}/\Msun = 10.16$), which should in principle scale linearly with the 3.6$\,\mu$m band brightness. The agreement lends us confidence that the radial profile measurements derived from our cluster stacking analysis can yield physically useful insight into the overall distribution of light, provided that measurement biases and uncertainties are well understood and fully taken into account in the analysis. 

\subsubsection{The change of cluster radial profiles with wavelengths and redshift}\label{subsec:spire_profile}

The profile shapes at near- and far-infrared wavelengths  probe past and current star formation activity by sampling stars and dust-obscured star formation, respectively, assuming the contribution from ICD is negligible (see \S~\ref{sec:dust_disc} for a broader discussion on this assumption). By comparing the two we may be able to examine how the cluster environment influences its member galaxies at a given cosmic epoch. Moreover, by measuring the redshift evolution of cluster radial profiles, we can trace the overall effect of its growth (via accretion of more halos, and star formation activity). In this subsection, we investigate how measured cluster profiles change with wavelength and redshift. 

\begin{figure*}
	\includegraphics[width=7in]{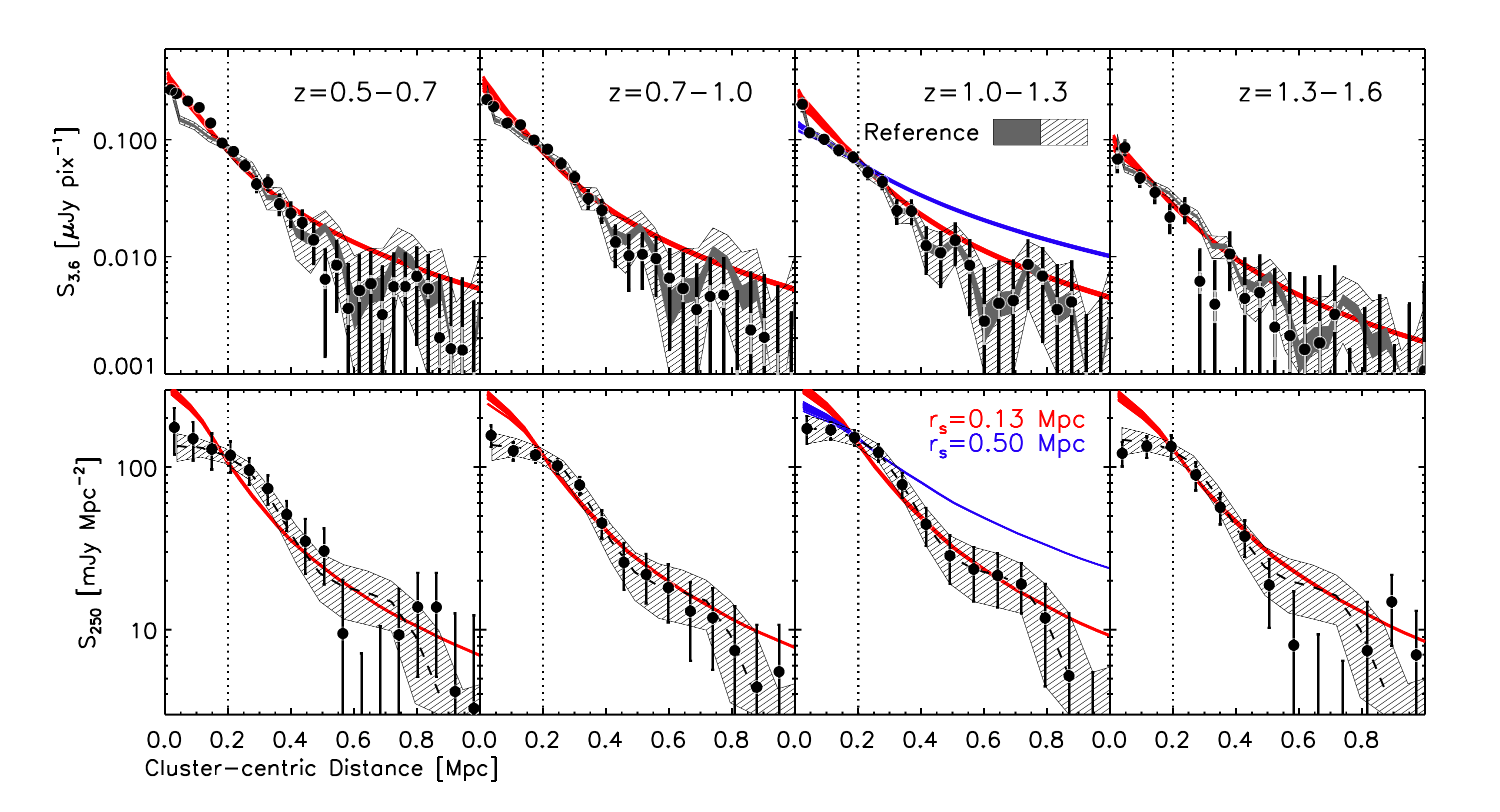}
    \caption{Radial SB profile measurements for the 3.6$\,\mu$m (top) and SPIRE 250$\,\mu$m  (bottom) bands are shown for our four redshift bins. In all panels, hatched regions represent our measurements for the $z=1.0-1.3$ bin for the given band shown  as a reference. For the IRAC profiles (top), we show two different ranges of uncertainties, namely, photometric errors (the fluctuations of the mean and rms of the sky, light gray errors bar for the individual profiles and shaded region for the reference) and possible variations in the mean sky across the full survey field. The two are added in quadrature (thick error bars). Two simulated NFW profiles, corrected for out centroiding uncertainties and beamsize -- $r_S=0.13$~Mpc (red) and 0.50~Mpc (blue) -- are shown to illustrate the steepness of the profile and the range of variations affecting the profile at the smallest scale due to the centroid uncertainties. All profiles are normalized at 0.2~Mpc.  
    }
    \label{fig:profile_zevol}
\end{figure*}

In Figure~\ref{fig:profile_zevol}, we show the measured profiles in the 3.6$\,\mu$m  and SPIRE 250$\,\mu$m bands for the four redshift subsamples, with the $z=1-1.3$ bin reproduced as a reference on all panels. Overlaid are two NFW profiles ($r_S=0.13$~Mpc and $0.50$~Mpc) convolved with the instrument PSF and centroid uncertainties (see \S~\ref{subsec:radprofile_sim}). The similarity of the measured profiles across the full redshift range is striking. In all cases, a highly concentrated NFW model ($c\gtrsim 7$) is preferred over a much shallower model ($c\lesssim 3$). % a paragraph discussing the IRAC profile
Physically the 3.6$\,\mu$m radial profiles are straightforward to interpret: in the ISCS cluster redshift range, the IRAC 3.6$\,\mu$m band samples  $\lambda_{\rm rest}$=$1.5-2.3 \,\mu$m, and thus is sensitive to total stellar mass content within the cluster. Taken at face value, our results suggest that the cluster stellar mass profile changes little over cosmic time. 
We postpone discussing this result in comparison with the literature to \S~\ref{sec:concentration_disc}; here we focus on evaluating its robustness. Specifically, if we are systematically oversubtracting the sky from the cluster signal, the effect can mimic a steep profile. This is conceivable for the near-IR wavelengths given that we have no choice but to estimate the local sky background at $\lesssim 250$\arcsec $-$ due to the distribution of cluster members identified via photometric redshifts (\S~\ref{subsec:wise_spitzer_processing}) $-$ where any NFW profile is expected to have a non-zero signal. The SPIRE stacking is not expected to be affected by local oversubtracting as the sky background is relatively uniform and subtracted on much larger scales.  

We test this possibility using the simulated NFW profiles spanning $r_S=0.1-0.3$~Mpc in scale-length at $z=1.15$. At 200--250\arcsec, the surface brightness level falls off to 0.5--2.0\% of that measured at 5.5\arcsec, which is the first angular bin at which our measurements are made. The higher SB at large radius corresponds to the larger $r_S$ (smaller $c$) value. Scaling from the observed profiles in the same band, the expected correction should be no larger than $2\times 10^{-4}$~$\mu$Jy~pix$^{-1}$. 

In addition to the above bias, we also assess our ability to estimate the sky background, which can also artificially alter the profiles at large radii to be steeper or shallower than the intrinsic one. As described in \S~\ref{wise_spitzer_radial_profile}, each individual image is processed to have `zero sky' before stacking. Since the number of clusters in all four redshift bins is comparable and the IRAC coverage is uniform across the survey field, the expectation is that the residual sky in the stacked images would also be similar. We have checked this and found that the sky background -- measured in the same angular annulus in all cases -- can vary within $(2-3) \times 10^{-4}$~$\mu$Jy~pix$^{-1}$. 

In Figure~\ref{fig:profile_zevol}, we have added  additional uncertainties of $4\times 10^{-4}$~$\mu$Jy~pix$^{-1}$ (i.e., slightly larger than the quadratic sum of the two sources of error) to the error budget to illustrate how such corrections would affect the NFW profile fit. These are shown as hatched regions for the reference sample (our $z=1-1.3$ redshift bin), and thick error bars for the data, respectively. The original errors are shown as dark grey shades and as light grey bars in the same figure. While local sky subtraction will have a tendency to artificially steepen the profiles, the effect is too small to significantly alter the inferred $r_S$ values. Thus, we conclude that the intrinsic profiles of cluster light must be steep.

We also examine how the profiles change with wavelength by comparing the  IRAC and SPIRE measurements. In the lower panels of Figure~\ref{fig:profile_zevol}, it is clear that at small scales ($r \lesssim 0.3$~Mpc), the SPIRE measurements fall below the best-fitting NFW model. In contrast, the IRAC data points are reasonably well described by the steep NFW profile at all scales. If the observed far- and near-IR flux trace dusty star-formation activity and existing stars, respectively, the comparison of the two profiles can tell us about how the specific (dusty) star formation rate varies with cluster-centric distance (modulo the SPIRE beam).

\begin{figure*}
	\includegraphics[width=7in]{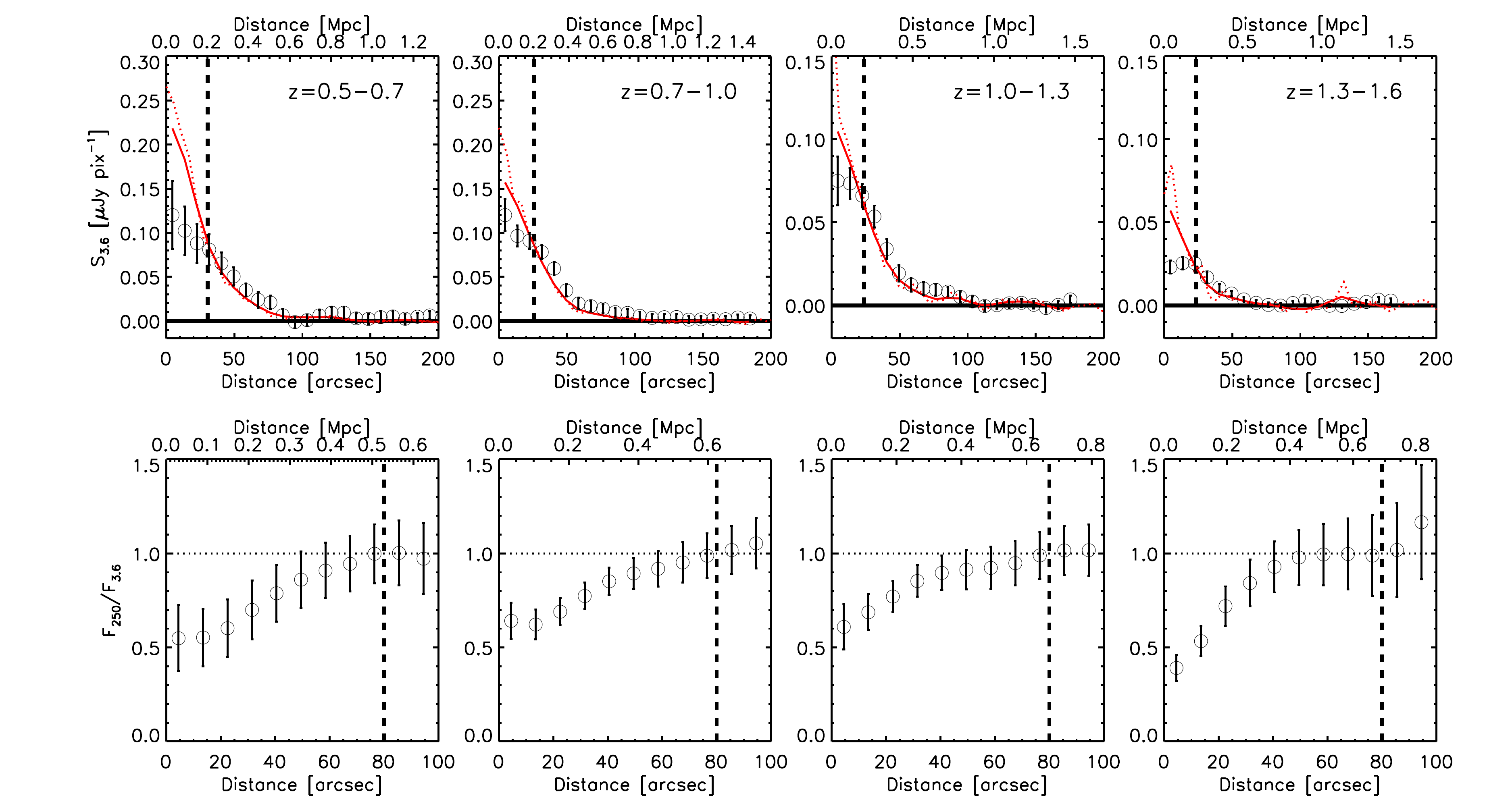}
    \caption{
    Comparison of the surface brightness and flux ratio measurements in the 3.6$\,\mu$m and 250$\,\mu$m bands. 
    {\it Top panels:} the IRAC profiles convolved with the SPIRE 250$\,\mu$m beamsize (red solid line) are compared with the 250$\,\mu$m data (open circles). The latter is rescaled to match the IRAC measurements at $r=0.2$~Mpc.  The unconvolved IRAC SB is shown as a red dotted line. The relative deficit of SPIRE flux at the cluster core is apparent. 
    {\it Bottom panels:} the ratio of the cumulative fluxes ($F_{250}/F_{3.6}$). Both fluxes are normalized at 80\arcsec. 
    }
    \label{fig:irac_vs_spire}
\end{figure*}

In order to study this behavior more clearly, we convolve the stacked IRAC images of each redshift sample with the SPIRE 250$\,\mu$m beam approximated by a 2D Gaussian kernel with FWHM=18.1\arcsec\ (Table~\ref{datatable}). The radial profile is measured directly from the convolved image. In the top panels of Figure~\ref{fig:irac_vs_spire}, we show the SPIRE surface brightness profile rescaled to match the IRAC profile at $r=0.2$~Mpc. 
Suppression of far-IR flux relative to IRAC near the cluster core is evident in all redshift bins.

The absolute magnitude of the deficit -- defined as  $\Delta S_{3.6-250} \equiv S_{3.6}-S_{250}$ -- is the greatest in the lowest-redshift bin ($z=0.5-0.7$). This is not surprising 
because they suffer the least amount of cosmological surface brightness dimming: i.e., given everything equal, both their luminosities and surface brightness (at mid-IR wavelength in the observer's frame) are higher than those of the higher-redshift clusters. Additionally, lower redshift clusters may be intrinsically more luminous due to the continued mass assembly process in clusters. 

An ideal measure to quantify the the suppression of far-IR emission in clusters free of these effects would be $\Delta S_{3.6-250}/S_{3.6}$, a fractional deficit normalized by the IRAC surface brightness at a given radial bin. 
However, such a measure proves problematic. First, both $\Delta S_{3.6-250}$ and $S_{3.6}$ asymptote to zero at large radii, making the measure very noisy. Imperfect sky subtraction can have an exacerbating effect by either exaggerating the noise or suppressing the trend.
As discussed earlier in this section, the over-subtraction of the sky background in the IRAC stack almost certainly plays a role. 

To circumvent these issues, we show the ratio of the cumulative fluxes, $F_{250}/F_{3.6}$, in the bottom panels of Figure~\ref{fig:irac_vs_spire}. Both fluxes are normalized to match at 80\arcsec. Unlike the SB ratios, the flux ratio is insensitive to where we normalize. The deficit of the SPIRE flux is still evident in all redshift bins although now it should be interpreted as the suppression of the dusty SF {\it averaged} within the given circular region, and not as a proxy for the SSFR at the given cluster-centric distance. 

There is a slight hint that the region of suppressed SF is widening with cosmic time. For the clusters in the $z=0.5-0.7$ bin, the flux ratio increases slowly and approaches unity at $\approx0.5$~Mpc. On the other hand, for the highest-redshift clusters, the  increase is a steeper function of cluster-centric distance, and reaches unity at $\approx0.3-0.4$~Mpc. The virial radii of these clusters are expected to be similar in both redshift bins \citep{bro07}. However, the trends seen in all redshift bins are comparable within the uncertainties. 

All in all, we find strong evidence for the suppression of dusty star formation activity in cluster cores in all probed redshift ranges. In order to better quantify the redshift evolution (in both magnitudes and physical scales of the SF suppression), larger cluster samples are needed, which will improve the precision with which we can measure the profile shapes and the sky background. 

\begin{table*}
\centering
\caption{Stacked total cluster fluxes in units of mJy within $R_{200}=1\,$Mpc. 
%\SA{Need to fix sig figs and maybe make sideways?}\KSL{I rearranged it but still looks weird. I sugugest we list log [flux]? }
}
%\resizebox{\textwidth}{!}{%
\begin{tabular}{lcccc}
\hline
\hline
Passbands & $z=0.5-0.7$ & $z=0.7-1.0$  & $z=1.0-1.3$ & $z=1.0-1.3$\\
 & ($z_{\rm mean}=0.57$) & ($z_{\rm mean}=0.86$) & ($z_{\rm mean}=1.13$) & ($z_{\rm mean}=1.44$) \\
%Sample  & $W1$  & $3.6\,\mu$m & $4.5\,\mu$m & $W2$ & $5.8\,\mu$m & $8.0\,\mu$m & $W3$ & $W4$ & $M24$ & $M70$ & $S250$ & $S350$ & $S500$  \\
\hline
unWISE $W1$ & $1.473\pm0.109$ & $0.727\pm0.051$ & $0.532\pm0.053$ & $0.168\pm0.056$\\
IRAC $3.6\,\mu$m & $1.433\pm0.070$  & $0.912\pm0.054$  &  $0.568\pm0.044$ & $0.133\pm0.044$ \\
IRAC $4.5\,\mu$m & $1.053\pm0.077$ & $0.6161\pm0.054$ & $0.531\pm0.047$ &  $0.209\pm0.050$ \\
unWISE $W2$ & $0.744\pm0.0950$ & $0.491\pm0.052$ & $0.449\pm0.053$ & $0.243\pm0.077$\\
IRAC $5.8\,\mu$m &  $0.945\pm0.224$  & $0.484\pm0.129$ & $0.523\pm0.229$ & $0.195\pm0.150$\\
IRAC $8.0\,\mu$m & $1.022\pm0.289$ & $0.424\pm0.110$  &  $0.337\pm0.156$ & $0.227\pm0.102$\\
WISE $W3$ & $1.0\pm0.3$ & $0.9\pm0.2$ & $0.8\pm0.2$ &  $0.2\pm0.2$\\
WISE $W4$ & $2.2\pm1.3$ & $2.7\pm0.5$  & $2.3\pm0.5$ & $0.7\pm0.4$\\
MIPS 24$\,\mu$m & $1.53\pm0.28$ & $1.87\pm0.16$ & $1.84\pm0.20$  & $1.00\pm0.17$\\
MIPS 70$\,\mu$m & $9\pm4$ & $9\pm3$ & $9\pm3$ & $7\pm3$ \\
SPIRE 250$\,\mu$m & $75\pm20$ & $74\pm11$ & $93\pm15$ & $60\pm17$ \\
SPIRE 350$\,\mu$m & $44\pm15$  & $56\pm10$ & 	$88\pm12$ & $63\pm13$\\ 
SPIRE 500$\,\mu$m & $22\pm10$ & $28\pm6$ & $45\pm8$ & $41\pm8$\\
  \hline
\end{tabular}
%}
\label{tab:stacked_fluxes}
\end{table*}
  
\subsection{Total Cluster Flux and Spectral Energy Distributions}\label{sec:sedfitting}

\begin{figure}
    \centering
    \includegraphics[width=3.3in]{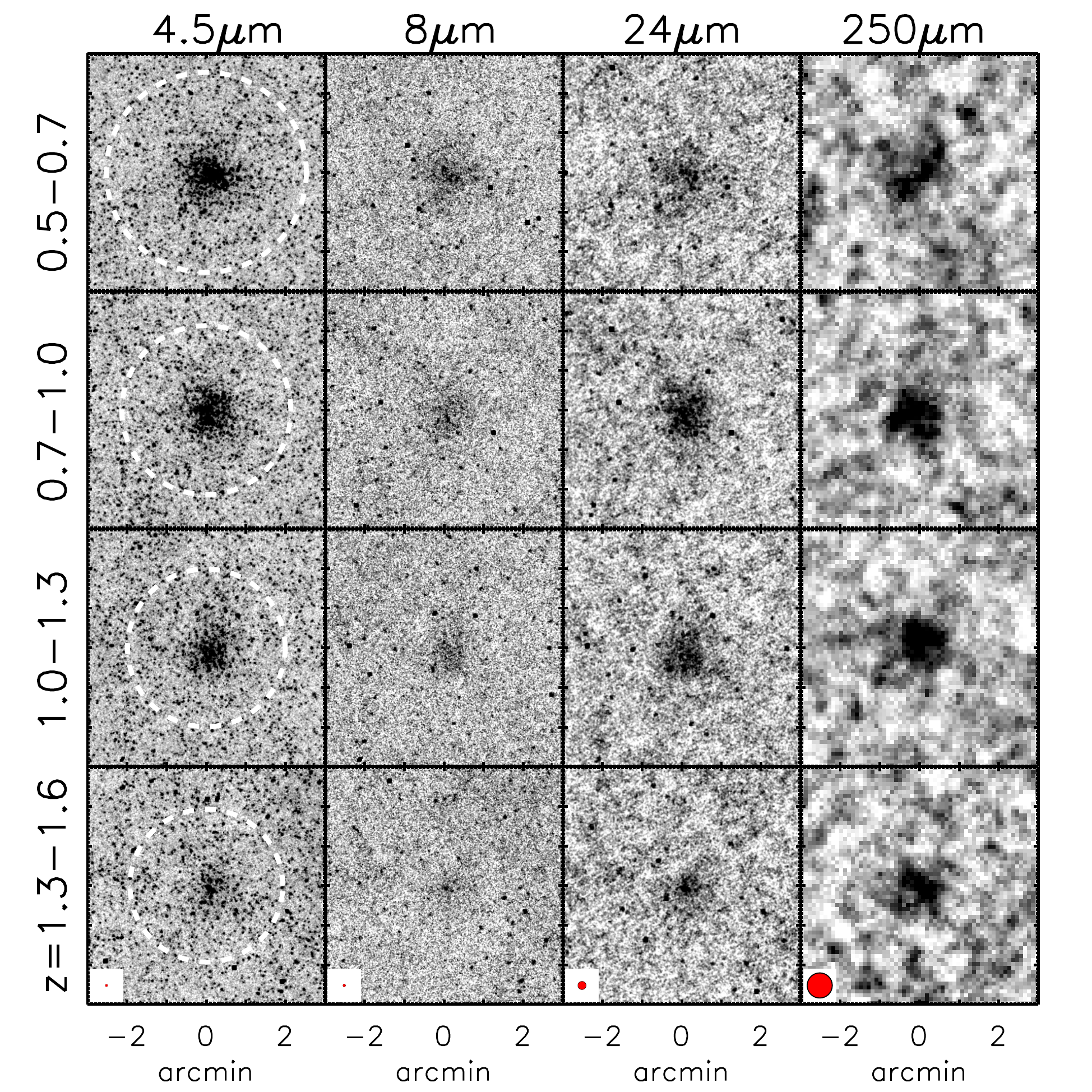}
    \caption{Image stacks in the IRAC 4.5$\,\mu$m, 8$\,\mu$m, MIPS 24$\,\mu$m and SPIRE 250$\,\mu$m bands are shown for four redshift bins. The brightening at far-infrared wavelength relative to the near-infrared with increasing redshift is apparent, perhaps indicating larger dust content and higher star formation activity at larger look-back time. In the first column, we also show a circular aperture corresponding to 1~Mpc in radius (white dashed circles), and the instrument FWHM (red circles in the bottom row).
    }
    \label{fig:rep_stacks}
\end{figure}

In this section, we combine our full stacking results and radial profile analysis in order to measure the total cluster flux as a function of wavelength and redshift.  The total fluxes are then used to build the total cluster light SED.  Figure~\ref{fig:rep_stacks} shows the cluster stacks in all redshift bins for a representative set of observed wavelengths: 4.5$\,\mu$m probes the stellar content of the clusters, 8 and 24$\,\mu$m probe the warm dust content and polycyclic aromatic hydrocarbon (PAH) features, and 250$\,\mu$m traces the far-IR cold dust emission reprocessed from star formation (plus any ICD component, if present).

To measure the total cluster flux, we choose a standard physical aperture of radius 1 Mpc for all redshift bins.  This aperture is the approximate virial radius ($R_{200}$) expected for the clusters in this sample \citep{bro07}.  This aperture is additionally motivated by our results: as demonstrated in \S~\ref{sec:radial_profiles}, the emission beyond 1 Mpc is undetected in all bands (Figure~\ref{fig:wise_radialprofile}) and the correction for beamsize and centroiding uncertainties is minimal (Figure~\ref{fig:profile_offset}).  The stacked total cluster flux densities and associated uncertainties are listed in Table~\ref{tab:stacked_fluxes}.

SED fitting is done using two methods: 1) using {\tt CIGALE} \citep{bur05, nol09, boq19}, a Bayesian SED modeling code which employs multi-wavelength energy balance and 2) a two-temperature modified blackbody fit to the far-IR data only. We note that the former assumes all emission originates from stellar processes in galaxies, while the latter makes no assumptions. Prior to SED fitting, we consider that the $W3$, $W4$, and MIPS 24$\,\mu$m bands probe the complex PAH emission spectrum in the mid-IR. Our stacks are performed in broad redshift bins ($\Delta z=0.3$) such that the PAH and absorption features at the redshifts of individual clusters  will be diluted when combined, with their distinct signatures on the mid-IR spectrum lost. We quantify this effect with a simple simulation assuming all clusters in a given stack have the same intrinsic luminosity and follow the same mid-IR SED of a star forming galaxy at $z\sim1$. We use the individual cluster redshifts to estimate the $W4$ and MIPS24 fluxes for each cluster and compare the mean and range of fluxes obtained for each redshift bin. We find that the simulated fluxes of individual clusters within a given redshift bin vary by factors of 1.2--2.5 due to the complicated shape of the mid-IR SED. This averaging over the complex mid-IR features is not anticipated by SED fitting routines such as {\tt CIGALE}.  Furthermore, despite the similar wavelengths and filter overlap, this effect can lead to differences between the $W4$ and MIPS $24\,\mu$m flux densities, at a level of $4-12\%$ under these simplifying assumptions. Indeed, we find small inconsistencies between these two bands, though typically within the uncertainties, confirming that these measurements are sensitive to the dilution of complex features in this part of the spectrum.  Given these expected variations in the measured stacked fluxes over the mid-IR range in our redshift bins, we exclude the $W3$, $W4$, and MIPS 24$\,\mu$m data points from our SED fitting.

{\tt CIGALE} SED fitting is performed with the stacked photometry, conservatively excluding the three bands as discussed above. 
We parameterize dust emission in {\tt CIGALE} following \citet{cas12} as a single-temperature modified blackbody $-$ tracing cold dust emission from the reprocessed light from young stars $-$ plus a mid-IR power law, which arises from warm to hot dust emission from starbursting regions and/or AGN. 
 Dust temperature is allowed to vary. The dust emissivity index ($\beta=1.5$) and mid-IR power law index ($\alpha=2.0$) are fixed.
Star formation histories are modeled using the {\tt CIGALE} module {\tt sfh2exp}, consisting of two decaying exponentials, representing a short burst and longer term SF components, assuming the \citet{bru03} simple stellar population models and a \citet{cha03} IMF. The resulting best-fit models all find zero contribution from the short burst term. From the best-fit {\tt CIGALE} SEDs, we derive the following parameters with their associated Bayesian uncertainties ($1\sigma$ from the probability distribution function): total stellar mass and infrared (obscured) SFR. The large stellar mass uncertainties are in part driven by the lack of an anchor at shorter wavelengths.  SFR$_{\rm IR}$ is measured from the best-fit total L$_{\rm IR}$ following SFR$_{\rm IR}$ = 1.5$\e{-10}$ L$_{\rm IR}$ \citep{mur11b}. These are listed in Table~\ref{tab:sed}.

To further quantify the far-IR SED, we fit a two-temperature component modified blackbody model to the far-IR data. This is motivated by the work of \citet[][hereafter K15]{kir15} where they found that high redshift galaxy SEDs were better fit to a two-temperature model. This also allows us to quantify the relative contribution from cold and warm dust to what is found in K15 for typical dusty SFGs at $z\sim1$. We fit to the model described in Equation 2 of K15 assuming the dust emissivity, $\beta=1.5$, and fix the two dust temperatures to their best-fit values ($T_{\rm{warm}}=62.4\,$K, $T_{\rm{cold}}=26.4\,$K), only allowing the normalizations of each dust component to vary. 

\begin{figure*}
    \centering
    \includegraphics[trim=10 10 0 0,clip,scale=0.9]{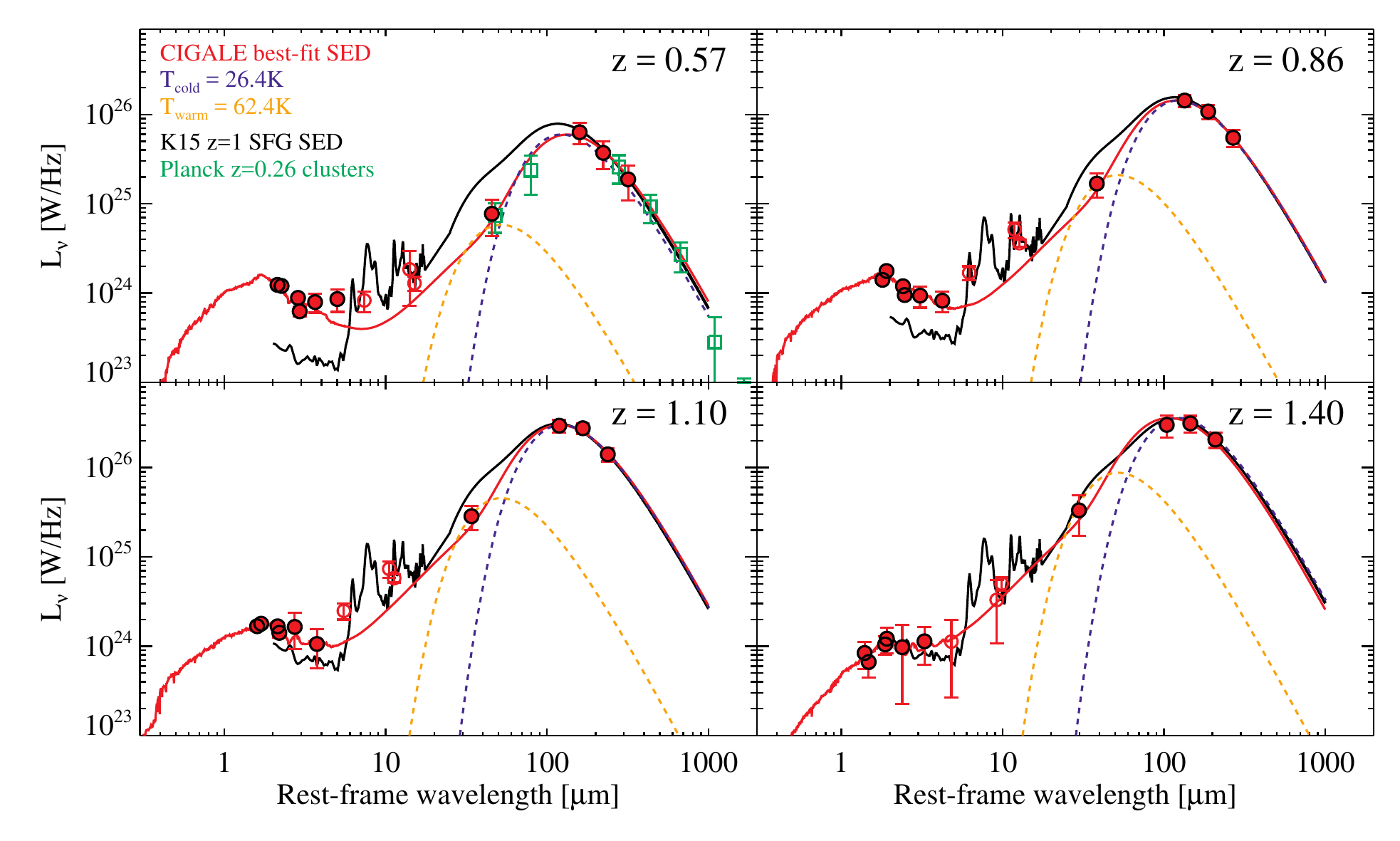}
    \caption{The best-fit CIGALE SEDs (red solid line) of our ``total light" cluster stacks (red data points) in four redshift bins. Closed datapoints were used in the SED fitting, while open datapoints are excluded.  For reference, we show the \citet{kir15} SED for SFGs at $z\sim1$ (black) normalized to the longest wavelength datapoint.  In the upper left panel, stacked photometry from the {\it Planck} cluster sample \citep[$z\sim0.26$,][]{Planck2016} scaled down by a factor of 6 are shown as open, green squares. The orange and blue dashed curves in each panel show the warm and cold dust components from a two-temperature modified blackbody fit where the temperatures are fixed to those used in the K15 model but the normalizations are free. The ratio of $L_{\rm{cold}}/L_{\rm{warm}}$ is 1.7--4.3 for the total cluster light compared to 1.4 for the K15 SED, suggesting more cold dust in the cluster stacks compared to typical massive galaxies. }
    \label{fig:seds}
\end{figure*}

The data and SED fits are presented in Figure~\ref{fig:seds} where the {\tt CIGALE} best-fit is shown in red and the two-temperature model is shown as the dashed curves. The total cluster SED has SFG-like near- to far-IR ratios and a far-IR spectrum well described by a modified blackbody, indicating stellar-heated thermal dust emission. There is no evidence for additional strong power-law dust emission in the mid-IR that would indicate a significant AGN contribution. 
The K15 dusty SFG SED displayed in black has been shown to well represent massive cluster galaxies at $z\sim1-2$ in this cluster sample \citep{alb16}.  The {\tt CIGALE} fit reveals a significant disparity between our total cluster SED and the empirical template in the three lower redshift bins: our total cluster SEDs are lacking in warm dust compared to a typical $z\sim1$ SFG. 
The two-temperature modified blackbody fit is used to quantify this effect. The blue and orange curves show the best-fit cold and warm dust components and the ratio of $L_{\rm{cold}}/L_{\rm{warm}}$ from these fits is given in Table~\ref{tab:sed}. For reference, the K15 SED has $L_{\rm{cold}}/L_{\rm{warm}}=1.4$, whereas in the total cluster SED the cold dust component contributes a factor of $>2$ more in all but the highest redshift bin. This is quantified in the effective dust temperature in Table~\ref{tab:sed} which is a luminosity-weighted average of these two components. The total cluster SED has $T_{\rm{eff}}=30$--36 K whereas $T_{\rm{eff}}=42$ K for the K15 SED which represents massive, luminous galaxies.

In the upper left hand panel of Figure~\ref{fig:seds}, we compare the SED for our lowest redshift bin to ``total light'' stacks for a cluster sample at $z\sim0.3$ from \citet{Planck2016}, scaled down by a factor of 6.  Selected via SZ signal \citep{Planck2014}, this work stacked 645 clusters at 60 and 100$\,\mu$m from IRAS and 350$\,\mu$m - 3 mm using {\it Planck} imaging.  Though this cluster sample is more massive than ours (M$_{200}\sim5\e{14}\,\Msun$ for the {\it Planck} sample) and at a lower redshift, we find excellent agreement in the shape of the far-IR SED, including the relative dearth of warm dust emission. We discuss the reason for the lack of warm dust, or excess of cold dust further in \S~\ref{sec:dust_disc}. 

\begin{table}
%\vspace{1in}
\begin{center}
\caption{SED parameters for our total cluster stacks. Stellar mass and SFR are from the {\tt CIGALE} SED fitting. The effective temperature is the luminosity-weighted temperature from a two-temperature modified blackbody fit. At the bottom, we show the derived values for the K15 SFG SED for comparison.}
\hspace{-0.2in}
\resizebox{\columnwidth}{!}{
\begin{tabular}{ccccc}
\hline
\hline
Average  & Stellar Mass & SFR$_{\rm IR}$  & T$_{\rm eff}$ & $L_{\rm cold}/L_{\rm warm}$  \\
 Redshift & [10$^{12}\,\Msun$] & [$\Msun$ yr$^{-1}$] & [K] &  \\
\hline
%old values from old MIPS measurements
%0.57 & 1.4$\pm$0.5 & 80$\pm$20 & 30 & $10\pm5$ \\
%0.86 & 0.9$\pm$0.4 & 250$\pm$40 & 35 & $3.3\pm0.6$ \\
%1.13 & 1.6$\pm$1.0 & 510$\pm$70 & 34 & $4.1\pm0.7$ \\
%1.44 & 0.7$\pm$0.8 & 630$\pm$100 & 36 & $2.7\pm0.3$ \\
%new values, May 26 from new MIPS measurements
0.57 & 1.4$\pm$0.6 & 98$\pm$21 & $30\pm3$ & $4.3\pm1.6$ \\
0.86 & 0.9$\pm$0.6 & 271$\pm$48 & $35\pm2$ & $2.9\pm0.6$ \\
1.13 & 1.5$\pm$1.5 & 586$\pm$87 & $33\pm1$ & $2.8\pm0.5$ \\
1.44 & 0.7$\pm$0.8 & 711$\pm$142 & $36\pm1$ & $1.7\pm0.2$ \\
 \hline
K15 SFG (z=1) & 0.04 & 57 & 42 & $1.36\pm0.01$ \\
 \hline
 \end{tabular}
 }
\label{tab:sed}
\end{center}
\end{table}
\renewcommand{\arraystretch}{1.0}

\subsection{Comparison of ``Total Light" Cluster Stacking to Individual Galaxy Stacking in the Far-Infrared} \label{sec:individual}

In the previous sections, we used a ``total light" stacking technique to push down the luminosity function to get a more complete accounting of the total infrared light coming from massive clusters.  In this section, we use the spectroscopic and robust photometric redshifts available for this cluster sample to tease out the contributions to the total far-infrared (i.e. the cold dust) from different cluster populations.  Namely, we performed stacking at 250$\,\mu$m on individual, known cluster members in bins of stellar mass.

Individual galaxy stacking is done following the procedure from \citet{alb14}, with minor updates which are summarized here.  Cluster membership is determined from a mass-limited galaxy catalog ($80\%$ complete at log $M_{\star}/\Msun=10.1$) using full photometric redshift probability distributions \citep{eis08, chu14}.  Cluster galaxies within 1 Mpc of the cluster centers are then sorted into the same redshift bins as our ``total light" stacking in two bins of stellar mass ($10.1<\mathrm{log}\,M_{\star}/\Msun<11$ and log $M_{\star}/\Msun>11$). Finally, the {\it Herschel}/SPIRE 250$\mu$m image is stacked at their positions.  A radius of 1 Mpc is chosen for this analysis in order to negate the need for beamsize and centroiding corrections when comparing to the total light stacks (\S~\ref{subsec:radprofile_sim}).

As discussed in \citet{alb14}, two corrections are then applied to the individual galaxy stacks.  First, a baseline correction is applied to account for flux boosting as a result of the large SPIRE beamsize.  This boosting is a strong function of galaxy density \citep{vie13}.  The baseline signal to be removed is determined through stacking random pixels along the line of sight to the clusters \citep[see][for more details]{alb14}.  We update this correction from how it was done in our previous study to incorporate an approximation of the cluster light profile determined from this work: the random pixels stacked are drawn from a normal distribution with a FWHM determined via fitting a Gaussian to our ``total light" cluster stacks (Figure~\ref{fig:stacks_highz}).  As discussed in \S~\ref{sec:spirephot}, the IR profile of our clusters has a weak or no dependence on redshift and a FWHM$\,=600\,$kpc is adopted for this correction in all bins.  The effect of this updated procedure is to increase the correction for flux boosting, particularly for stacks out to large radii (the correction is increased relative to \citet{alb14} by a factor of 2-3 for an aperture of $r\sim1\,$Mpc), by more properly weighting the galaxy distribution toward the center of the cluster (see Appendix~\ref{appendix_spire} for more details).  Secondly, a field correction, which mitigates contamination from field galaxies due to photometric redshift uncertainties, is applied as described in \citet{alb14}. 

\begin{figure*}
    \centering
    \includegraphics[scale=0.7]{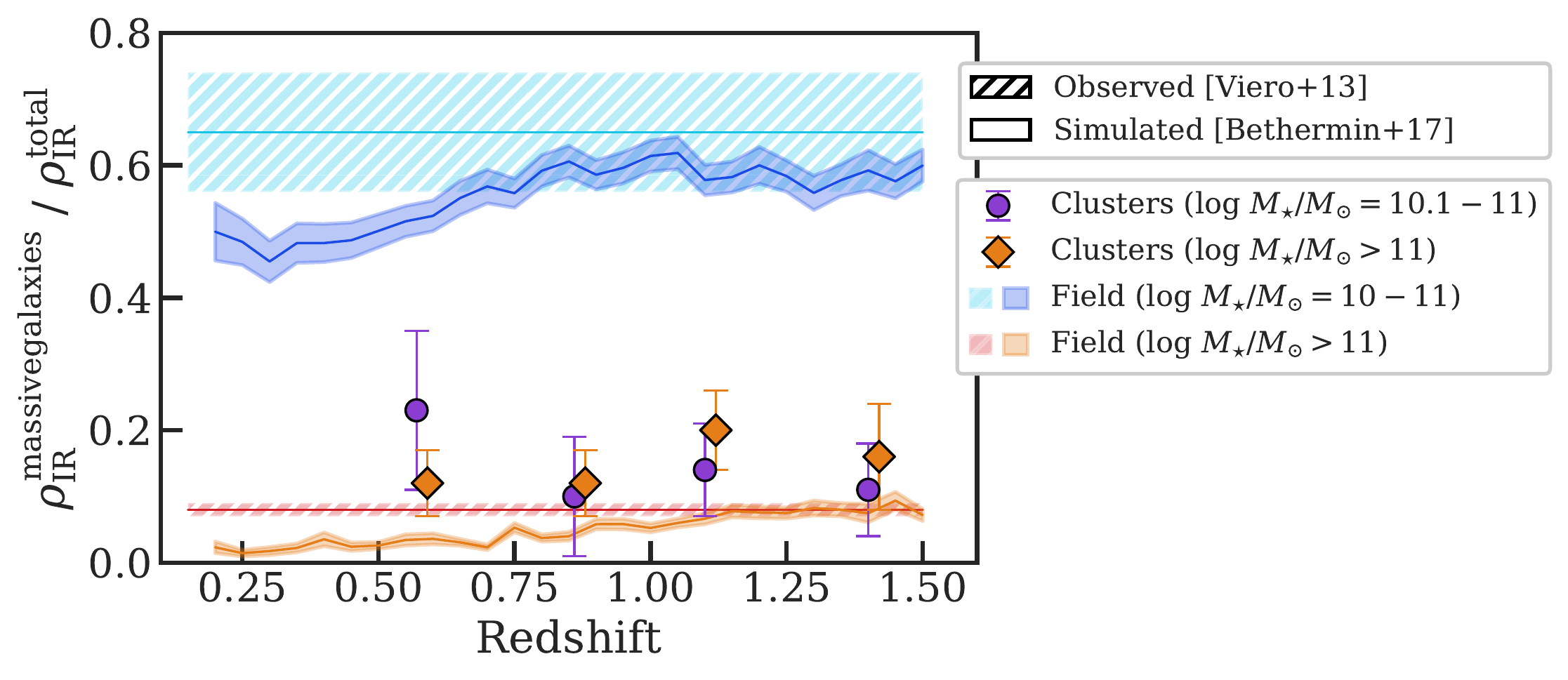}
    \caption{The ratio of the infrared luminosity density contributed from intermediate ($10.1<\mathrm{log}\,M_{\star}/\Msun<11$; purple circles) and high (log $M_{\star}/\Msun>11$; orange diamonds) mass galaxies to the total IR derived from ``total light" stacking for cluster galaxies as a function of redshift.   We compare our cluster results to the field values in the same mass bins derived from deep {\it Herschel} imaging \citep[teal and red lines;][]{vie13} and the SIDES simulation \citep[blue and orange lines;][]{bet17}.  The shaded/hatched regions represent the uncertainty and scatter on this ratio; for the SIDES simulation, this is based on the known uncertainties on the simulation inputs and is likely an underestimate of the true scatter (see \S~\ref{sec:lowmass} for further details).  Although the high mass cluster galaxies agree well with the field ratios at comparable mass, the intermediate mass cluster galaxies sit well below the expected contribution to the total IR from similar mass galaxies in the field.
    }
    \label{fig:ind}
\end{figure*}

These corrected stacks measure the contribution to the cold dust emission at observed 250$\,\mu$m from massive (log $M_{\star}/\Msun>10.1$) cluster galaxies.  By ratioing these stacks to the ``total light" cluster stacks (Figure~\ref{fig:ind}), we find that high mass (log $M_{\star}/\Msun>11$) and intermediate mass ($10.1<\mathrm{log}\,M_{\star}/\Msun<11$) cluster galaxies {\it each} make up 10-20$\%$ of the total cluster light with no discernible redshift trend given our uncertainties. This analysis demonstrates that studies of massive cluster galaxy populations are probing only a fraction of the dust-obscured activity in massive clusters.  Further implications of this result are discussed in \S~\ref{sec:quenching_disc}.

\section{Discussion} \label{sec:disc}

\subsection{Total dust emission in clusters}\label{sec:dust_disc}

In Figure~\ref{fig:seds}, we showed that the total cluster SEDs have relatively more cold dust than warm dust in the far-IR compared to what is seen in massive galaxies in our three lower redshift bins, with a slightly warmer SED the highest redshift bin. At $z\sim0.6$, our SEDs are consistent with the on average cooler stacked SEDs found in an independent study of 645 low-z SZ-selected clusters from the {\it Planck} team \citep{Planck2016}. These total cluster stacks contain all galaxies, low and high mass, as well as ICD. We expect from numerous studies that the ICD component is minimal \citep[][]{che07, bai07, gia08, mcg10,gui14, gui17, lom20}; for example, \citet{bia17} measured an upper limit on the 250$\mu$m emission in Virgo as $I_{250}\sim0.1\,$ MJy sr$^{-1}$, which would be a negligible contribution to our stacks. This leaves the galaxy component to dominate the SEDs.  \citet{alb16} found that massive cluster galaxies fit well to the warmer K15 empirical template found for similarly massive field galaxies and in Figure~\ref{fig:ind} we demonstrate that the massive (log $M_{\star}/\Msun>10$) dusty galaxies contribute 20--30\% of the total IR emission. This indicates that 70-80\% of the IR light is likely originating from lower mass (log $M_{\star}/\Msun<10$) cluster galaxies. Given this, a colder spectrum is expected from the known dust luminosity-temperature relation for the field: $T_{\rm{dust}} \propto L_{\rm IR}^{0.06}$ for nearby IR luminous galaxies \citep{cas14}. This relation implies that galaxies which are 100 times less luminous in the IR will have dust temperatures that are 1.3 times lower. The luminosity-weighted dust temperature of K15 SED is $\sim42$K whereas we get 30--36K for the cluster stacks (Table \ref{tab:sed}), with a possible trend toward warmer temperatures in our highest redshift bin which may be due to i.e. increasing AGN activity \citep{mar13, alb16} and/or changing dust properties. Therefore, the cooler SEDs for the cluster stacks are consistent with the dominant contribution from the lower luminosity, cooler galaxies. While we do not need to invoke ICD to explain the cooler SED, we cannot rule out a contribution from ICD to the total cluster SEDs. 

This result for the average SED in $z<1.6$ clusters can be compared to a recent study of 179 Subaru-selected proto-clusters at $z\sim4$ \citep{kub19}, where a similar total light measurement was made. While the $z<1.6$ clusters show a prominent cool dust component with less warm dust than even typical (high mass) high-$z$ star forming galaxies, the $z\sim4$ proto-clusters exhibit an excess in the mid-IR which would require intense starbursts and/or AGN heating. Simulations predict that proto-cluster growth at $z\sim4$ is characterized by active growth with a high contribution to the total SFR density at those epochs \citep{chi17}. Approximately 4 billion years later these proto-clusters could have evolved into the clusters in our sample.  Though we have evidence that AGN activity within the massive cluster galaxy population increases with redshift up to $z\sim2$ \citep{mar13, alb16}, which may be influencing the warmer SED in our highest redshift bin, in general the gas needed to fuel these intense starbursts and AGN $-$ and propel them to dominate the total cluster light $-$ will have diminished from the proto-cluster to the cluster regime \citep[e.g.][]{liu19}. %\textbf{YK: Liu+19, https://arxiv.org/abs/1910.12883}. 

We can gain further insight into what is dominating the far-IR light in high-$z$ clusters by looking at its distribution relative to the stellar light.  In Figures~\ref{fig:profile_zevol} and \ref{fig:irac_vs_spire}, we showed via the (corrected) radial profiles that the dust emission is similarly extended as the stellar mass distribution, but that the dust emission is suppressed in the core relative to the much stronger peak in the stellar light.  This is expected at lower redshifts \citep[i.e.][at $z\sim0$]{chu11}, where, for example, \citet{rod19} found a suppression of star-formation and a decrease in the SSFR of massive galaxies in the cores of clusters from $z=0.2-0.9$, which they attributed to slow environmental processes such as strangulation.  Extending this analysis for obscured star formation in massive cluster galaxies up to $z=1.75$, \citet{bro13} and \citet{alb16} found that this suppression appears to reverse, on average, at $z\sim1.4$, suggesting a transition epoch dominated by a fast quenching mechanism.  These studies, however, also noted strong cluster-to-cluster variation within relatively small cluster samples ($\sim10-15$), stressing the need for measurements averaged over large cluster samples.  In this work, we have averaged the total light $-$ again dominated by the low mass cluster constituents (Figure~\ref{fig:ind}) $-$ over $\sim40-80$ clusters per redshift bin, finding that the IR suppression is evident, on average, up to $z\sim1.6$. 

Even though we find that the dust emission is, on average, less centrally concentrated than the stellar light in our  cluster sample, the far-IR radial profiles are still best-fit by a relatively high concentration NFW profile.  This appears to be inconsistent with studies that find that blue, star forming galaxies usually have {\it lower} concentrations than their quiescent counterparts \citep{vdb14, hennig17}.  Since these studies probed massive galaxies, one explanation could be that lower mass SFGs, which dominate the total IR signal in the clusters, are more centrally concentrated, while higher mass SFGs are less concentrated, potentially due to preferential quenching at small radii.  To test this, we repeat the analysis discussed in \S~\ref{sec:individual} $-$ stacking massive galaxies individually, then ratioing this to the total light cluster stacks $-$ while masking out the central 200-300 kpc, to determine the fraction of cluster IR light contributed by massive galaxies relative to the total outside of the regime where we see IR suppression.  If massive cluster SFGs are less centrally concentrated, we would expect to find a higher fractional contribution to the total cluster IR light with the core masked. However, we find that the fraction contributed outside of the core is consistent with the fraction including the core, within the uncertainties. It is therefore unlikely that the IR suppression in the cluster cores is driven solely by the preferential quenching of massive cluster SFGs.  Centrally concentrated low mass cluster SFGs are also likely being quenched, in agreement with recent studies which show environmental quenching of low mass (log $M_{\star}/\Msun<10$) galaxies ramps up quickly at $z<1.5$ \citep{kaw17}.
We note a potential alternative explanation for the difference in concentration: previous works that found a low concentration for cluster SFGs based on (optically) blue colors were not accounting for the obscured star-formation component, which dominates in high mass galaxies at these redshifts \citep{mad14,whi17}, and may be particularly relevant in overdense environments, even at low masses (see \S~\ref{sec:lowmass}).

\subsection{The decline of total obscured star formation activity in clusters} \label{sec:quenching_disc}

\begin{figure*}
    \centering
    \includegraphics[scale=0.7]{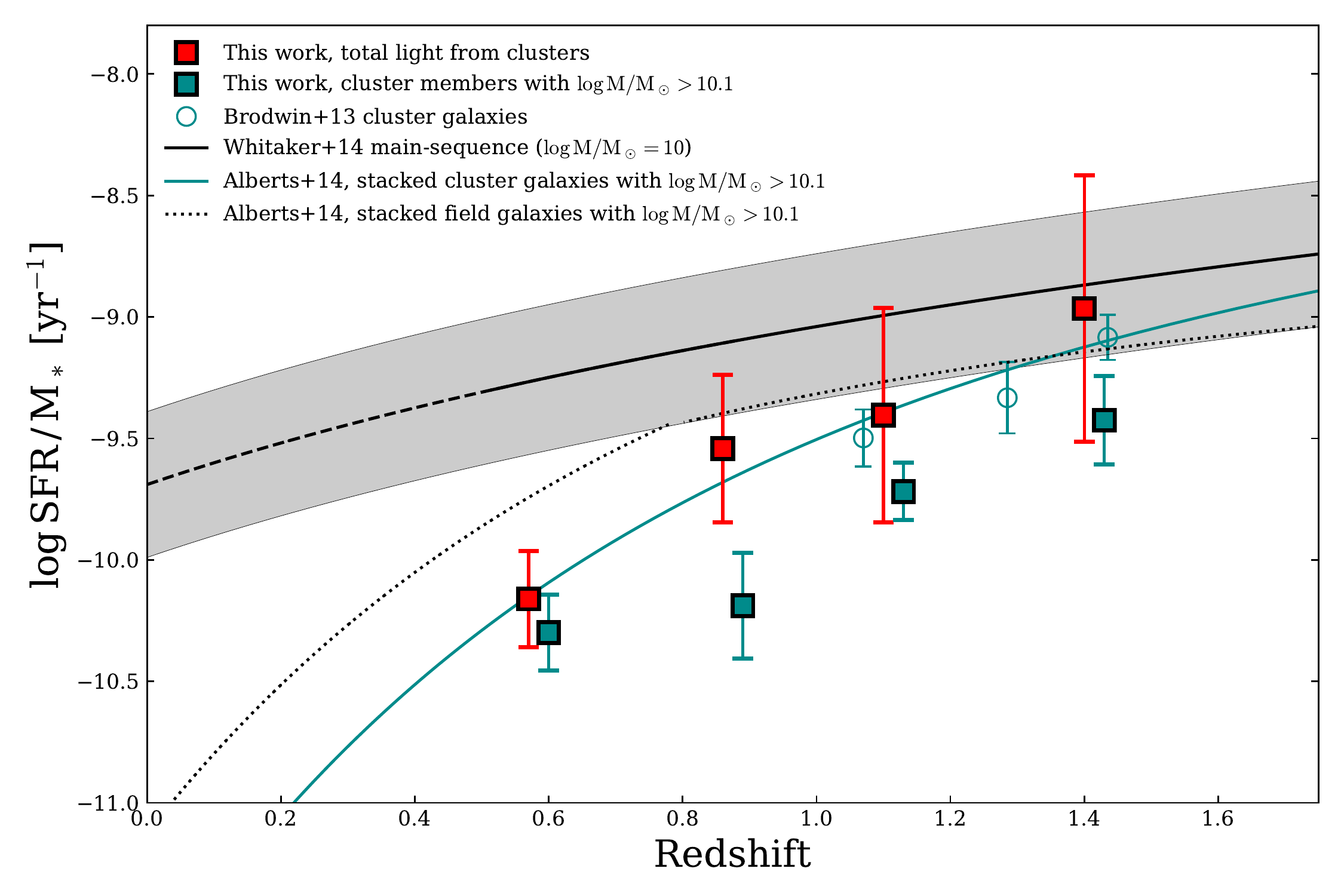}
    \caption{The SSFR enclosed in $1\,$Mpc as a function of redshift, derived from our ``total light" stacks using {\it Spitzer} and {\it Herschel} data to measure the total stellar mass and obscured SFRs, respectively. We compare these results to trends found for massive cluster galaxies (log $M_{\star}/\Msun>10.1$) determined via SPIRE stacking \citep[teal solid line and squares;][this work]{alb14} and from individual detections via deep mid-IR imaging \citep[teal open circles;][]{bro13}.
    The SSFR-$z$ relation for field galaxies with log $M_{\star}/\Msun>10.1$ is shown as the dotted black line \citep{alb14}.  The shaded region denotes the Main Sequence \citep{whi14} for field galaxies at fixed stellar mass  (log $M_{\star}/\Msun=10$) for reference.  The ``total" cluster SSFRs have a steeper evolution than comparable field populations, but reach field-like levels at $z\sim1.5$, in agreement with studies of massive cluster galaxies.  This shows that the high and low mass cluster members are undergoing similar evolution in clusters at this epoch.}
    \label{fig:ssfr}
\end{figure*}

In \S~\ref{sec:sedfitting}, we derived the total SFR$_{\rm IR}$ enclosed in the virial radius ($R_{200}=1\,$Mpc) from our best-fit SEDs (Table~\ref{tab:sed}), assuming a likely negligible contribution from ICD based on the literature, which has obtained mostly upper limits \citep[][but see \citet{lom20}]{che07, bai07, gia08, mcg10,gui14, gui17}. 
We found a dramatic 8-fold decrease in the obscured star formation from $z\sim1.4$ to $z\sim0.6$, a strong evolution that mirrors the evolution of the average SFR$_{\rm IR}$ found for massive cluster galaxies \citep{alb14}.  

Figure~\ref{fig:ssfr} combines the total stellar mass with SFR$_{\rm IR}$ to examine the evolution of the ``total light" SSFR as a function of redshift. We compare this evolution to that measured for massive galaxies only through both mid-infrared detections of individual cluster galaxies \citep[teal open circles;][]{bro13} and far-infrared stacking \citep[teal solid line and squares;][this work]{alb14}.  We find that there is a remarkably similar evolution between the total and massive galaxy only SSFRs, with a similar rate of decline toward lower redshifts.  Figure~\ref{fig:ssfr} shows that the SSFRs of massive cluster members (teal squares) are on average a factor of 2 below the total light SSFRs (red squares).  This implies that the log $M_{\star}/\Msun<10$ cluster galaxies will be on average a factor of 2 higher.  This is roughly consistent with the trend of SSFR with stellar mass observed for field galaxies \citep{whi14}, with the caveat that the field measurements were made for SFR$_{\rm UV+IR}$, with the UV component expected to be significant for the low mass galaxies \citep{whi17}.  We additionally note that the bootstrapped uncertainties for the ``total light" SSFR are large ($0.2-0.5$ dex), which likely includes the significant cluster-to-cluster variation that has been observed before \citep[i.e.][]{bro13, alb16}.  

Figure~\ref{fig:ssfr} further compares the evolution of the total and massive cluster SSFRs to field measurements:  stacked field galaxies with log $M_{\star}/\Msun>10.1$ \citep[dotted black line;][]{alb14} and the Main Sequence at fixed mass \citep[log $M_{\star}/\Msun = 10$, solid/dashed black line and shaded region;][]{whi14}.  Relative to the field, the total cluster measurements are comparable in star formation activity per unit stellar mass at the highest redshift bin probed ($z\sim1.3-1.6$), with rapid quenching below the field level thereafter.  This result reveals that the transition epoch around $z\sim1.4$ for massive galaxies in clusters from abundant star formation to significant quenching found in previous studies \citep{bro13, alb14, alb16} also holds for the lower mass galaxies probed by the total cluster measurements.  

Which cluster galaxies are driving the evolution in the total obscured cluster activity?  In addition to massive cluster galaxies SSFRs being roughly a factor of two lower than that of the total cluster SSFRs, in Figure~\ref{fig:ind} we showed that intermediate ($10.1 < \mathrm{log\,} M_{\star}/\Msun < 11$) and high (log $M_{\star}/\Msun > 11$) mass cluster galaxies together only account for a minority ($\sim20-30\%$) of the far-infrared emission in these clusters at any redshift probed in this work.  Given the similar evolution with redshift between the total and massive cluster galaxies in SFR and SSFR, this indicates that low mass (log $M_{\star}/\Msun\lesssim10$) cluster galaxies are experiencing quenching alongside their massive cousins during this epoch.  This result is consistent with recent literature which has examined the quenched fraction and environmental quenching efficiency $-$ the fraction of cluster galaxies that would be forming stars in the field $-$ down to similarly low masses.  Work based on nearest neighbor techniques in the ZFOURGE survey \citep{str16} found that the environmental quenching of low mass (log $M_{\star}/\Msun\sim9.0 \,(9.5)$ at $z\sim1.3\,(2)$) galaxies in overdense regions is very inefficient at $z\gtrsim1.5$, but ramps up quickly at lower redshifts \citep{kaw17}.  Using the same sample and studying the stellar mass function (SMF) of quiescent cluster galaxies, \citet{pap18} demonstrated that environmental quenching must be mass dependent at high redshift, with the efficiency of quenching low mass galaxies in overdense regions growing between $z\sim1.5$ and $z\sim0.5$, in order to match the faint end of the SMF.  A study of more massive clusters (log $M_{200}/\Msun\sim14-15$) found a similar ramp up in the faint end of the red sequence (RS), with a distinct deficit in faint RS galaxies at $z\sim1-1.3$ followed by a factor of 2 growth by $z\sim0.6$ \citep{cha19}.  This ramp up may complete by $z\sim0.6$; \citet{vdb18} found that the environmental quenching efficiency in massive clusters was independent of stellar mass at fixed radius down to log $M \sim9.5$. Together with this work, the emerging picture is that the environmental quenching of low mass (log $M_{\star}/\Msun <10$) galaxies at $z\lesssim1.5$ is a significant driver of cluster galaxy evolution.
%AP: I followed most of this and it flows nicely except for the sentence that goes "A study of more massive clusters..factor of 2 growth by $z\sim0.6$." is that referring to the van der burg study mentioned in the next sentence? it is not clear that you need this sentence since the point of the vdB study is clear from the sentence that goes "At z~0.6..M~9.5." 

\subsection{The low mass contribution to the total far-infrared emission}\label{sec:lowmass}

To further shed light on the nature of low mass cluster galaxies, we compare the contribution from massive galaxies to the total IR light to two studies which quantify this ratio for the field (Figure~\ref{fig:ind}).  First, we compare to deep blank field {\it Herschel} observations.  \citet{vie13} used unbiased stacking methods on $K$-band selected galaxy catalogs to estimate the contribution to the Cosmic Infrared Background (CIB) from various populations. This includes mass-limited galaxy samples down to log $M_{\star}/\Msun=9$, below which low mass {\it field} galaxies are expected to contribute little to the CIB \citep[$<30\%$ of their SFR is obscured;][]{whi17}.  The observational results of this field study are that intermediate mass ($10 < \mathrm{log\,} M_{\star}/\Msun < 11$) field galaxies dominate the CIB at 250$\,\mu$m ($65\pm9\%$), with high mass galaxies contributing a minimal amount ($8\pm1\%$), up to $z\sim4$.

Secondly, we compare to the predicted field properties from the publicy available Simulated Infrared Dusty Extragalactic Sky \citep[SIDES;][]{bet17}.  This simulation provides a 2 deg$^2$ blank field of log $M_{\rm halo}/\Msun \sim 10-12$ DM haloes, which are populated with galaxies via abundance matching and observed stellar mass functions.   Galaxy infrared properties are then assigned using the 2SFM galaxy evolution model \citep{sar12, bet12, bet13} and the \citet{mag12} SED library.  The resulting catalog\footnote{\url{http:cesam.lam.fr/sides}} provides the 250$\,\mu$m flux densities for galaxies down to log $M_{\star}/\Msun = 8$.  The SIDES predicted ratio of the luminosity density contributed by intermediate and high mass galaxies to the total is shown in Figure~\ref{fig:ind}, with the shaded regions representing the combined input scatter into the simulation for the MS \citep[0.3 dex;][]{sch15, ilb15} and the SED library \citep[0.2 dex in $<U>$;][]{mag12}, which will directly affect the 250$\,\mu$m flux.  We caution that this simulation extrapolates to low mass and low $L_{\rm IR}$, a regime that is not well constrained observationally, and so the uncertainty is likely larger than depicted.  To first order, the SIDES simulation agrees with the observed contributions to the CIB from \citet{vie13} at each of these two stellar mass bins.

This comparison to the field reveals that high mass ($\mathrm{log\,} M_{\star}/\Msun > 11$) cluster and field galaxies have similar (minimal) contributions to the total IR budget. With our dataset, it is difficult to disentangle mass- and environmental-quenching in this case. The difference between the cluster and field intermediate mass ($10.1 < \mathrm{log\,} M_{\star}/\Msun < 11$) galaxies, on the other hand, is striking and suggests significant environmental differences.  Compared to the percentage in the field ($65\pm9\%$), intermediate mass clusters galaxies only account for $15\pm5\%$ (averaged over our redshift bins) of the total infrared output, a factor of 4 deficit.  This low fractional contribution at all redshifts is puzzling.  By $z\sim0.6$, we expect that a large fraction of intermediate mass cluster galaxies have quenched \citep[i.e.][]{muz12, nan17}; however, as discussed earlier, lower mass cluster galaxies are also quenching, such that at least one study finds no mass dependence for the quenching efficiency at this epoch \citep{vdb18}.  At the high redshift end of our sample, the low fractional contribution suggests that intermediate mass galaxies largely quenched earlier than  $z\sim1.5$.  This has not, however, been found to be the case in the literature.  A recent look at the environmental quenching efficiency as a function of redshift for log $M_{\star}/\Msun>10.3$ cluster galaxies \citep{nan17} found that while the quenched fraction is $80\%$ by $z\sim1.3$, rising to $88\%$ at $z<1.1$, it is only $42\%$ at $z\sim1.6$, consistent with the quenched fraction in the field \citep[see also][]{vdb13}.  This decrease in environmental quenching efficiency is congruous with the findings of field-like SF activity at these high redshifts \citep{bro13, alb14, alb16} and together these different approaches suggest that intermediate mass galaxies should contribute significantly to the IR budget in clusters.

Further complicating the picture at high redshift is that the total cluster SSFR is within a factor of $\sim3$ (given the $1\sigma$ uncertainty) of the field activity, suggesting that minimal SF activity is ``missing", assuming there is no strong reversal of the SFR-density relation.  Yet if the intermediate (and high) mass cluster galaxies are quenched at earlier times, then at $z\sim1.4$ the observed SF must be provided by a population that does not provide it in the field.  This could be solved with a significantly different ratio of low to intermediate/high mass SFGs in clusters than in the field; however, studies of the star forming SMF in low vs high density environments find them to be remarkably similar \citep{vdb13, pap18}.  Alternatively, low mass cluster SFGs could have enhanced SFRs.  Studies of low mass cluster galaxies, however, typically find they reside on or below the Main Sequence \citep{old20}, though we note that these studies often do not directly trace the obscured component.  

There is a third possibility: the mode of star formation in cluster galaxies is different than in the field without being enhanced; that is, that cluster galaxies are dustier than their field equivalents \citep{koy11, sob16, hat16}.  Overdensities of DSFGs are increasingly being found to be signposts for proto-clusters at $z\gtrsim2$ \citep[i.e.][]{kat16, cas16}; however, there have been few studies looking at the dust content in clusters relative to field galaxies at the epoch relevant to this study.  \citet{hat16} found an excess of red galaxies in the main cluster (log $M_{200}/\Msun=13.76$) forming in a proto-cluster at $z=1.6$.  Using SED fitting and MIPS 24$\,\mu$m photometry, they determined that this excess is comprised of both passive cluster members and DSFGs, with the DSFGs occurring at a rate of 3 times that found in group and field environments.  Though they did not establish a mass dependence for this red excess of DSFGs, they found it extended to their low mass cutoff (log $M_{\star}/\Msun>9.7$).  Accounting for their selection effects $-$ biased particularly against faint, red galaxies $-$ they estimated an upper limit for the red fraction at their low mass end, which could accommodate an additional factor of 2 in low mass DSFGs in the cluster environment.  This finding of excess dust may be indirectly corroborated by studies showing enhanced gas fractions  \citep{mok16, nob17, nob19, hay18}, assuming similar dust/gas ratios, and enhanced gas-phase metallicities in overdense environments both locally and at high redshift \citep{ell09, pen14, mai19a, mai19b, fra20}.  

In summary, the behavior of the total cluster SSFR, particularly at high redshift, and the low fractional contribution of intermediate mass cluster galaxies to the total IR budget likely indicates that multiple effects are at play. We posit that mass-dependent environmental quenching which evolves with redshift plus a more obscured mode of SF in lower mass cluster galaxies may be required to explain the behavior of the total far-IR emission in clusters.  We note that throughout this section, we have relied on the obscured SFR and SSFR, meaning the unobscured component is unaccounted for.  If low mass cluster galaxies have largely unobscured SF like field galaxies \citep{whi17}, this component may be significant and boost the SSFR even more relative to the field at high redshift.  Measuring the unobscured component in addition to the dust mass directly are key to disentangling these effects.  We discuss the prospects for this in \S~\ref{sec:implications}.

\begin{figure*}
    \begin{center}
         \includegraphics[width=0.97\textwidth]{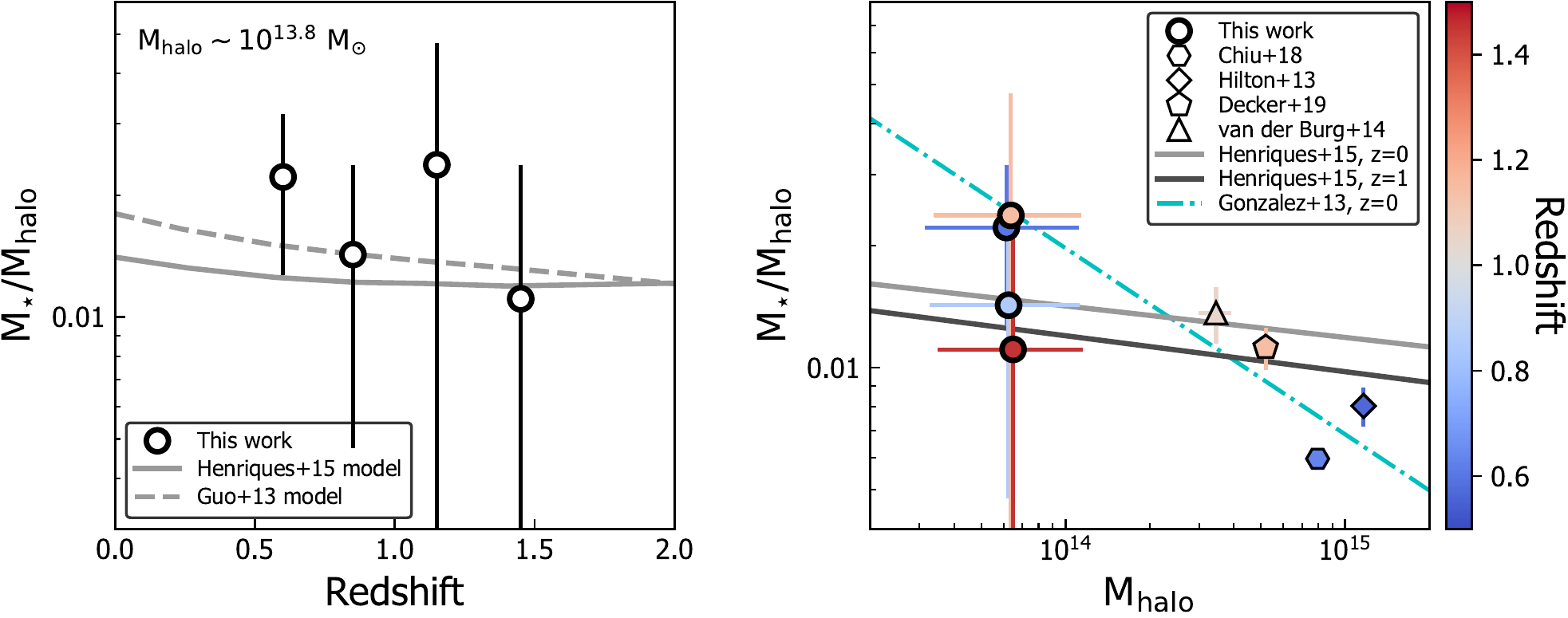}
    \end{center}
    \caption{The average total stellar to halo mass ratios $M_{\star}/M_{\rm halo}$ as function of redshift (left panel) and halo mass $M_{\rm halo} = M_{200}$ (right panel). Circles in both panels show the measurements for our cluster sample of roughly constant halo mass (log $M_{200}/\Msun = 13.8$) at four redshift bins, with the total stellar masses constrained buy the SEDs of the ``total light'' stacks (Figure~\ref{fig:seds}; Table~\ref{tab:sed}). For comparison, we show the semi-analytic model predictions from \citet{hen15} and \citep{guo13} in the left panel, and \citet{hen15} in the right panel. Cyan dashed-dotted line on the right panel shows the observed relation for an X-ray selected cluster sample at $z=0$ from \citet{2013ApJ...778...14G}. Literature measurements for several more massive cluster samples at $0.5\lesssim z \lesssim 1$ are also shown on the right panel with various symbols as labeled.}
    \label{fig:ms_mh}
\end{figure*}

\subsection{Stellar to halo mass ratios}
\label{sec:sims_disc}

Here we discuss the stellar mass content of clusters with respect to their dark matter content, a ratio determined by the balance between the assembly of dark matter and externally formed stars, as well as {\it in situ} star formation. As mentioned in \S~\ref{sec:sample}, based on clustering measurements, the mean halo mass of our cluster sample is about $10^{13.8}~\Msun$ with no significant redshift evolution \citep{bro07, lin13, alb14}; in other words, our cluster samples should be treated as roughly fixed mass, rather than as an progenitor sample. Accordingly, the total stellar mass per cluster based on the full SED fitting (constrained primarily by the near-IR {\it WISE} and {\it Spitzer} bands) is also roughly constant over the four redshift bins (Table~\ref{tab:sed}). This gives a total stellar-to-halo mass ratio, $M_{\star}/M_{\rm halo}$, of about 1--3$\%$ for our clusters as shown in Figure~\ref{fig:ms_mh}. The left panel shows the lack of a significant redshift evolution of this quantity for our clusters, while the right panel shows the trend of $M_{\star}/M_{\rm halo}$ with halo mass, with additional measurements from more massive cluster samples over a similar redshift range \citep{2013MNRAS.435.3469H, vdb14,2018MNRAS.478.3072C, 2019ApJ...878...72D}. 
%that, when combined with measurements of more massive cluster samples over the similar redshift range \citep{2013MNRAS.435.3469H, vdb14,2018MNRAS.478.3072C, 2019ApJ...878...72D}, $M_{\star}/M_{\rm halo}$ appears to decrease with increasing halo mass. 

As our stacking measurements capture the stellar mass content in not only the central but also satellite galaxies and intracluster stars, it is insufficient to compare our result with the often referenced stellar mass--halo mass relations for central galaxies \citep[e.g.,][]{2010ApJ...710..903M}. Instead, we extract and show in Figure~\ref{fig:ms_mh} the total stellar mass--halo mass ratios predicted in the semi-analytic galaxy models of \citet[][solid lines]{hen15} and \citet[][dashed line]{guo13} coupled with the Millennium suite of cosmological $N$-body simulations \citep{spr05}. In the left panel, the model predictions are extracted by selecting the most massive clusters in the $\sim(500/h\ \rm Mpc)^3$ simulation box down to a redshift-dependent mass threshold chosen such that the mean halo mass of the sample is a constant at log $M_{200}/\Msun=13.8$. This selection is designed to mimic that of our observations. Overall, we see slightly different $M_{\star}/M_{\rm halo}$ versus redshift for the mass-matched clusters in the two models due to different implementations of quenching, while both models are still consistent with our measurements for observed clusters within the errors. We also extract the halo mass dependence of $M_{\star}/M_{\rm halo}$ predicted in the \cite{hen15} model at $z=1$ and $z=0$ shown in the right panel of Figure~\ref{fig:ms_mh} using dark- and light-gray solid lines, respectively. 

In addition to the $M_{\star}/M_{\rm halo}$ measurements shown from this work and the literature, we also compare the $M_{\star}/M_{\rm halo}$ versus halo mass relation estimated for a sample of X-ray selected clusters at $z\sim 0$ from \cite{2013ApJ...778...14G} in the cyan dashed-dotted line. The \cite{hen15} model predicts a relatively weak relation between $M_{\star}/M_{\rm halo}$ and halo mass compared to that observed at $z=0$ in \cite{2013ApJ...778...14G}.  From the data compiled it is difficult to distinguish these cases; the constraining measurements from the literature shown suggest a steep slope but unfortunately our cluster sample supports both scenarios with large uncertainties.  Providing constraining anchors at additional halo masses will have important implications for how environmental quenching is implemented in the models.  Improvements in the empirical measurements will come from wide-area cluster surveys that span very large dynamical ranges in halo mass.  For instance, the all-sky MaDCoWS survey \citep{gon19} contains thousands of $z\sim1$ groups and clusters with both IRAC and {\it WISE} imaging spanning a halo mass range of $13.2 \la \log{M_{\rm halo}/M_\odot} \la 15.2$.  This will allow for uniform measurements of both (richness-based) halo and stellar masses, thus minimizing systematic errors along the halo mass axis.  The 1500 deg$^2$ SPT-3G cluster survey \citep{ben14}, though limited to masses
above $\sim 10^{14} M_\odot$, will provide accurate SZ masses for all systems.
These and other wide-area surveys will much better constrain the
$M_{\star}/M_{\rm halo}$ versus halo mass at relation at high-$z$.

%and perhaps also the higher redshift data points that we compile. This suggests that the environmental quenching in the model might not be implemented accurately and could potentially be improved by using the observed trend to constrain the set of efficiency parameters of difference quenching mechanisms in the model.

\subsection{The concentration of the total stellar mass in intermediate mass clusters at $z\sim0.5-1.6$} \label{sec:concentration_disc}

In \S~\ref{sec:radial_profiles}, we presented the ``total light" averaged radial flux profiles for our cluster stacks in four redshift bins and across our wavelength bands.  We compare our results to a fiducial model, a projected NFW profile \citep{nav96}, which has been shown to describe the cluster stellar mass distribution \citep[e.g.][]{lin04, muz07, vdb14, hennig17, lin17}. From fitting an NFW profile to the (corrected) 3.6$\,\mu$m radial profile $-$ which probes $\lambda_{\rm rest}=1.5-2.3$ for our sample and therefore is expected to scale linearly with stellar mass $-$ we found our profiles heavily favor scale lengths of $r_s=0.13-0.15\,$Mpc, corresponding to a concentration of $c\approx7$, with no significant redshift evolution across our sample.

Only a handful of concentration parameter measurements exist in the literature at similar redshifts, based on individual galaxy distributions.  \citet{vdb14} used a $K$-band selected catalog to look at the stellar mass radial profile in ten $z\sim1$ clusters with a typical mass of log $M_{200}/\Msun\sim14.3$.  They found a similar high concentration of $c\sim7$ for massive (individually detected) galaxies.  Using SPT-SZ \citep{sto13} clusters, \citet{hennig17} found hints of high concentrations ($c\approx5-10$) in their stacked number density radial profiles at $z\sim0.8-1$ for log $M_{200}/\Msun \sim 14.8$ clusters, and noted only statistically weak trends with redshift and no significant trend with halo mass as well as a large scatter in individual cluster profiles. In contrast, \citet{lin17} found $c\sim3$ for $z\sim1$ HSC-SSP clusters with comparable masses to the \citet{vdb14} sample.  Additional studies at $z\sim1$ find concentrations of $c\approx3-4$ for more massive cluster samples \citep[log $M_{200}/\Msun \gtrsim 14.5-15$;][]{capozzi12, zenteno16}.

In this work, we have used a uniformly-selected cluster sample to produce radial profiles at near fixed halo mass (\S~\ref{sec:data},\ref{sec:sedfitting}) over a range of redshifts.  We note that averaging via stacking reduces the influence of sub-structure on the measured concentration \citep{gao08}. The uniformity of our radial profiles and high measured $c$ values at a range of redshifts suggests that halo mass, rather than redshift, is the dominant factor in determining the stellar mass concentration.  This is consistent with progenitor studies $-$ which indicate $c$ evolves strongly with cluster growth \citep{vdb15} $-$ and with the generally lower concentrations found for more massive clusters at low redshift \citep[i.e.][]{lin04, muz07, budzynski12} and up to $z\sim1$ \citep{capozzi12, zenteno16, lin17}.

Observations of the dark matter profiles of clusters via lensing and X-ray imaging have found a steep dependence of $c_{\rm DM}$ on the halo mass \citep{ser13, amo16}.  This dependence, however, has not been been reproduced in the studies of stellar mass or galaxy number density concentrations where a range of halo masses are available \citep[i.e.][]{budzynski12}. Simulations have similarly not yet converged on what detailed effects baryons have on the total concentration in high mass systems \citep[i.e.][]{gne04, duf10, deb13}.  A strict halo mass dependence would, for example, put our results at odds with the \citet{vdb14} results; however, this is again not a fair comparison as ``total light'' stacking is probing a different population than that which requires individual galaxy detections.  In order to reconcile observed cluster concentrations and inform simulations, more uniform cluster selection and depth of observations is needed to compare like populations and dynamical states over a range of halo masses. Techniques such as stacking further minimize cluster-to-cluster variations such as sub-structure, stressing the importance of large samples.

\subsection{Implications and prospects for future studies} \label{sec:implications}

\subsubsection{The total unobscured star formation in massive clusters}
\label{sec:implications_uv}

Throughout this work, we have probed the far-IR as a proxy for (obscured) star formation.  For massive field galaxies near cosmic noon, the obscured component dominates the SFR budget in SFGs; however, lower mass field galaxies are found to have a lower fraction of their star formation obscured by dust \citep[$f_{\rm{obscured}}=\rm{SFR_{IR}\,/\,SFR_{UV+IR}}=0.5$ for galaxies with $\mathrm{log}\,M_{\star}/\Msun=9.4$;] []{whi17}. Our results (Figure~\ref{fig:ind}) show that the total far-infrared cluster stacks are dominated by lower mass, lower luminosity galaxies \citep[$\mathrm{log}\,M_{\star}/\Msun<10$, equivalent to SFR$\,\lesssim10\,M_{\odot}/\rm{yr}$ for main sequence galaxies at $z\sim1$,][]{whi14}. If this low mass cluster population follows the field in terms of dust properties, this predicts a significant unobscured SF component which must be added to the total SF budget.  If, on the other hand, cluster galaxies are dustier than their field counterparts (as discussed in \S~\ref{sec:lowmass}), this component will be limited.  An adaption of our stacking technique to work with {\it GALEX} UV data, which directly probes unobscured SF, will be presented in future work for the Bo\"{o}tes cluster sample and other (proto-)cluster surverys. 

\subsubsection{Quantifying the dust mass and gas content in large (proto-)cluster samples}
\label{sec:implications_gas}

The results in this work demonstrate that the far-IR component provides vital information about cluster evolution for high-$z$ clusters.  However, as shown in Figure~\ref{fig:seds}, our current data does not extend much beyond $250\,\mu$m in the rest-frame for clusters at $z>1$. The longer (sub-)millimeter wavelengths are a very clean tracer of the total dust mass, a proxy for the total ISM mass assuming constant gas-to-dust ratios \citep[e.g.][]{sco14}. While ALMA has the sensitivity to detect massive, IR-luminous galaxies individually in high redshift (proto-)clusters \citep[i.e.][]{nob17, zav19}, it is inefficient at mapping the large areas needed to survey a large sample of (proto-)clusters. The Large Millimeter Telescope (LMT) in Mexico will soon welcome a powerful new instrument, TolTEC, which will simultaneously image the sky at 1.1, 1.4 and 2.1$\,$mm \citep{bry18}. The project is planning to complete four public legacy surveys in 2020--2022, including two extragalactic surveys \citep{pop19}. The Large Scale Structure Survey will cover 50 square degrees over several multi-wavelength data rich fields including the Bo\"{o}tes field. With 5 arcseconds spatial resolution at 1.1$\,$mm, the confusion noise is greatly reduced compared to previous single-dish millimeter surveys. We can apply our total light clusters stacking technique on the TolTEC maps to provide an anchor on the Rayleigh-Jeans tail of the dust distribution and constrain the dust masses. This will allow us to investigate how the total dust mass and radial dust mass distribution evolves in this Bo\"{o}tes cluster sample and compare to other cluster samples mentioned previously in this section.

\subsubsection{Application to other (proto-)cluster samples}
\label{sec:implications_samples}

We have developed techniques to measure and compare the total light from samples of galaxy clusters over a large wavelength range. Our initial application of this technique is presented in this work for a large sample of clusters in the Bo\"{o}tes field, uniformly selected based on photometric over-densities of galaxies. This sample represents an excellent test case given the well studied nature of this cluster sample, including an existing census of the massive dust-obscured galaxies \citep{bro13, alb14, alb16}. Galaxy cluster surveys have long sought to understand the selection effects of various cluster samples from SZ, X-ray, and optical selections, and how a cluster's halo mass is related to its evolutionary state. Our total light stacking technique can be applied to samples of clusters selected in different ways and with different masses to determine if the galaxy populations $-$ including low mass components $-$ have similar radial distributions and obscured star formation properties (see discussions in \S\ref{sec:dust_disc}-\ref{sec:lowmass}). Specifically, we plan to apply our technique to the large samples of clusters from the SPT \citep{sto13, ben14} and MaDCoWS \citep{gon19} surveys. These samples will greatly increase the dynamic range in halo mass analyzed, reducing systematics when tracing the effects of halo mass on total stellar mass and concentration (see discussions in \S~\ref{sec:sims_disc}-\ref{sec:concentration_disc}).  Furthermore, as was done in \citet{kub19}, this technique can extend the wavelength baseline into the near-IR for higher redshift proto-cluster samples such as those selected from the {\it Planck} survey \citep{Planck2014, Planck15}. The results of these additional analyses will be presented in future papers.

\section{Conclusions} \label{sec:conc}

``Total light" cluster stacking $-$ the (background subtracted) summation of all light in a sample of clusters without regard to previously identified constituents $-$ provides a look into cluster evolution which is complementary to the ongoing efforts which identify and study in detail (individually detected) sub-populations in small cluster samples.  ``Total light" stacking provides an averaged view, marginalizing over sources of cluster-to-cluster variation \citep[i.e.][]{gea06, alb16} such as sub-structure \citep{gao08}, while being significantly less sensitive to detection limits and the need for expensive spectroscopy.  Specifically, this stacking technique currently provides the {\it only} method of analysing the lower mass, lower luminosity cluster components for large samples of (proto-)clusters; large samples being necessary to span a range of redshifts, halo masses, and dynamical states.

In this work, we presented our ``total light" stacking techniques for multiple wavelength regimes: the near- and mid-infrared (via {\it WISE} and {\it Spitzer}) as well as the far-infrared (via {\it Herschel}).  We applied our techniques to a well-studied cluster sample at $0.5<z<1.6$ in the Bo\"{o}tes field, using existing photometric redshifts and characterizations of the massive cluster population to check the robustness of our stacking and look at cluster populations over a range of redshifts.

Our main conclusions are as follows:

\begin{enumerate}
    \item ``Total light" stacking is a powerful tool for recovering extended cluster emission from the near- to far-IR even in the case of minimal ancillary data.  Through existing large surveys, particularly all-sky coverage with {\it WISE}, these techniques are applicable to a much wider range of existing and future cluster surveys.  For datasets where a significant fraction of sources are above the detection limit (i.e. the short wavebands of {\it WISE} and {\it Spitzer}), ``total light" stacking provides a lower limit on the total emission, which can then be corrected given cluster membership information.  We provide our correction factors for {\it WISE} in Table~\ref{tab:missing_fraction_wise}, which can be used as a guideline for other cluster samples given the uniformity of the all-sky {\it WISE} coverage.
    \item We measure the near-IR stellar light profiles, a direct proxy for the stellar mass distribution, using {\it WISE} and {\it Spitzer} for our clusters in four redshift bins spanning $z=0.5$ to $z=1.6$.  The near-IR profiles drop off steeply, with the majority of the stellar emission enclosed in the virial radius ($R_{200}\sim1 \,$Mpc). We find good agreement between the overlapping {\it WISE} and {\it Spitzer} filters.  After applying appropriate corrections for centroiding uncertainties and beamsizes, we find that the near-IR profiles are well described by an NFW profile with a high concentration factor at all redshifts.  We speculate that the uniform halo mass in our cluster sample is a driving factor behind this uniform concentration; however, a comparison with the literature highlights that the effect of baryons on the concentration is likely more complicated than a simple halo mass scaling relation.
    \item Stacking in the far-IR reveals extended dust emission from (observed) 250--500$\,\mu$m with a profile that is remarkably similar to the stellar mass distribution except for the inner core ($\lesssim0.3\,$Mpc), where the total dust emission is suppressed at all redshifts.  
    \item The ``total light" SED can be well described by a SFG model with a  dearth of warm dust relative to massive field galaxies at $z\lesssim1.3$, with a slightly warmer SED in our highest redshift bin.  This is in general consistent with the SED being dominated by low mass, low luminosity galaxies which are are found in the field to have colder SEDs than their massive cousins.  No intracluster dust is required to model the ``total light" SEDs; however, a minor contribution from ICD cannot be ruled out.
    \item SSFRs derived from the ``total light" SEDs are found to evolve strongly with redshift, mirroring the evolution of the massive cluster members found in previous studies for this sample \citep{bro13, alb14, alb16}.  In our highest redshift bin, this SSFR is comparable to that expected in the field.  A direct comparison of the IR luminosity density from massive (log $M_{\star}/\Msun\gtrsim10$) galaxies to the ``total light"  shows that massive galaxies make up a minority ($20-30\%$) of the total cluster emission, consistent with the relatively cold SED.  In particular, intermediate mass (log $M_{\star}/\Msun=10-11$) cluster galaxies contribute significantly less to the total IR light than is predicted for field galaxies.  We discuss the possible origins of this difference, including quenching and excess dust in cluster environments.
    \item The total stellar mass measured in our four redshift bins is flat with redshift.  This is expected given the nature of our cluster sample, which, due to the selection technique, has a uniform halo mass at the different epochs explored.  We compare the total stellar mass to halo mass ratio of our sample to predictions from semi-analytic models \citep{guo13, hen15} and $N$-body simulations \citep{spr05}.  We find good agreement with the predicted $M_{\star}/M_{\rm halo}$ value of $1-3$\%.
\end{enumerate}

``Total light" stacking can be extended to other wavelength regimes.  The dominance of the low mass cluster galaxy population in the total dust emission measured in this work may imply that a substantial portion of the cluster star formation is unobscured and thus best measured in the UV.  Complementary to this, large (sub-)millimeter imaging surveys will more directly probe the total cluster dust mass, enabling a comparison of the dust properties in clusters vs field and providing a proxy for gas content.  Expanding the wavelength regimes covered and applying these stacking techniques to additional (proto-)cluster samples will be presented in future work.

\section*{Acknowledgements}

The authors thank Denis Burgarella for advice on {\tt CIGALE}, Carlotta Gruppioni for discussions about her IR luminosity functions, and Matthieu Bethermin for discussion about the SIDES mock catalog.  The authors additionally thank George Rieke, Mattia Vaccari, and David Shupe for insight into the B\"{o}otes MIPS imaging and Aaron Meisner, Dustin Lang, and Ned Wright for discussion regarding the WISE imaging. The authors acknowledge financial support from NASA
through the Astrophysics Data Analysis Program (ADAP), grant number 80NSSC19K0582. 
This publication makes use of data products from the {\it Wide-field Infrared Survey Explorer}, which is a joint project of the University of California, Los Angeles, and the Jet Propulsion Laboratory/California Institute of Technology, funded by the National Aeronautics and Space Administration. This work is based [in part] on observations made with the Spitzer Space Telescope, which was operated by the Jet Propulsion Laboratory, California Institute of Technology under a contract with NASA.
%Herschel
The {\it Herschel} spacecraft was designed, built, tested, and launched under a contract to ESA managed by the {\it Herschel}/{\it Planck} Project team by an industrial consortium under the overall responsibility of the prime contractor Thales Alenia Space (Cannes), and including Astrium (Friedrichshafen) responsible for the payload module and for system testing at spacecraft level, Thales Alenia Space (Turin) responsible for the service module, and Astrium (Toulouse) responsible for the telescope, with in excess of a hundred subcontractors. This research has made use of the NASA/IPAC Infrared Science Archive, which is funded by the National Aeronautics and Space Administration and operated by the California Institute of Technology.

\section*{Data Availability Statement}

The {\it WISE}, {\it Spitzer}/IRAC, and {\it Herschel} data underlying this article are available in the NASA/IPAC Infrared Science Archive at \url{irsa.ipac.caltech.edu}.  The {\it Spitzer}/MIPS data were provided by the MIPS AGN and Galaxy Extragalactic Survey (MAGES) team; access may be requested through the corresponding author and granted with permission from the MAGES team. Other data products generated in this article will be shared on reasonable request to the corresponding author.

%%%%%%%%%%%%%%%%%%%%%%%%%%%%%%%%%%%%%%%%%%%%%%%%%%



\bibliographystyle{mnras}
\bibliography{refs} % Entries are in the "refs.bib" file

%\begin{thebibliography}{99}
%\bibitem[Alberts et al.(2014)]{2014MNRAS.437..437A} Alberts, S., Pope, A., Brodwin, M., et al.\ %2014, \mnras, 437, 437 
%\bibitem[Ashby et al.(2009)]{2009ApJ...701..428A} Ashby, M.~L.~N., Stern, D., Brodwin, M., et %al.\ 2009, \apj, 701, 428 
%\bibitem[Lang(2014)]{2014AJ....147..108L} Lang, D.\ 2014, \aj, 147, 108 
%\bibitem[Meisner et al.(2017)]{2017AJ....154..161M} Meisner, A.~M., Lang, D., \& Schlegel, D.~J.\ %2017, \aj, 154, 161 
%\bibitem[Vaccari(2015)]{2015fers.confE..27V} Vaccari, M.\ 2015, The Many Facets of Extragalactic Radio Surveys: Towards New Scientific Challenges, 27 
%\bibitem[Wright et al.(2010)]{2010AJ....140.1868W} Wright, E.~L., Eisenhardt, P.~R.~M., Mainzer, %A.~K., et al.\ 2010, \aj, 140, 1868 
%\bibitem[Xue et al.(2017)]{2017ApJ...837..172X} Xue, R., Lee, K.-S., Dey, A., et al.\ 2017, \apj, %837, 172 

%\end{thebibliography}

%\appendix % if needed

\appendix 
\renewcommand\thesection{\Alph{section}}
\renewcommand{\thefigure}{\thesection\arabic{figure}}

\section{Comparison of Different Stacking Methods for {\it WISE}}\label{appendix_a}

We give a full description of the various tests we have conducted to quantify how different image processing procedures and stacking techniques can impact the stacked flux measurements. The same behavior is observed in all bands, and here, we illustrate our results in the $W2$ band. 

\begin{figure*}
	\includegraphics[width=7in]{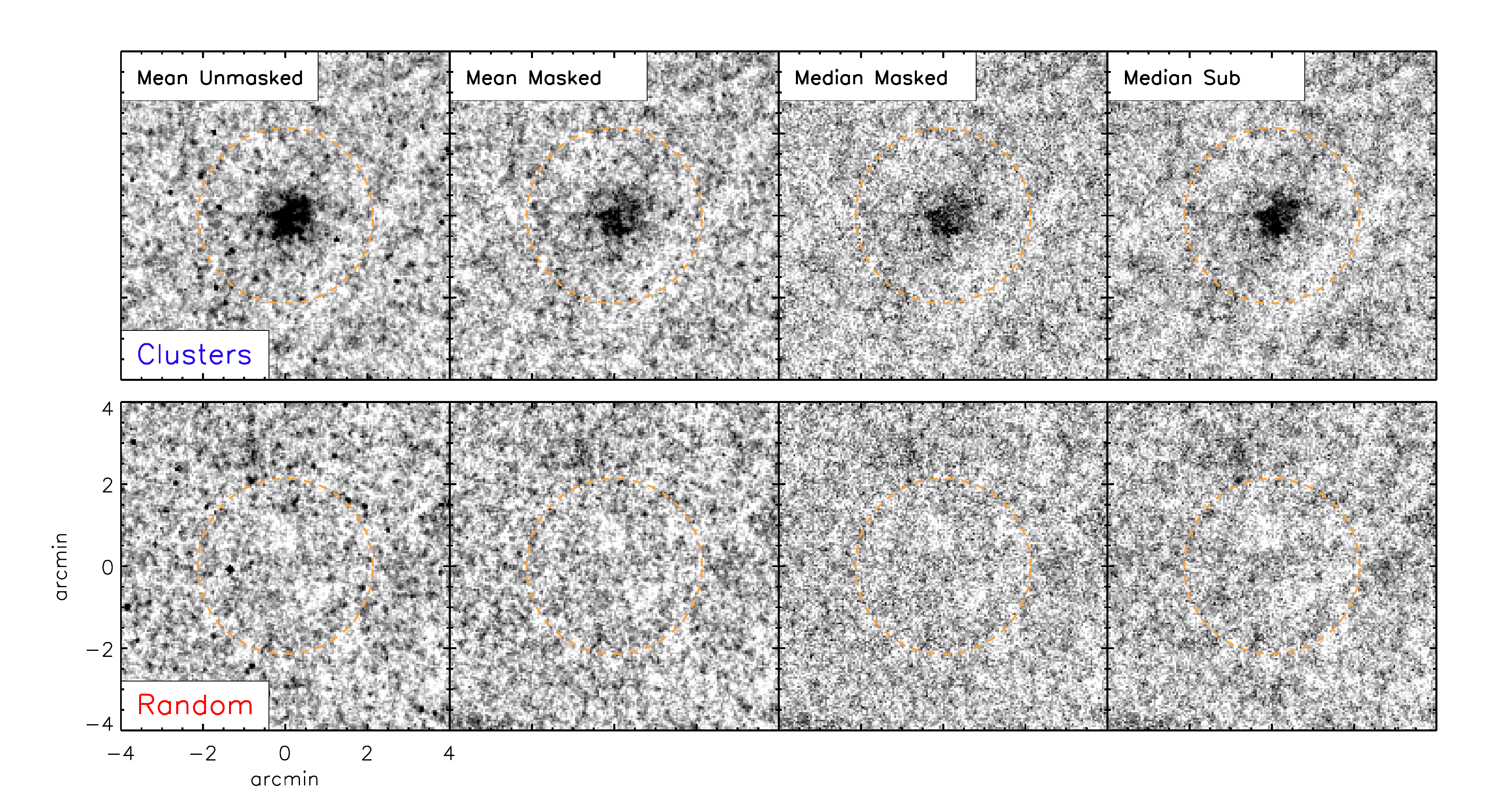}
    \caption{
    Results of our {\it WISE} $W2$ stacking analyses using the $z=0.7-1.0$ sample are shown to illustrate the three stacking techniques we employ in this work as described in \S~\ref{subsec:wise_spitzer_processing} and \S~\ref{subsec:wise_spitzer_stacking}. Top and bottom panels show the results of each method, labelled atop, performed on the cluster positions (top) and random positions (bottom). 
    Each image is 8\arcmin\ on a side. In each panel, we show a circle of  1~Mpc radius at $z=0.86$ as a dashed line. 
    }
    \label{fig:wise_imagestack}
\end{figure*}

In Figure~\ref{fig:wise_imagestack}, we show four different image stacks of the {\it WISE} W2 band on the cluster and off-cluster positions, all of which are performed on the $z=0.7-1.0$ sample. The leftmost panels are {\tt mean\_unmasked} stacks, which are created by taking a pixel-wise mean of the {\tt \_unmasked} images. The second left column shows the {\tt mean\_masked} stacks where we mask out all sources present in the image. The third and fourth columns show the {\tt median\_masked} and {\tt median\_sub} images, respectively. The latter use the {\tt \_sub} images which are simply sky-subtracted images where all sources are present therein.

In Table~\ref{table1_appendix}, we list the aperture fluxes measured from these images. We also show the sky properties measured in the cluster and off-cluster stack. For the latter, we list the mean value after 500 random measurements. We do not give the spread of the $\sigma_{\rm p}$ because it is consistent with zero in the adopted precision. It is evident that in all but the {\tt mean\_unmasked} stack the measured fluxes are consistent with one another within 15\%.  The fluxes at large radii are somewhat sensitive to small changes in the assumed sky level. In particular, the agreement of the {\tt median\_sub} fluxes with the others is remarkable considering that the source density in the former is much higher than that in the latter. As noted previously (see \S~\ref{sec:understanding}),  the {\tt mean\_unmasked} fluxes are expected to be higher than other versions as it should represent the {\it total} flux including those contributed by the photo-$z$ cluster members.

\begin{table*}
\vspace{1in}
\begin{center}
\begin{minipage}{12in}
\scriptsize
\caption{$z$=0.7--1.0 clusters:  $W2$-band fluxes and noise properties of four different image stacks }
\begin{tabular}{ccccc}
\hline
\hline
Radius & {\tt mean\_unmasked} & {\tt mean\_masked} & {\tt median\_masked} & {\tt median\_sub} \\
%Aperture & Masked   & Repaired  & Masked  & Masked   & Repaired  & Masked  & Masked   & Repaired  & Masked  & Masked   & Repaired  & Masked \\
%radius &  median  &  median &  mean &  median  &  median &  mean &  median  &  median &  mean &  median  &  median &  mean\\

[\arcsec] &  [$\mu$Jy]  &  [$\mu$Jy]  &  [$\mu$Jy]  &  [$\mu$Jy]    \\
\hline
%sky (cluster)& 0.247 &0.209  & 0.111 & 0.209   \\
%sky (off-cluster) & $0.229 \pm 0.002$ & $0.201\pm 0.002$ & $0.201 \pm 0.001$ & $0.201 \pm 0.001$ \\
$\sigma_{\rm p}$ (cluster) & 0.148 & 0.130 & 0.140 & 0.147 \\
$\sigma_{\rm p}$ (off-cluster) & 0.149 & 0.131 & 0.131 & 0.131\\
\hline
   5.50 &       14.0 (  0.5)	 &        8.1 (  0.5)	 &        8.2 (  0.5)	 &       10.3 (  0.5)\\ 
  11.00 &       47.8 (  1.0)	 &       27.1 (  0.9)	 &       26.4 (  1.0)	 &       36.8 (  1.0)\\ 
  16.50 &       92.9 (  1.6)	 &       52.1 (  1.4)	 &       54.5 (  1.5)	 &       70.1 (  1.5)\\ 
  22.00 &      143.7 (  2.1)	 &       85.5 (  1.8)	 &       84.6 (  2.0)	 &      111.6 (  2.0)\\ 
  27.50 &      188.5 (  2.6)	 &      118.4 (  2.3)	 &      115.8 (  2.5)	 &      149.5 (  2.5)\\ 
  33.00 &      229.2 (  3.1)	 &      146.9 (  2.7)	 &      145.9 (  2.9)	 &      185.4 (  3.1)\\ 
  38.50 &      267.9 (  3.7)	 &      181.9 (  3.2)	 &      171.8 (  3.4)	 &      215.2 (  3.6)\\ 
  44.00 &      303.9 (  4.2)	 &      209.0 (  3.7)	 &      192.2 (  3.9)	 &      235.0 (  4.1)\\ 
  49.50 &      323.9 (  4.7)	 &      224.6 (  4.1)	 &      206.5 (  4.4)	 &      245.7 (  4.6)\\ 
  55.00 &      337.5 (  5.2)	 &      238.9 (  4.6)	 &      212.8 (  4.9)	 &      249.3 (  5.1)\\ 
  60.50 &      347.0 (  5.8)	 &      249.6 (  5.0)	 &      218.7 (  5.4)	 &      253.5 (  5.6)\\ 
  66.00 &      360.5 (  6.3)	 &      266.2 (  5.5)	 &      229.1 (  5.9)	 &      258.5 (  6.1)\\ 
  71.50 &      376.8 (  6.8)	 &      281.6 (  5.9)	 &      237.6 (  6.4)	 &      260.0 (  6.6)\\ 
  77.00 &      388.1 (  7.3)	 &      287.5 (  6.4)	 &      245.1 (  6.9)	 &      260.0 (  7.1)\\ 
  82.50 &      395.0 (  7.9)	 &      290.7 (  6.9)	 &      249.9 (  7.4)	 &      261.7 (  7.6)\\ 
  88.00 &      400.9 (  8.4)	 &      299.1 (  7.3)	 &      258.6 (  7.8)	 &      266.6 (  8.1)\\ 
  93.50 &      407.6 (  8.9)	 &      311.5 (  7.8)	 &      260.8 (  8.3)	 &      264.1 (  8.7)\\ 
  99.00 &      412.1 (  9.4)	 &      324.2 (  8.2)	 &      258.1 (  8.8)	 &      257.3 (  9.2)\\ 
 104.50 &      416.1 ( 10.0)	 &      330.1 (  8.7)	 &      266.2 (  9.3)	 &      257.4 (  9.7)\\ 
 110.00 &      417.5 ( 10.5)	 &      333.8 (  9.1)	 &      265.4 (  9.8)	 &      259.1 ( 10.2)\\ 
 115.50 &      425.3 ( 11.0)	 &      329.0 (  9.6)	 &      273.6 ( 10.3)	 &      263.2 ( 10.7)\\ 
 \hline
\end{tabular}
\label{table1_appendix}
\end{minipage}
\end{center}
\end{table*}

\section{Comparison of {\it WISE} vs {\it Spitzer} Results}\label{appendix_wise_v_irac}

As listed in Table~\ref{datatable}, the imaging depths and pixel scales of overlapping {\it Spitzer} and {\it WISE} filters differ substantially. Nevertheless, the agreement between various measurements from these datasets is impressive, showcasing the prospect of utilizing all-sky {\it WISE} coverage  to constrain faint infrared emission from cluster galaxies. 

\begin{figure*}
	\includegraphics[width=6.3in]{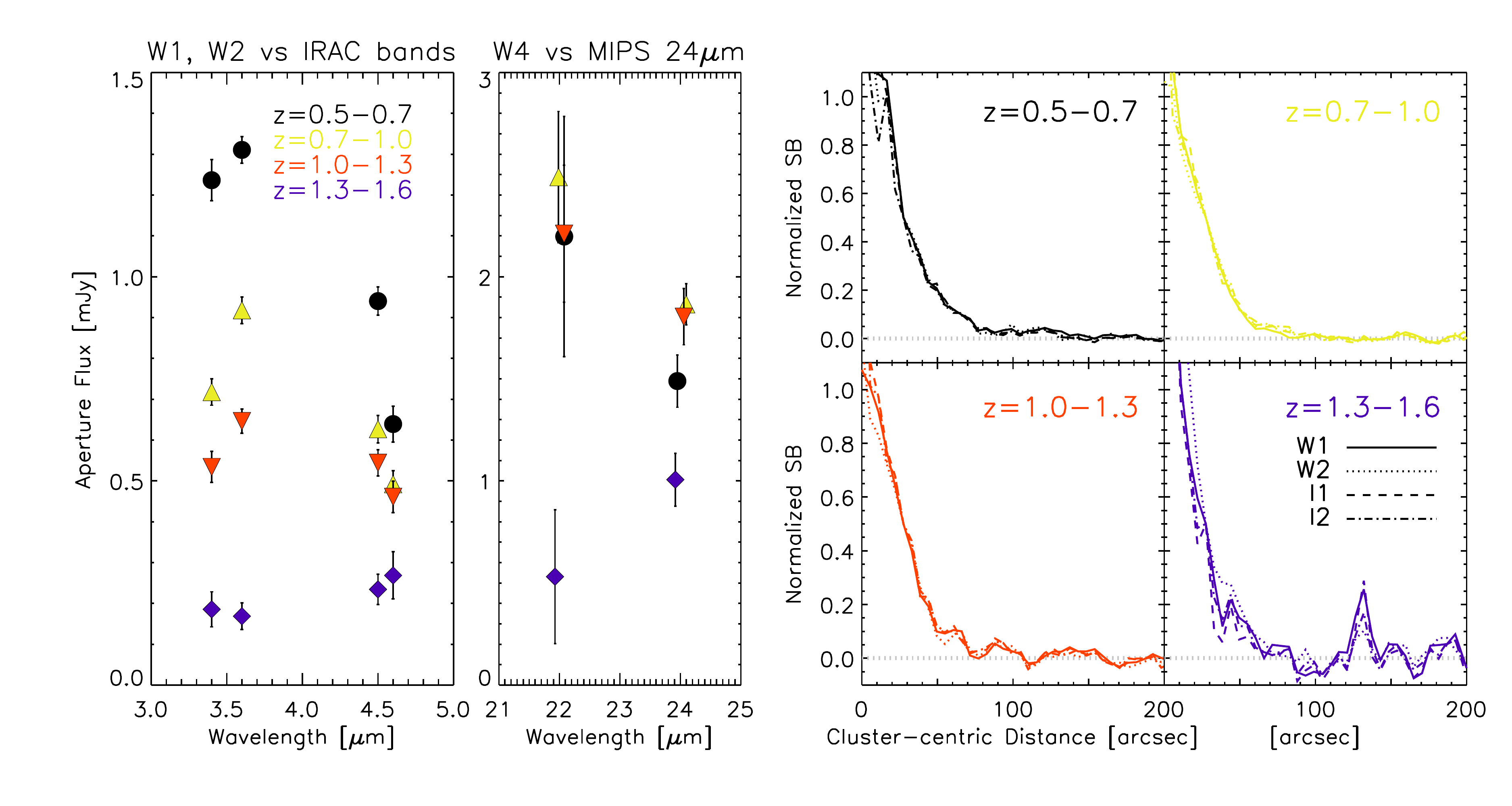}
    \caption{
   Fluxes measured in 100\arcsec\ apertures in the WISE $W1$, $W2$, $W3$, IRAC 3.6$\,\mu$m and 4.5$\,\mu$m, and MIPS 24$\mu$m bands. Different color symbols denote four redshift bins in which the measurements are made. Whenever applicable, the wavelengths are shifted slightly for clarity. {\it Left:} In comparing $W1$ vs 3.6$\,\mu$m and $W2$ vs 4.5$\,\mu$m, their flux measurements agree with each other within 15\% in all redshift bins. {\it Middle:} the measured flux in the $W4$ band is also in good agreement with the MIPS 24$\,\mu$m band, with a slight tendency toward a higher total flux.
    Possible reasons for this minor disagreement are discussed in Appendix~\ref{appendix_wise_v_irac} and \S~\ref{sec:sedfitting}. {\it Right:} Normalized surface brightness measured in the  WISE $W1$, $W2$, $W3$, IRAC 3.6$\,\mu$m and 4.5$\,\mu$m bands are shown for each redshift bin. All are normalized such that the value is 0.5 at 27.5\arcsec (10 native WISE pixels). Despite the differences in angular resolution, the agreement is excellent at nearly all scales.   
   }
    \label{fig:compare_wise_spitzer}
\end{figure*}

In the left panel of Figure~\ref{fig:compare_wise_spitzer}, we compare the fluxes measured in the {\it WISE} $W1$/$W2$ bands and IRAC 3.6/4.5$\,\mu$m bands within 100\arcsec\ radius apertures. Different colors represent four redshift bins as defined in Table~\ref{tab:sample}. The mean deviation of the {\it WISE} fluxes relative to their IRAC counterparts is 15\%; smaller aperture sizes improve the agreement by a few per cent. The largest discrepancy (30\%) occurs in the lowest redshift bin between the 4.5$\,\mu$m and $W2$ bands where both bands sample the steeply declining end of the stellar emission ($\lambda_{\rm rest}\approx 2.7-2.8~\mu$m). 

The flux errors as shown in Figure~\ref{fig:compare_wise_spitzer} fully account for the uncertainties from Poisson noise and from large-scale fluctuation of sky background  (see \S~\ref{wise_spitzer_radial_profile}). However, there may be two systematic uncertainties unaccounted for in our error estimate. First, the imaging sensitivity, to a large extent, determines the surface density of sources within a given image including numerous faint sources. Together with the source detection setting (i.e., how aggressively we reject them by masking in the pre-stack image processing), these factors can modulate the precision of sky background determination in a way that has yet to be explored. We plan to investigate this issue thoroughly in the future by comparing the results from different detection settings as well as through simulated images spanning a range of added sky rms noise. 

In the right panel of Figure~\ref{fig:compare_wise_spitzer}, we show the IRAC and WISE-measured radial profiles in all four redshift bins. Evidently, the agreement is excellent in all but the smallest scales where the effects of PSF sizes and angular resolution are expected.

The agreement  between the {\it WISE} $W4$  and {\it Spitzer} MIPS 24$\,\mu$m bands is also reasonable, as illustrated in the middle panel of Figure~\ref{fig:compare_wise_spitzer}.
%Figure~\ref{fig:compare_wise_spitzer} additionally shows the discrepancy.  
As discussed in \S~\ref{sec:sedfitting}, these bands are wide ($\Delta \lambda$=4--5$\mu$m) and probe a complex portion of the spectrum including several prominent PAH and absorption features. Variations of flux densities up to a factor of a few are expected in individual clusters within a given redshift bin, which certainly contribute to the difference.  In addition, a systematic bias may play a role. The 24$\mu$m flux is always  lower than the corresponding $W4$ flux, with the exception of the highest redshift bin where the detection significance is low.

At the wavelength range which the MIPS instrument samples,  any data reduction process necessarily includes a more aggressive spatial filtering  than is customary at optical and near-IR wavelengths. Median filtering is applied to homogenize the sky background, thereby optimizing for point source detection. The same process unfortunately can eliminate low-level diffuse signals such as that originating from cluster light, leading to a systematic underestimation of both surface brightness and total flux. While such flux loss may be corrected for (through simulations or otherwise) provided that the precise detail of the filtering scheme is known, we wish to stress that spatial filtering can  have a nonnegligible impact on both flux and SB measurements. 

In this work, we have analyzed two different versions of the MAGES data: the official data product and that released as part of the Spitzer Data Fusion project. The latter used a median filter (box with 110\arcsec\ on a side: D. Shupe, in private communication), much smaller than the former. The identical procedure is followed for both to create image stacks and make flux and SB measurements. We find substantial difference: the use of the aggressively filtered data  leads to the smaller angular size of the signal (SB falls to the field level at 50\arcsec\ compared to 70\arcsec-80\arcsec) and significantly lower fluxes (by $\approx$60\%). 

Our analysis shows that the effect is less dramatic ($\approx$15--20\% in total flux densities) at 70$\mu$m. Our result is in line with the expectation that the larger filter size used for the 70$\mu$m data to accommodate its much larger beamsize (Table~\ref{datatable}) leads to a less exacerbating effect on altering the faint cluster signal. 

All in all, our results demonstrate the impressive agreement between the WISE, IRAC, and MIPS measurements, showcasing the potential of the WISE all-sky data to further elucidate the stellar population and star formation activities in galaxy clusters.

\section{Testing {\it Herschel}/SPIRE Stacking}\label{appendix_spire}

\begin{figure}
    \centering
    \includegraphics[width=\columnwidth]{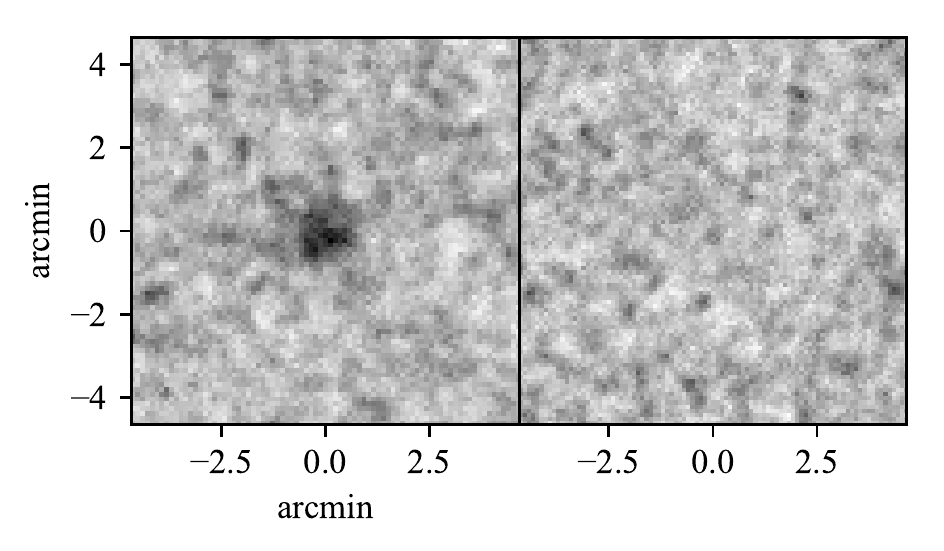}
    \caption{``Total light" stacking on the SPIRE 250$\,\mu$m map centered on 53 $1<z<1.3$ clusters (left) and on 53 random positions offset from the clusters (right).
    }
    \label{fig:spire_background}
\end{figure}

In this appendix, we describe the checks and simulations used to verify the far-infrared stacking techniques used in this work.  Stacking performed on {\it Herschel}/SPIRE maps which are calibrated to have a zero mean should require no additional background noise subtraction or correction for the contribution from the field galaxy population.  To verify this, we stacked 53 random cutouts placed on positions offset from any known cluster. This background stack is compared to a stack of 53 clusters in Figure~\ref{fig:spire_background} and is dominated by (confusion) noise.  An appropriate aperture (125\arcsec or $\sim1\,$Mpc at $z\sim 1$) recovers a value of zero within the measured uncertainty, indicating that the field population has been correctly ``zeroed" out.

As a further test, we create simulated maps by injecting artificial clusters composed of point sources arrayed in a normal distribution with a FWHM$\,=\,600$ kpc, approximately representing the radial profiles of our real clusters (\S~\ref{sec:spirephot}), into a blank map and our real SPIRE 250$\,\mu$m map. The point sources have a uniform, known flux of 3.3 mJy, the typical flux of a massive cluster galaxy \citep{alb14}, below the confusion limit \citep[5.8 mJy, $1\sigma$, at 250$\,\mu$m][]{ngu10}.  We then apply our ``total light" stacking technique (\S~\ref{sec:spirestacking}) to our artificial clusters.  The stacking performs the same on the blank and real maps, confirming that our cluster signal is not being boosted by the field population.  For the artificial clusters injected in the real map, we recover the total stacked flux within $10-15\%$, within the real bootstrapped uncertainties that are driven by instrument noise, confusion noise, and population variance.

As discussed in \S~\ref{sec:spirephot}, when centroiding the real cluster stacks, we found a systematic pixel offsets of $\sim1-2$ pixels that persisted across different cluster sub-sets.  Injecting and stacking fake hot pixels into the real maps was used to verify that this offset was not artificially introduced during stacking. A chance offset driven by a few bright, outlier clusters is ruled out by randomly rotating the cutouts, which eliminates the centroiding offset.  This suggests a systematic in the data, perhaps driven by the scan pattern.  To be conservative, we randomly rotate all cutouts during stacking.

In \S~\ref{sec:individual}, we discuss an update to the baseline correction first presented in \citet{alb14}.  In this update, pixels previously chosen at random are now selected from a two dimensional normal distribution with a FWHM$\,=\,600$ kpc, a representative approximation of the extent of the clusters in this study.  We test this updated procedure on the simulated maps described above, stacking this time on the known positions of fake cluster members injected into the real map.  This test confirms the that updated baseline correction more accurately corrects for flux boosting and clustering signal due to the crowded cluster field and large SPIRE beam out to a radius of $\sim1\,$Mpc, recovering the average flux of the fake cluster sources within the uncertainties. The impact of this update is minimal at $0.5\,$Mpc (up to a factor of $\sim1.3$), but increasingly more important when look at cluster galaxies at larger radii (up to a factor of $\sim2.6$ at $1\,$Mpc).

%%%%%%%%%%%%%%%%%%%%%%%%%%%%%%%%%%%%%%%%%%%%%%%%%%
% Don't change these lines
\bsp	% typesetting comment
\label{lastpage}
\end{document}